\documentclass[fleqn]{2017SCGE}
\usepackage{graphicx,epsfig,latexsym,overpic,amssymb,color}
\usepackage{bm}
\usepackage{ctex}
\usepackage[normalem]{ulem}

\begin{document}


\ensubject{subject}

\ArticleType{Review}
\Year{2023}
\Month{xxx}
\Vol{XX}
\No{X}
\DOI{xxx/xxx}
\ArtNo{000000}
\ReceiveDate{xxx, 2023}
\AcceptDate{xxx, 2023}
\title{Advancing Nuclear Physics with Machine Learning and Artificial Intelligence}

\author[1,2]{Wanbing HE}{hewanbing@fudan.edu.cn}
\author[3,4]{Qingfeng LI}{liqf@huznu.edu.cn}
\author[1,2,5]{Yugang MA}{mayugang@fudan.edu.cn}
\author[6]{Zhongming NIU}{zmniu@ahu.edu.cn}
\author[7,8]{Junchen PEI}{peij@pku.edu.cn}
\author[4,8,9]{\\Yingxun ZHANG}{zhyx@ciae.ac.cn}

\AuthorMark{W. B. He}
\AuthorCitation{W. B. He, Q. F. Li, Y. G. Ma, Z. M. Niu, J. C. Pei, Y. X. Zhang}

\address[1]{Key Laboratory of Nuclear Physics and Ion-beam Application (MOE), Institute of Modern Physics, Fudan University, Shanghai 200433, China}
\address[2]{Shanghai Research Center for Theoretical Nuclear Physics, NSFC and Fudan University, Shanghai 200438, China}
\address[3]{School of Science, Huzhou Normal University, Huzhou 313000, China}
\address[4]{Department of Nuclear Physics, China Institute of Atomic Energy, Beijing 102413,  China}
\address[5]{School of Physics, East China Normal University, Shanghai 200241, China}
\address[6]{School of Physics, Anhui University, Hefei 230601, China}
\address[7]{State Key Laboratory of Nuclear Physics and Technology, School of Physics,
Peking University, Beijing 100871, China}
\address[8]{Southern Center for Nuclear-Science Theory (SCNT), Institute of Modern Physics, Chinese Academy of Sciences, Huizhou 516000,  China}
\address[9]{Guangxi Key Laboratory of Nuclear Physics and Technology, Guangxi Normal University, Guilin 541004, China}


\abstract{Machine learning (ML) and artificial intelligence (AI) are becoming powerful tools in scientific research across various disciplines. In this review, we summarize recent progress in AI-assisted studies of  nuclear structure and reaction observables, heavy-ion collisions and dense nuclear matter properties, many-body wave functions,   experimental facilities and data analysis. 
This review focus on new progress since the last review in 2023, and machine learning in nuclear physics is evolving from purely data inferences  to physics informed learning.
Future directions on  the integration of physical knowledge with modern learning architectures, large foundation models and other emerging methods are also reviewed. These developments suggest that AI is enabling and advancing new approaches towards most challenging nuclear physics problems.
}


\maketitle

\begin{multicols}{2}
\section{Introduction}\label{section1}



\begin{figure*}[t]
	\centering
	\includegraphics[width=0.9\textwidth]{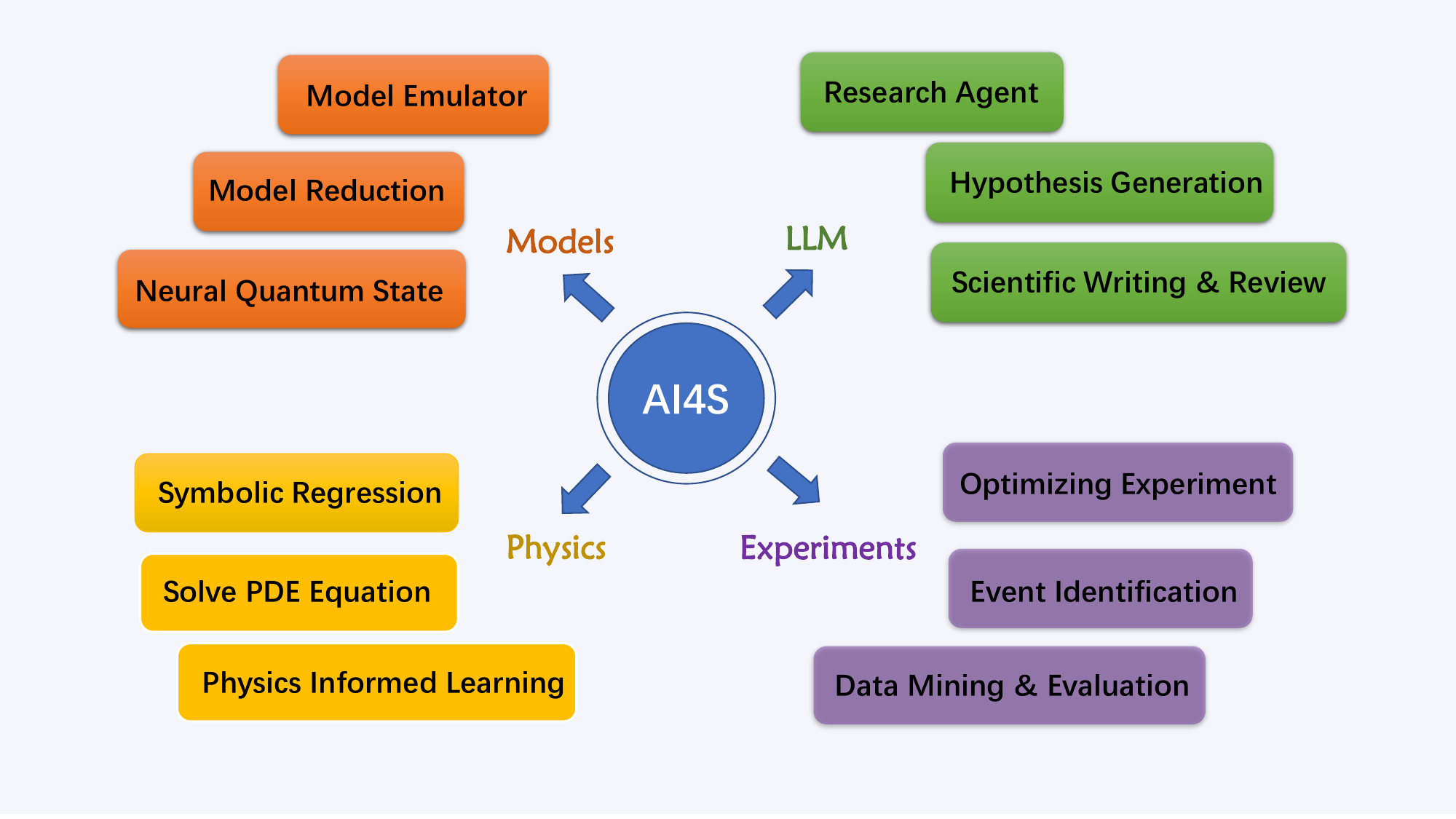}	
	\caption{ The conception of AI4S across various application disciplines, including but not limited to AI-assisted theoretical models, the integration of physics and AI, experimental facilities and data analysis,  and the usage of large language models. }
	\label{Fig1}
\end{figure*}

In the past decade, we have witnessed explosive growth in artificial intelligence (AI)  and machine learning (ML), characterized by significant advances in algorithmic techniques, infrastructures, and applications across various disciplines \cite{book,WangYF_2026}. In particular, the advent of deepseek has greatly boosted the developments of AI in China \cite{deepseek}. It is now more and more convincing that AI and ML will help scientific research in a transformative way. The landmark event is that both Nobel prizes in physics and chemistry were awarded to AI in 2024 
~\cite{Nobel}.
AI for science (AI4S) 
demonstrated rapid progress in  enhancing and accelerating scientific research (see Fig.\ref{Fig1}), helping scientists to analyze and interpret data, emulate costly computations and experiments, and generate insights and hypotheses. 
While AI will not ultimately replace scientists, those who embrace it will gain substantial advantages and new opportunities.

AI refers to a broad concept that simulates human cognitive functions, and ML is the method belonging to AI by learning from data. The applications of ML and AI rely on the powerful representation capabilities of neural networks. 
The information of huge high-dimensional data can be approximately represented by neural networks using a much smaller number of parameters. Subsequently, the interpolation and extrapolation of the trained neural networks can be performed in negligible time, even the training data are incomplete, imperfect, noisy and heterogeneous \cite{heterogeneous}. The logic behind physics models lies in their ability to capture universal principles through a concise set of equations. In practice, the real process or systems are often too complex involving multi-scale couplings, or tremendous degrees of freedom, or underlying correlations, so that conventional numerical methods might fail to solve such problems. Therefore, ML and AI provide a new paradigm to tackle these complex problems. 

As well, the applications of ML and AI in nuclear physics have grown rapidly~\cite{RevModPhys.94.031003,He2023,MaYugang2022}. For example, there have been applications in the inferences of nuclear structure and reactions observables where experiments are not accessible, solving ill-posed inverse problems, emulators of costly theoretical models and experiments, efficient solutions of ab initio many-body problems, and analysis of experimental data and optimizing experimental design. 
In particular, in addition to studies on traditional low-energy nuclear structure observables, recent developments have been made in high-energy heavy-ion collisions, where observables in the final-state momentum space are used to infer the initial-state structure of atomic nuclei (see, e.g., Refs.~\cite{ZhangS_2017,Ma_2023,HeJJ_2021,STAr_2024,JiaNST,GiacaloneNST,SchenkeeNST,FangNST}, which has greatly enriched our understanding of nuclear structure.
There are an increasing number of ML methods employed including various neural networks
~\cite{transformer,GNN,Hashemi2024}
the regression and classification methods
~\cite{EOSclassfication,neutronHaloClass,regression},
the gaussian process
~\cite{Yuksel2024PRC,Ye2025PRC},
decision trees~\cite{Cai2023,Cubist,CSDT}, the transfer learning
~\cite{CSTL,TLdeformation},
the reinforcement learning~\cite{Radaideh2025MultistepCS,Kaiser2024}, etc. Meanwhile, there are still some significant questions to be addressed in nuclear physics \cite{MaYG_arxiv}.
For example, studies of dense nuclear matter and short range correlations are crucial to understand heavy-ion collision mechanism~\cite{Yong2017PRC}, nuclear quark effects and neutron stars~\cite{SHANG2025139976, ZHEN2025139350}. The synthesis of superheavy new elements is extremely challenging, while theoretical inferences have large uncertainties~\cite{QIANG2024139057, zhangmh}. 
Ab initio calculations of heavy nuclei are notoriously difficult, but are important for probing underlying fundamental symmetries
and astrophysical scenarios~\cite{ye2025physics,Jiang:2026gci,Jin:2025dvf,Ma:2025nex,Ma:2025ulw}. The supply of high quality nuclear data is essential for sustainable nuclear energy production using new types of fuel, and more accurate theories and experiments in nuclear fission are anticipated~\cite{wangza1,Shang2025Nst}. It is expected that these relevant theoretical and experimental studies can be greatly improved by leveraging the powerful AI and ML techniques.

Since the previous review of machine learning in nuclear physics~\cite{RevModPhys.94.031003,He2023,MaYugang2022}, there have been many new publications in this field. In particular, scholars in China are very active and contribute a lot to this disciplinary research. These publications are evolving from simple data inferences in the early days to improved performances by integrating with physics information. This means that in reality, ML can indeed help and advance nuclear physics research. 
Since 2023, there are already numerous exciting progress in machine learning for nuclear physics.
We find it urgently needed to publish a timely review of the latest developments in machine learning and AI in nuclear physics to further boost this field. 

In recent years, new techniques such as the large language models (LLM), symbolic regression (SR), Kolmogorov-Arnold networks (KAN), physics-informed machine learning, AI-based partial differential equation (PDE) solvers, transferable representations and foundation models, and quantum machine learning (QML) have attracted strong interest. The LLM can generate emergent knowledge when the model is sufficiently large~\cite{wei2022emergentabilitieslargelanguage}. The SR can reveal new formulas and equations from data~\cite{AbdusSalam:2024obf}. While the KAN is different from conventional neural networks and is promising towards inherently interpretable AI~\cite{liu2025kankolmogorovarnoldnetworks}. The physics-informed ML with physics-informed priors, or physics-constrained loss functions, physics-guided feature inputs, or physics-aligner can avoid the drawbacks of purely data-driven ML~\cite{PIML2021}, since it can incorporate comprehensive physics conservation laws and quantum effects, and is particularly suitable to exploit the maximum values of sparse nuclear data. 
AI-based PDE solvers, including physics-informed neural networks and operator-learning methods, can accelerate the solution of nuclear structure and reaction equations~\cite{Wen2025,Lu2021DeepONet,Li2021FNO}. Transferable representations and foundation models aim to reuse learned knowledge across nuclear observables, simulations, and experiments~\cite{park2026fmnpp,wiesner2026physicsfoundationmodel}. QML uses quantum algorithms and quantum hardware for potential applications in nuclear calculations and data analysis~\cite{Wang:2026vti,Zhang:2025rzl,Fang:2024ple}.

This paper is organized as follows. Section~\ref{Applications} reviews recent applications of ML and AI in nuclear physics. 
Section~\ref{ML-methodology} discusses emerging trends in AI and ML for nuclear physics. Finally, a summary and outlook are given in Section~\ref{summary}, highlighting both the opportunities and challenges of AI-driven nuclear physics research.

\section{Recent applications of ML and AI in nuclear physics}
\label{Applications}
\subsection{Nuclear properties}
\subsubsection{Nuclear structure properties}

The neural network method was first applied to nuclear structure studies in Ref.~\cite{Gazula1992NPA} to explore whether the ML can learn correlations in nuclear data, make predictions, or extract new physical insights. Since then, many ML methods have been used to investigate various nuclear structure properties, including nuclear masses, charge radii, and half-lives~\cite{He2023, RevModPhys.94.031003, Bedaque2021EPJA}. Compared with earlier applications, recent studies since 2023, have mainly focused on incorporating more physical information into ML methods, thereby improving their predictive accuracy and extrapolation ability~\cite{Huang2025PRC}. The extrapolation ability of ML methods is usually tested using the data excluded from the training set or newly measured data. However, such tests are often relatively easy for most ML methods, since the testing data are usually close to the training region. A long-range extrapolation test can be performed by comparing the $r$-process simulations based on ML predictions with the solar $r$-process abundances, as in Ref.~\cite{Li2024PLB}.

About 2500 nuclear masses have been measured to date~\cite{Wang2021CPC, Qu2025NST}, and the associated uncertainties are steadily improving as well, which constitutes the data basis for the studies of nuclear masses with ML approaches. In recent years, many ML approaches have been employed to describe nuclear masses, including the neural networks, kernel regression, tree-based methods, and symbolic regression. The artificial neural network (ANN) has strong expressive power and its predictive performance can be further improved by introducing physical input features
~\cite{Huang2025PRC, Niu2018PLB, Le2023NPA, Zeng2024PRC, Dai2025CPC, ChoiS2026PRC},
optimized activation functions~\cite{Kim2026PRC, Liu2026PRC}, the constraints from the Garvey-Kelson (GK) constraints~\cite{Mumpower2022PRCL} and the nucleon separation energies in the loss function~\cite{Wang2026APS}. Moreover, the Bayesian neural networks (BNN)~\cite{Niu2022PRCL, Qu2025CPC}, complex-valued product-unit networks (CPUN)~\cite{Dellen2024PLB}, mixture density networks~\cite{Lovell2022PRC}, and convolutional neural networks (CNN)~\cite{Lu2025PRC}, have also been applied to nuclear mass predictions. Recent developments of kernel regression methods include the improvements of radial basis function (RBF) based on mirror-nuclei symmetry~\cite{LiT2026PLB, LiT2026CPCEnergy}, kernel ridge regression (KRR) applications to RCHB mass predictions~\cite{Wu2024PRC, Guo2024PRC}, development of anisotropic KRR (AKRR)~\cite{Wu2024PRC1, Tian2025PRC}, mass studies with support vector regression (SVR) and Gaussian process (GP) ~\cite{Yuksel2024PRC, Yuan2024NST, Ye2025PRC, Jalili2025EPJA, HuangWJ2026NST, Ye2026PRC}. The tree-based methods, including random forest (RF), gradient boosted decision trees (GBDT), extreme gradient boosting (XGBoost), light gradient boosting machine (LightGBM), and categorical boosting (CatBoost), have also been used to refine nuclear mass models~\cite{LiuGP2025PRC, Guo2025PRC, ZhangXY2026CPC}. The symbolic regression (SR) can provide interpretable analytical expressions that reveal underlying patterns in the data. The Kolmogorov-Arnold network (KAN) as a simplified SR approach has been applied to nuclear mass predictions, which yields an analytical binding-energy expression consistent with the classical liquid drop model~\cite{LiuH2025PRC}. The model averaging methods, including the principal component analysis (PCA)~\cite{Wu2024SCPMA, Wu2026PLB}, ensemble Bayesian model averaging (EBMA)~\cite{Saito2024PRC}, naive Bayesian model averaging (NBMA)~\cite{Zhang2024NPA}, and power-moderated mean (PMM)~\cite{Zhang2024PRC}, can combine the advantages of different mass models to improve nuclear mass predictions. 

\begin{table*}[htbp]
  \centering
  \scriptsize
  \setlength{\tabcolsep}{3pt}
  \caption{Comparison of ML methods for nuclear mass predictions. The ``AME20\#: $80\%$" in the learning dataset column indicates that $80\%$ of the AME20 mass data were randomly selected as the learning dataset, while the remaining $20\%$ served as the interpolation testing dataset. Here, ``\#" denotes that extrapolated masses in AME are also included besides the measured masses. In the extrapolation testing dataset column, ``AME16(AME03)-20" means that the nuclei from AME16 (AME03) were used as the learning dataset, while those newly appearing nuclei in AME20 were used as the extrapolation testing dataset. The last three columns denote the rms deviations $\sigma_{\rm rms}$ in MeV between the ML mass predictions and the experimental data for the corresponding datasets.}
  \label{tab:nuclear_mass_ml}
  \begin{tabular}{p{2.1cm}
                  p{2.4cm}
                  p{2.4cm}
                  p{2.4cm}
                  >{\centering\arraybackslash}p{1.5cm}
                  >{\centering\arraybackslash}p{2.5cm}
                  >{\centering\arraybackslash}p{3.0cm}}
    \hline
    \hline
    ML methods & learning dataset & interpolation testing dataset & extrapolation testing dataset & $\sigma_{\rm rms}$ for learning dataset & $\sigma_{\rm rms}$ for interpolation testing dataset & $\sigma_{\rm rms}$ for extrapolation testing dataset \\
    \hline
    ANN~\cite{Huang2025PRC}
    & AME20\#: 80\%
    & AME20\#: 20\%
    & AME16-20
    & 0.052
    & 0.122
    & 0.191 \\
    MDN~\cite{Mumpower2022PRCL}
    & AME16: 450
    & AME16
    & AME20
    & 0.186
    & 0.316
    & 0.336 \\
    BNN~\cite{Niu2022PRCL}
    & AME16
    & ---
    & AME16-20
    & 0.084
    & ---
    & 0.170 \\
    CPUN~\cite{Dellen2024PLB}
    & AME20: 1693
    & AME20: 726
    & AME20: $Z>100$
    & 0.328
    & 0.394
    & 0.600 \\
    CNN~\cite{Lu2025PRC}
    & AME20\#: 80\%
    & AME20\#: 20\%
    & AME16-20
    & 0.095
    & 0.167
    & 0.211 \\
    RBF~\cite{LiTao2025PRC}
    & AME20
    & leave-one-out
    & AME20: $Z = 104-110$
    & ---
    & 0.254
    & 0.256 \\
    AKRR~\cite{Tian2025PRC}
    & AME20
    & ---
    & AME03-20
    & 0.055
    & ---
    & 0.106 \\
    GP~\cite{Ye2026PRC}
    & AME20: 70\%
    & AME20: 30\%
    & 21 newly measured masses after AME20
    & 0.086
    & 0.210
    & 0.221 \\
    SVR~\cite{Yuksel2024PRC}
    & AME20: 75\%
    & AME20: 25\%
    & AME16-20
    & 0.23
    & 0.39
    & 0.74 \\
    LightGBM~\cite{LiuGP2025PRC}
    & AME20: 80\%
    & AME20: 20\%
    & boundary nuclei in AME20
    & 0.048
    & 0.160
    & lose model-repair ability when $d>6$ \\
    KAN~\cite{LiuH2025PRC}
    & AME20\#: 2856
    & AME20\#: 600
    & FRDM12 mass predictions
    & 0.25
    & 0.30
    & reasonably agree with FRDM12 predictions\\
    PCA~\cite{Wu2024SCPMA}
    & AME20: $Z\leqslant 60$
    & ---
    & AME20: $Z>60$
    & 0.584
    & ---
    & 0.506 \\
    NBMA~\cite{Zhang2024NPA}
    & AME20
    & ---
    & AME16-20
    & 0.293
    & ---
    & 0.381 \\
    PMM~\cite{Zhang2024PRC}
    & AME20
    & ---
    & FRDM12 mass predictions
    & 0.358
    & ---
    & $\sim$2 MeV when $d=30$ \\
    \hline
    \hline
  \end{tabular}
\end{table*}
Table~\ref{tab:nuclear_mass_ml} summarizes the root-mean-square (rms) deviations $\sigma_{\rm rms}$ between the ML mass predictions and the experimental data for the learning, interpolation testing, and extrapolation testing datasets. The $\sigma_{\rm rms}$ of WS4 model for the AME20:~$80\%$, AME20:~$20\%$, and AME16-20 datasets are $0.294$ MeV, $0.297$ MeV, and $0.356$ MeV, respectively, which can be viewed as the baseline to check the performance of ML methods. These ML methods generally show much lower $\sigma_{\rm rms}$ than the baseline for the learning dataset except the CPUN and model averaging methods, and those of ANN, BNN, CNN, AKRR, and LightGBM are lower than 100 keV by combining nuclear mass models~\cite{Huang2025PRC, Niu2022PRCL, Lu2025PRC, Tian2025PRC, LiuGP2025PRC}. The interpolation abilities of ML methods can be verified by randomly splitting the dataset into a learning dataset and an interpolation testing dataset. The results show that $\sigma_{\rm rms}$ for the interpolation testing dataset of ML methods is larger than that for learning dataset, and the ratio between them can be employed to verify the overfitting problem, which is typically within about 2 times with the largest factor of $3.3$ for the LightGBM. The AME16-20 test is usually employed to check the extrapolation abilities of ML methods. The $\sigma_{\rm rms}$ of ANN, BNN, CNN, RBF, AKRR, and GP methods are still much lower than the baseline for extrapolation testing dataset. However, the differences of $\sigma_{\rm rms}$ between this short-distance extrapolation test and interpolation test are not significant since the newly measured data are generally close to the original dataset. Long-distance extrapolation tests can be performed by gradually removing the boundary nuclei from the dataset~\cite{NiuZM2019PRCb}. For example, the LightGBM method based on WS4 loses its model-repair ability when the extrapolation distance $d > 6$~\cite{LiuGP2025PRC}. This boundary-removal test may provide valuable information for the extrapolation ability, while the long-distance test would lose many learning data since the boundary data would be separated into the testing dataset. Although only the nuclei with $Z \leqslant 60$ are used as the learning set, the PCA has even lower $\sigma_{\rm rms}$ for $Z > 60$ nuclei than that for the learning dateset~\cite{Wu2024SCPMA}. However, the extrapolation ability of PCA is mainly determined by the performance of the relevant nuclear models, and there is no remarkable improvement in the prediction accuracy comparing with the best nuclear model and its $\sigma_{\rm rms}=0.506$ MeV is larger than $\sigma_{\rm rms}$ of most ML methods for the extrapolation testing dataset. In addition, the uncertainty quantification is an important part of theoretical predictions, which can also help us to understand the extrapolation abilities of ML methods if it is evaluated properly. By comparing the uncertainty quantification evaluated through repeated calculations with different hyperparameters and datasets used in most ML methods, the BNN and GP can provide more reasonable uncertainty quantification. The predictions of nuclear model can also be used as pseudo-experimental data to test the extrapolation ability of ML methods~\cite{Niu2022PRCL, LiuH2025PRC, Zhang2024NPA}. However, this test depends on the selected nuclear model and is generally used to test obviously non-physical extrapolation results, and cannot provide a strict extrapolation test. A long-range extrapolation test based on $r$-process simulations is also limited by the complexity of $r$-process simulations and can only analyze the influence of ML predictions on the $r$-process simulations~\cite{Li2024PLB}. Therefore, the extrapolation test of ML methods remains challenging in nuclear physics.

In recent years, the ML methods have also become powerful tools for globally describing nuclear charge radii with high accuracy
~\cite{Liu2025NST,TangLu2024NST}.
The BNN method has achieved great success in the study of charge radii due to its excellent performance~\cite{XXDong2023PLB, ZYXian2025PLB, XZhang2024PRC, YuanCX2026EPJA}. A Monte Carlo dropout variant incorporating quadrupole and hexadecapole deformations, together with a modified Casten factor to avoid issues at closed shells has been proposed. This approach significantly improves predictive performance for charge radii, reducing the rms deviation of the training set with respect to the experimental charge radii 
to 0.0084 fm~\cite{ZYXian2025PLB}. By including the information of charge density distributions, the ML methods along with calibration to experimental charge radii not only accurately reproduce the experimental charge radii but also provide reliable charge density distributions~\cite{ZXYang2023PRC, ZXYang2023PLB, Shang2024PRC}. In addition, other ML methods including the CNN~\cite{YYCao2023NST, Su2023Symmetry}, RBF~\cite{LiT2026CPCRch}, GP~\cite{Ye2026PRC, ZLLi2025PRC, Maheshwari2026PLB}, decision trees~\cite{ZLLi2025PRC, WFLi2024PS}, and continuous Bayesian probability estimator~\cite{LiuJ2025NST} have also achieved remarkable success in the study of charge radii, demonstrating the advantages of ML methods in describing nuclear charge radii.

In addition to nuclear masses and charge radii, the ML methods have also been applied to study many other nuclear properties recently, such as ground-state spin~\cite{Wen2023APS, Liu2024NST}, low-lying spectra~\cite{LiZL2026CPC, Gao2024JPG, Lv2024PLB, Lv2025PRC, Liu2025NST, Zhang2026EPJP}, nuclear level densities~\cite{Du2024PRC, Wang2024CPC, Zhao2026ApJ}, binding energy~\cite{Yuan2024NST}, neutron skin~\cite{Wei:2022iuy}, $\alpha$-decay half-lives~\cite{Jyothish2025PRC, Shree2025EPJAa, Shree2025EPJAb, You2025NST, Shree2026NPA, Yang2026PRC}, $\beta$-decay half-lives
~\cite{Li2024JPG, Li2025NST, Jalili2025PRC},
and giant dipole resonance parameters~\cite{Bairwa2025PS}. Beyond the conventional ML methods, Large language models (LLMs) have achieved great success in many fields including natural language understanding, content generation, logical reasoning and industrial intelligent applications. The pre-trained DeepSeek-R1-1.5B model has been employed for the description of nuclear structure observables~\cite{GuoSJ2026arXiv}, whose performance can be remarkably improved by including more relevant input and output features. Specifically, the LLMs achieve interpolation accuracies of approximately 100 keV, 170 keV, 200 keV, 0.02 fm for decay energies, binding energies, separation energies, and charge radii, respectively, which are comparable to those of dedicated ML models. These findings validate that LLMs can serve as an efficient framework for multi-task regression analysis of nuclear structure properties.

\begin{figure*}[htbp]
    \centering
    \includegraphics[width=0.4\textwidth]{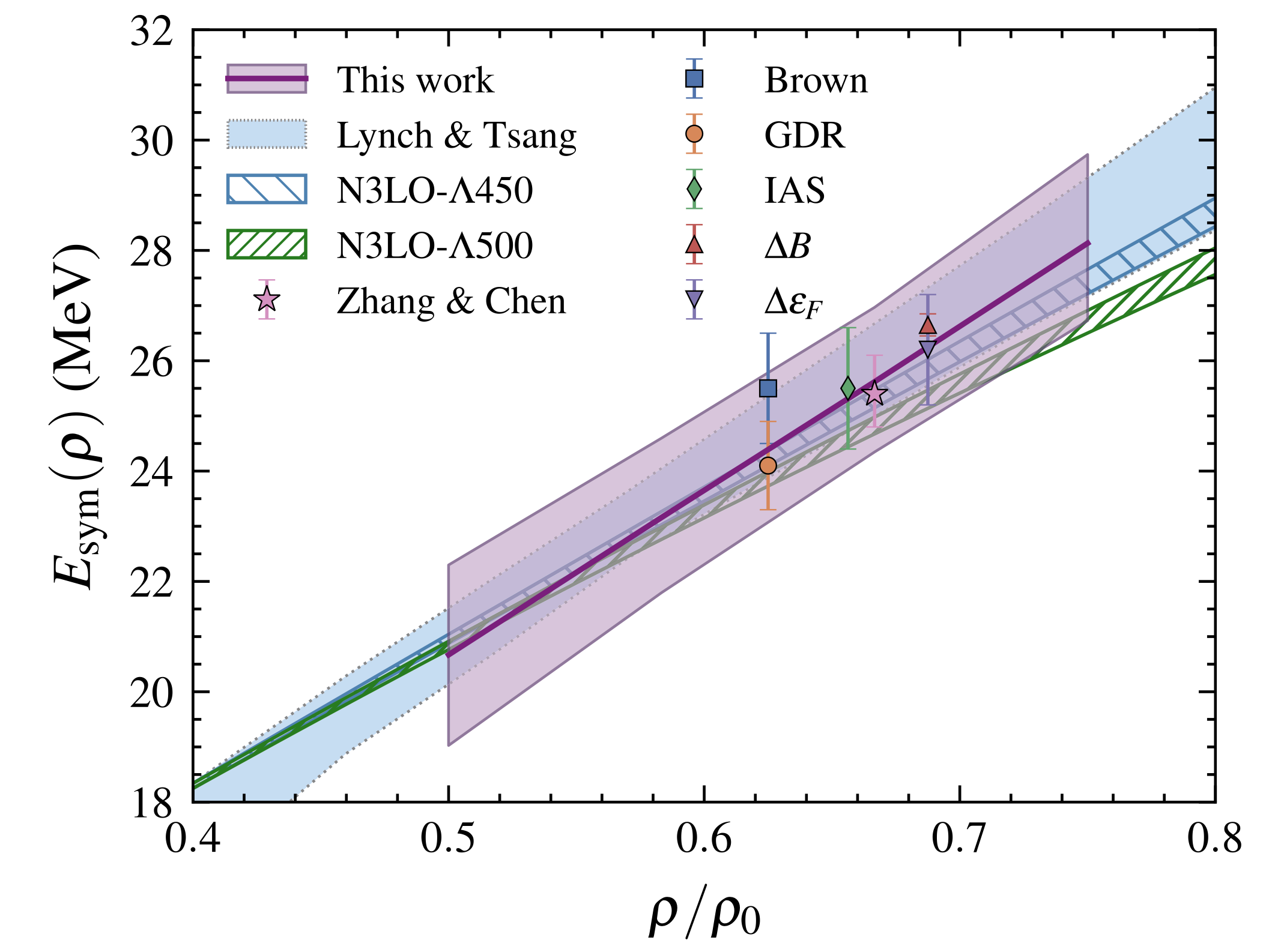}
    \caption{Constraints on the symmetry energy as a function of density $\rho$ from Bayesian model averaging of Skyrme and RMF model predictions. Various experimental constraints and theoretical predictions are shown for comparison. Figure taken from Ref.~\cite{MYQiuPLB24}. }
    \label{Fig:CLW}
  \end{figure*}

The quantification of uncertainties in the inferred physical quantities from nuclear properties is another key point for understanding the properties of the EoS.
It has become an integral part of nuclear physics research, as both the design of nuclear physics experiments and the extraction of crucial physical information from experimental data heavily rely on nuclear theoretical models 
~\cite{Paquet:2023rfd,Svensson:2025jde}.
The Bayesian model averaging (BMA) is a statistical approach that combines predictions from multiple models weighted by their ability to reproduce experimental data, thereby providing more reliable inferences than any single model. In Ref.~\cite{MYQiuPLB24}, the nuclear symmetry energy \(E_{\mathrm{sym}}(\rho)\) at subsaturation density around \(2\rho_{0}/3\) (with \(\rho_{0}\) the nuclear saturation density) is extracted from the effective proton-neutron chemical potential difference \(\Delta\mu_{\mathrm{pn}}^{*}\) measured in five doubly magic neutron-rich nuclei: \(^{48}\mathrm{Ca}\), \(^{68}\mathrm{Ni}\), \(^{88}\mathrm{Sr}\), \(^{132}\mathrm{Sn}\), and \(^{208}\mathrm{Pb}\). The theoretical methods include both non-relativistic Skyrme energy density functionals and nonlinear relativistic mean-field (RMF) models. Gaussian process emulators are constructed to map the strong correlation between \(\Delta\mu_{\mathrm{pn}}^{*}\) and \(E_{\mathrm{sym}}(2\rho_{0}/3)\). The key innovation is the application of BMA to systematically address model dependence---i.e., discrepancies between the two theoretical frameworks---and to provide a statistically rigorous unified constraint. Fig.~\ref{Fig:CLW} presents the constraints on the nuclear symmetry energy \(E_{\mathrm{sym}}(\rho)\) as a function of nucleon density \(\rho\), obtained from the Bayesian model averaging (BMA) of Skyrme energy density functional and nonlinear relativistic mean field (RMF) model predictions \cite{MYQiuPLB24}. The purple solid line and shaded band denote the median value and the 1\(\sigma\) credible interval from the BMA analysis, respectively. For comparison, the figure also includes various experimental constraints and theoretical predictions from the literature, such as those from Lynch \& Tsang (cubic polynomial fit), chiral effective field theory calculations (N3LO with different cutoffs), and analyses based on doubly magic nuclei properties, giant dipole resonance, binding energy differences, and neutron-proton Fermi energy differences. The BMA results are in good agreement with these independent constraints, demonstrating their statistical robustness and reliability for describing the symmetry energy around subsaturation densities.
The BMA analysis yields \(E_{\mathrm{sym}}(2\rho_{0}/3) = 25.6^{+1.4}_{-1.3}\)~MeV at the 68.3\% confidence level, which is consistent with microscopic predictions and other isovector indicators, demonstrating the power of BMA for reliable uncertainty quantification in nuclear EoS studies.

\subsubsection{Nuclear many-body calculations}

Machine learning approaches have been extensively applied to nuclear quantum many-body calculations, which can be categorized into three main paradigms: neural quantum states for wave-function representation, physics-informed neural networks and energy density functional constructions, and reduced-order models for parameter scans.

\begin{figure*}[t]
    \centering
    \includegraphics[width=0.7\textwidth]{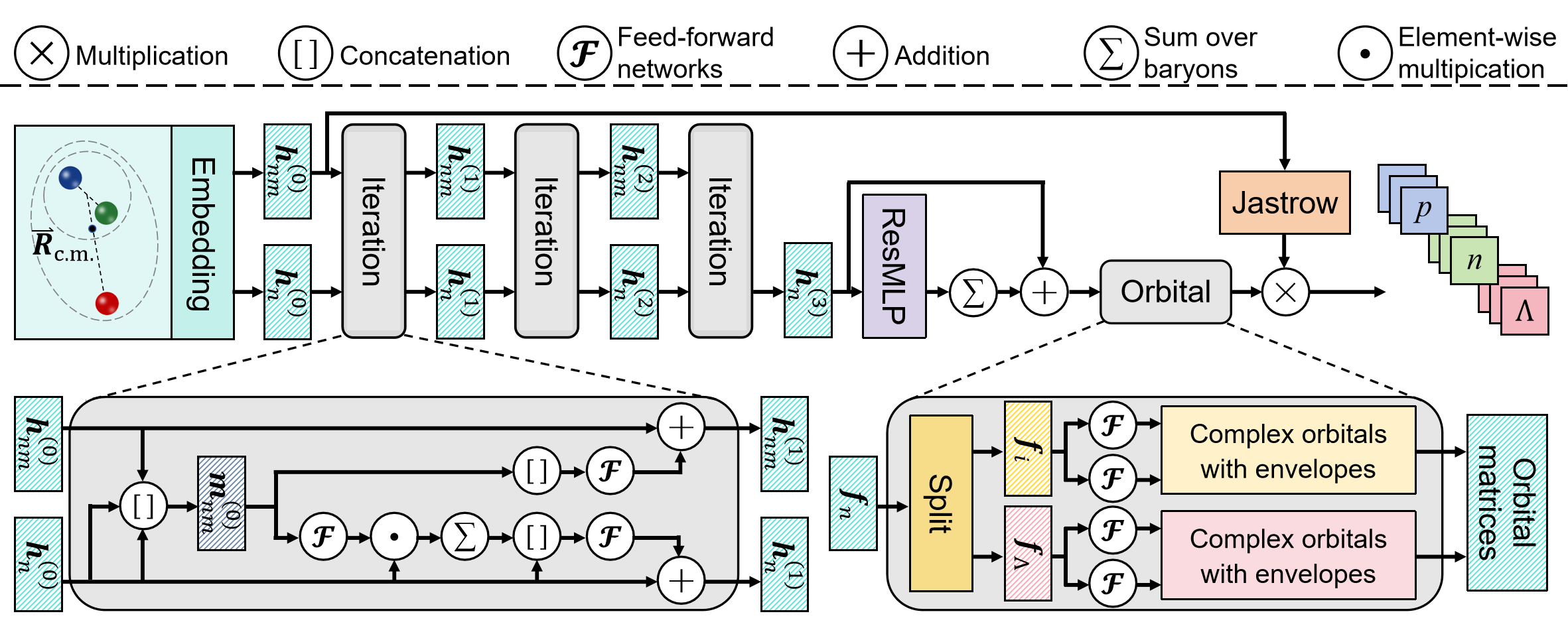}
    \caption{Representative architecture of neural-network quantum states for hypernuclear many-body calculations. Baryonic configurations are encoded into latent neural representations and optimized variationally to construct correlated many-body wave functions for nuclei and hypernuclei. 
    }
    \label{Fig:SGNQS}
  \end{figure*}

\begin{figure*}[htbp]
    \centering
    \includegraphics[width=0.4\textwidth]{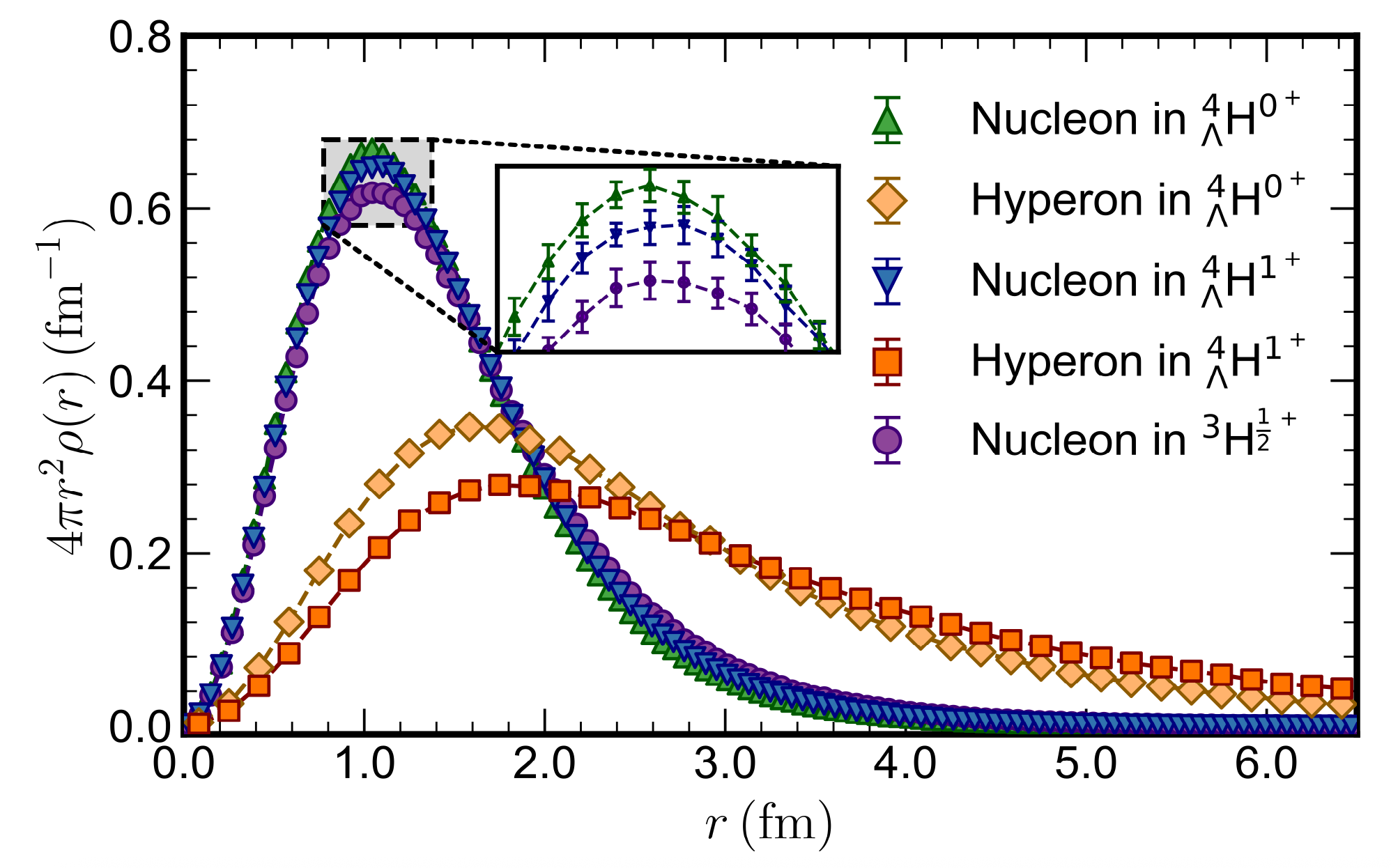}
    \caption{The spatial distributions of $^3H$, $^4 _\Lambda H^{0+}$ and $^4 _\Lambda H^{1+}$ obtained with VMC-NQS-SG method. Green upper triangles, blue lower triangles and purple circles denote the point-nucleon distributions. The $\Lambda$ orbits are presented as yellow diamonds and orange squares for ground and excited states. Figure taken from Ref.~\cite{zhang2026}}
    \label{Fig:ZZX}
  \end{figure*}

Neural networks have been applied within the Variational Monte Carlo (VMC) framework. The Restricted Boltzmann Machine was first proposed as a variational representation of the quantum wavefunction. 
They interpreted the wavefunction as a computational black box and learned the mapping between input configurations and complex amplitudes through neural networks. This proved that NQS could achieve state-of-the-art accuracy in 1D and 2D spin models while effectively circumventing the phase problem in dynamical evolution.
Subsequently, researchers began applying NQS to more complex ab initio nuclear calculations. A minimal single-layer feed-forward neural network was employed to successfully solve the deuteron bound state problem in momentum space, proving that high precision could be reached with very few parameters. 
A NQS based on the Deep Sets architecture, utilizing intrinsic coordinates to eliminate spurious center-of-mass motion, successfully extended the calculation scale from light nuclei to open-shell nuclides with $A=6$.
To handle more strongly correlated medium-mass nuclei, NQS architectures were further refined. The "Hidden Nucleons" concept was introduced, which considerably augmented expressivity by adding fictitious fermionic degrees of freedom in the Hilbert space, thereby improving the nodal surface of traditional trial wavefunctions.
FeynmanNet was developed successfully, implementing multi Backflow transformations through neural networks~\cite{Yang2023}. This allows orbitals to explicitly depend on the variables of all nucleons, successfully achieving high-precision variational calculations for $^{16}\text{O}$ and surpassing the energy accuracy of traditional Diffusion Monte Carlo (DMC) methods. The NQS is first applied to hypernuclear systems containing hyperons~\cite{zhang2026}. Through the Spinor Grouping (SG) method and spin purification schemes, they overcame statistical errors and spin contamination issues in weakly bound systems, achieving a unified treatment across different baryon species, as illustrated in Fig.~\ref{Fig:SGNQS}. Fig.~\ref{Fig:ZZX} illustrates the spatial density distributions of the nucleon core and the \(\Lambda\) hyperon for the hypernuclear states $^4_\Lambda H^{0+}$ and $^4_\Lambda H^{1+}$ obtained with the VMC-NQS-SG method ~\cite{zhang2026}. The point-nucleon distributions are shown as green upward triangles, blue downward triangles, and purple circles, while the \(\Lambda\) orbits are represented by yellow diamonds (ground states) and orange squares (excited states). A clear shrinkage effect is observed, where the nuclear core becomes slightly compressed due to the presence of the \(\Lambda\) hyperon. The excited states predominantly arise from changes in the hyperon orbital, consistent with the picture of a weakly interacting hyperon moving in a relatively frozen nuclear core.

In parallel with standard variational wave functions, recent research has also integrated microscopic nuclear cluster models with deep neural network technologies to overcome the computational bottleneck caused by the exponential expansion of model space in describing cluster phases and separation thresholds.
An efficient AI-based nuclear theory calculation method, termed the Control Neural Network (Ctrl.NN)~\cite{cheng2025evidence, zhu2026microscopic, tian2024lambda, liu2025structural, tian2025hypernuclear, tian2026cluster}, achieves the efficient optimization of nuclear microscopic many-body wave functions and greatly accelerates the quantum many-body calculation process. Its core logic involves taking the massive parameters of superposed wave functions (such as generator coordinates, high-momentum nucleon pairs, and spin configurations) as network inputs to predict their variations, while constructing a physics-informed loss function based on properties like system energy, root-mean-square radius, and spatial symmetry. Leveraging a just-in-time training strategy~\cite{cheng2025evidence}, the network guides the wave function through an iterative cooling evolution, thereby accurately capturing the physical features of the target quantum states within a minimal model space.

The universal approximation theorem enables the ANN to describe the wave function in quantum mechanics, and hence it can be employed to solve quantum many-body problems by using the unsupervised machine learning without training datasets~\cite{Mehta2019PRp, Hermann2023NRC,Wang:2024ynn}. This method has been successfully used to solve the Schr\"{o}dinger equations by minimizing the violation to the Schr\"{o}dinger equation for several classical problems in quantum mechanics, including the Woods-Saxon potential widely used in nuclear physics. It has also been extended to solve the nucleonic Driac equation, and the variational collapse problem due to the Dirac sea can be overcome by employing the inverse Hamiltonian method~\cite{Wang2025EPJA} or minimizing the violation to the Dirac equation~\cite{Du2026CTP}. Moreover, the ML method can also be used to construct the nuclear energy density functional (EDF), which has been proven to exist but still unknown. The KRR approach has been successfully applied to construct a robust and accurate orbital-free EDF not only for the spherical nuclei~\cite{Wu2022PRCDFT} by also for the deformed nuclei~\cite{Wu2025CP}, which can reproduce the exact ground-state densities and total energies from the orbital-dependent Kohn-Sham solutions with high accuracy. The covariant EFT has also been constructed using the ANN, which exhibits significantly better extrapolation abilities than traditional ML methods for binding energies~\cite{LiWF2026PLB}.

Reduced-order methods address the computational cost of repeated many-body calculations across large parameter or model spaces. Eigenvector continuation (EC) and related projection methods construct a low-dimensional solution manifold from selected high-fidelity calculations, thereby accelerating parameter scans while retaining a direct connection to the underlying Hamiltonian~\cite{Frame_2018,Duguet_2024,Melendez2022MOR,Drischler2023}. For neutrinoless double-beta decay, ML and many-body emulators have accelerated generator-coordinate kernels, nuclear-matrix-element calculations, and uncertainty propagation~\cite{Zhang2023GCMML,Belley2024NMEUQ,Zhang2025SPCDFT}. These approaches emulate specified many-body frameworks, so uncertainties associated with interactions, model-space truncations, and transition operators must remain explicit.

\subsection{Nuclear reactions}
\subsubsection{Low energy nuclear reactions}

\begin{figure*}[htbp]
    \centering
    \includegraphics[width=0.8\textwidth]{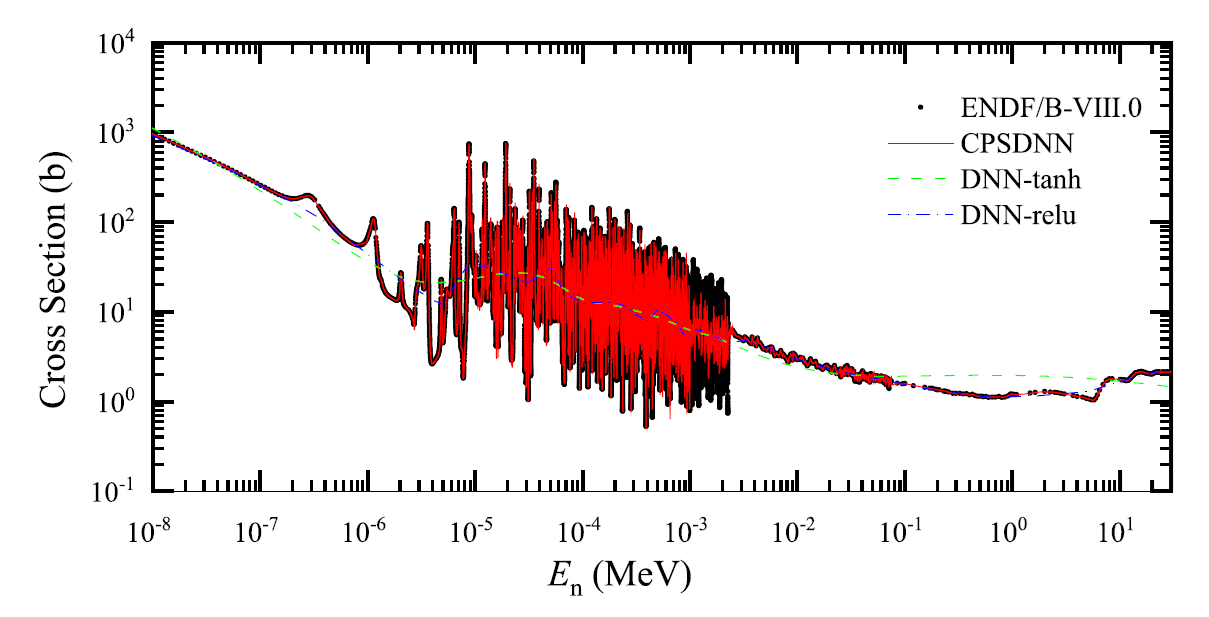}
    \caption{Comparison between the raw data with the predicted results of DNNs and
CPSDNN for the $^{235}$U(n,f) reaction. The black points are the evaluated cross
sections derived from ENDF/B-VIII.0. The dashed green and the dashed-dotted
blue lines are the predicted results of DNNs with the tanh and relu activation
functions, respectively. The red solid line is the predicted results of CPSDNN. Figure taken from Ref.\cite{XING2024138825}}
    \label{fig:CSPPNN}
\end{figure*}

Nuclear reaction data  generally  have large uncertainties compared to nuclear structure data.  In particular, the neutron induced reaction data are rather imperfect, but are key for energy productions and applications~\cite{Shang2025Nst}. 
The nuclear fission product yields, key nuclear data associated with multiple post-fission observables, still suffer from insufficient accuracies of theoretical models and experimental measurements. The evaluation of fission yields
is only available at thermal neutron energies, 0.5 and 14 MeV in major nuclear data libraries. In this respect, ML of incomplete fission yields can be helpful and a promising evaluation method~\cite{wangza1,heterogeneous}. 
Recently, physics guided feature inputs have been incorporated into BNN to improve the interpolation of energy dependencies in fission yields~\cite{Qiao2026NST,Chen2024JNST,chen2026pebnn} . In particular, the evaluation of the fission yields of $^{232}$Th is of great interest, as it can be the new type of fuel for next generation energy production~\cite{Qiao2026NST}.
 Inferences of fission yields using complicated physics guided feature inputs can  provide interpretable insights for modeling fission processes~\cite{j5dj-htxj}. 
In addition, the Bayesian mixture density network also demonstrated improved performances in such evaluations by capturing the odd-even effects~\cite{gd8l-zzmy}. 
Very recently, physics-informed Bayesian machine learning using physics model pre-trained priors has been newly developed, which can incorporate comprehensive physics knowledge such as conservation laws and quantum effects to overcome overfitting issues and data scarcity in purely data-driven learning approaches~\cite{liu2026PIBM}. The informed Bayesian prior is a well-known conception, but it is the first time that it is applied to real nuclear data.

\begin{figure*}[t]
    \centering
    \includegraphics[width=0.9\textwidth]{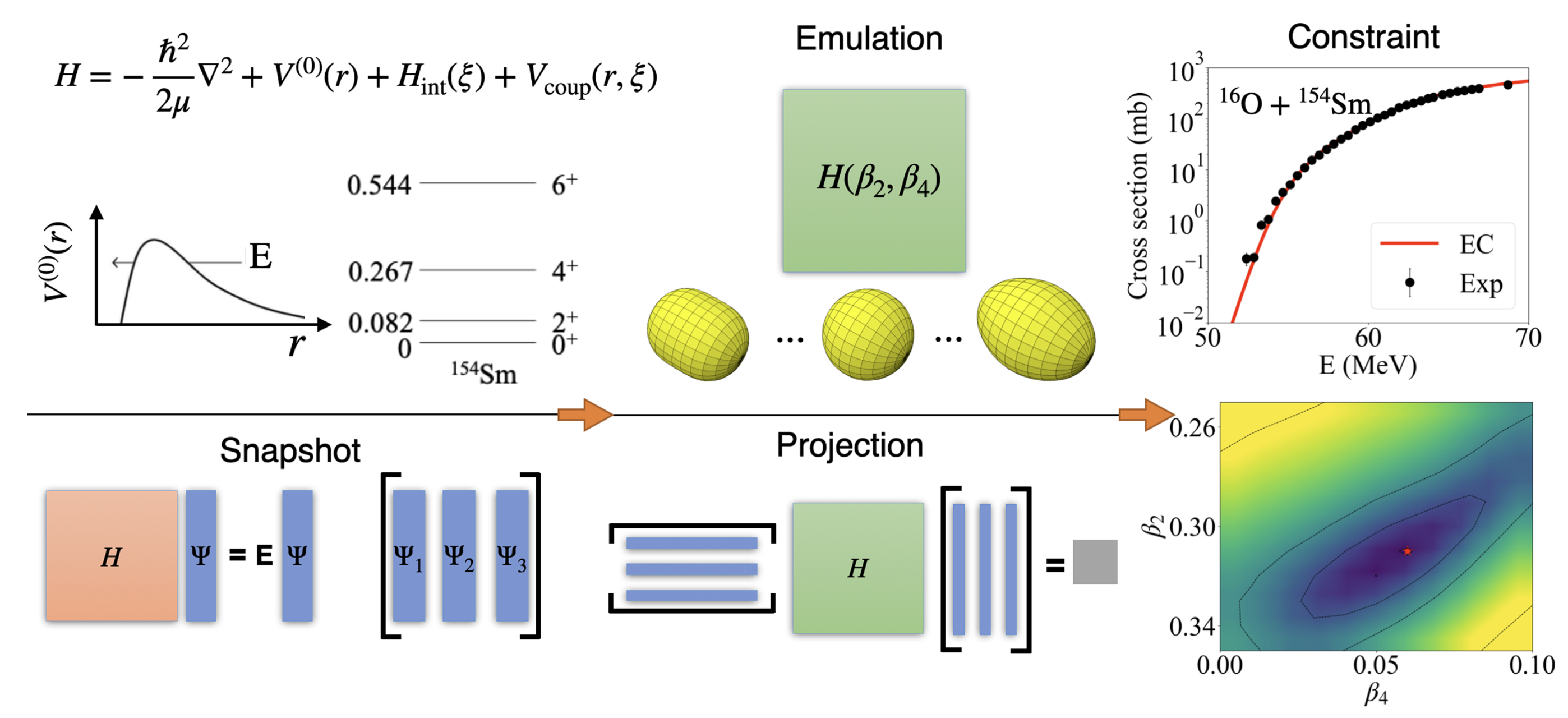}
    \caption{The workflow of the emulator for determining the nuclear shape. High-fidelity eigenfunctions $\Psi$ of the coupled-channels
equations are first obtained for a selected set of parameters (Snapshot). The full Hamiltonian $H$ is projected onto the subspace formed by these eigenfunctions (Projection) to build an emulator, which facilitates rapid predictions of fusion cross sections (Emulation). The results of the emulator are then compared with experimental data to perform a $\chi^2$ analysis, with which optimum values of the parameters are extracted (Constraint). Figure taken from Ref.~\cite{liao_2025}. 
    }
    \label{Fig:ZhuL}
  \end{figure*}

In addition to fission yields, ML has also been applied to infer neutron induced reaction cross sections~\cite{PhysRevC.109.044616, XING2024138825,Huang:2024acg}.
A novel Phase Shift Deep Neural Network (PSDNN) framework was proposed to accurately reproduce neutron resonance cross sections in~\cite{XING2024138825}, to address the long-standing difficulty of modeling high-frequency oscillatory structures in the reaction $^{235}U(n,f)$. Conventional deep neural networks (DNNs) fail in this regime due to the frequency principle, which biases learning toward low-frequency components. To overcome this limitation, the authors introduce the Parallel Phase Shift DNN (PPSDNN) and its more efficient variant, the Coupled Phase Shift DNN (CPSDNN)(Fig.~\ref{fig:CSPPNN}), which transform high-frequency components into low-frequency ones via phase-shift operations in the Fourier domain. This enables effective learning across the full frequency spectrum.
Similarly, the understanding of reaction spectroscopic factors remains a challenge~\cite{Li2026SCPMA}.  
In addition, BNN has been successfully applied to predict reaction cross sections ($\gamma$, $n$)~\cite{Sun2025PhotonuclearBNN} and fusion cross sections of light nuclei~\cite{Cheng2025NST}.

Beyond data-driven evaluation of reaction data, reduced-order methods can accelerate repeated evaluations of physics-based reaction models. The first EC emulator for scattering observables was developed by Furnstahl \textit{et al.}~\cite{Furnstahl}, and was subsequently combined with the Kohn variational principle to construct efficient reaction emulators with Bayesian uncertainty quantification~\cite{Drischler2021}. General introductions to model-order reduction and emulation in nuclear physics are provided in Refs.~\cite{Melendez2022MOR,Drischler2023}. These approaches have since been extended to optical-model calculations through the reduced-order scattering emulator ROSE~\cite{Odell} and to coupled-channels reaction calculations~\cite{Catacora}.

A direct application is the extraction of nuclear deformation parameters from reaction data, which typically requires repeated coupled-channels calculations and can become prohibitively expensive in extensive parameter scans or Bayesian analyses~\cite{GUPTA2020,GUPTA2023}. An EC emulator constructed from a small set of full-model solutions can reproduce reaction observables over a broad deformation-parameter range at substantially reduced computational cost~\cite{liao_2025}. Combined with experimental fusion data, this framework enables efficient inference of nuclear deformation properties. Fig.~\ref{Fig:ZhuL} shows the emulator-based workflow for determining nuclear shape proceeds in four stages. First, high-fidelity eigenfunctions $\Psi$ of the coupled-channels equations are computed for a chosen set of parameter values, constituting the snapshot stage. Next, the full Hamiltonian $H$ is projected onto the subspace spanned by these eigenfunctions--this projection step constructs the emulator, enabling rapid predictions of fusion cross sections. In the emulation phase, the emulator efficiently evaluates cross sections across the parameter space. Finally, the emulator outputs are compared against experimental data through a 
$\chi^2$ analysis, and the parameters that minimize the discrepancy are identified as the optimal values during the constraint stage. In that study \cite{liao_2025}, axial quadrupole and hexadecapole deformations were incorporated into the coupled-channels formalism for deformed nuclei. It is worth noting that the theoretical framework is not limited to these shapes; rather, it can be straightforwardly generalized to accommodate more complex deformation degrees of freedom, such as triaxiality ($\gamma$) and octupole ($\beta_3$) deformations. 

\subsubsection{Heavy-Ion collisions} 

Heavy-ion collisions (HICs) across low, intermediate, and relativistic energies provide important probes of nuclear synthesis, nuclear structure, dense matter, and QCD phenomena. In this section, we review recent applications of ML to these topics across different collision-energy regimes.

Low-energy HICs mainly focus on synthesizing the super-heavy nuclei by fusion reactions. This process involves colliding heavy-ions to form a highly excited compound system, which subsequently cools down through neutron emission, leading to the production of super-heavy nuclei.  
With more than one thousand excitation functions already measured, these data can be used to train a ML model for precise prediction of the FCS, see, e.g., Refs. \cite{Li:2023ukd,li}. 
This approach drastically improves prediction precision and outperforms several conventional models. Notably, the FCSs predicted by these ML models demonstrate comparable quality to those from advanced microscopic theories. For example. as shown in Ref.~\cite{Li:2023ukd}, even for the very neutron-rich systems $^{40,48}$Ca + $^{78}$Ni, the FCSs predicted by the ML model are consistent with those calculated using the density-constrained time-dependent Hartree-Fock (DC-TDHF) approach, one of the most advanced microscopic frameworks for modeling heavy-ion fusion. Similarly, Ref.~\cite{PhysRevC.109.024601} also applies ML methods to low-energy heavy-ion fusion reactions, but with a primary focus on constraining the key parameters of the nucleus–nucleus interaction potential. 

\begin{figure*}[t]
    \centering
    \includegraphics[width=0.9\textwidth]{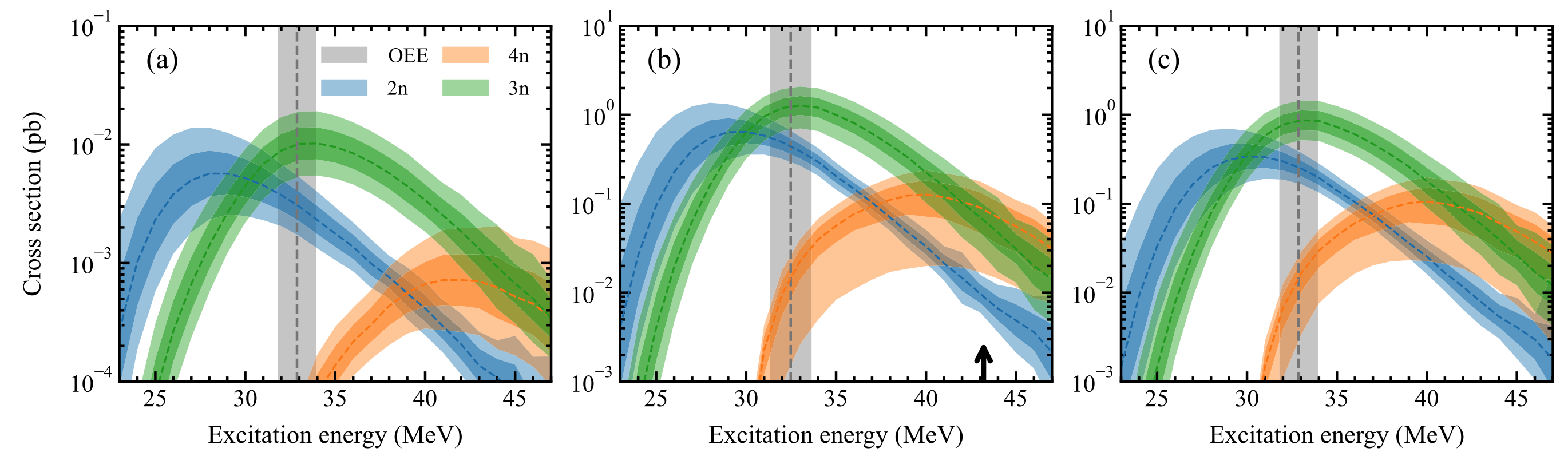}
    \caption{The ERCS and the OEE in reactions (a) $^{54}$Cr + $^{243}$Am, (b) $^{50}$Ti + $^{249}$Bk, and (c) $^{51}$V + $^{248}$Cm. The dark and light shaded bands represent the 1$\sigma$ and 2$\sigma$ confidence levels for the ERCS, respectively. The grey coloured bands are the 2$\sigma$ confidence levels of the OEE. The dashed lines are obtained by taking the mean of ERCS and OEE. The black arrow in (b) denotes the incident energy used in the GSl for the reaction $^{50}$Ti + $^{249}$Bk \cite{SHE119}. Figure taken from Ref.~\cite{FANG2024139069}. 
    }
    \label{Fig:ZhuL2}
  \end{figure*}

Recently, Bayesian analysis has been incorporated into the DNS model to constrain key parameters and predict evaporation residue cross sections for superheavy-element synthesis~\cite{FANG2024139069}. Inferred parameter correlations significantly reduce the uncertainties of the predicted cross sections and enable reliable estimates for prospective $Z=119$ production reactions.
As shown in Fig.~\ref{Fig:ZhuL2}, the Bayesian-constrained DNS-sysu model provides predictions for the evaporation residue cross sections (ERCS) and optimal excitation energies (OEE) for three candidate reactions to synthesize element \( Z = 119 \): \( ^{54}\mathrm{Cr}+^{243}\mathrm{Am} \), \( ^{50}\mathrm{Ti}+^{249}\mathrm{Bk} \), and \( ^{51}\mathrm{V}+^{248}\mathrm{Cm} \) \cite{FANG2024139069}. The \( 1\sigma \) and \( 2\sigma \) confidence bands for the ERCS remain within about one order of magnitude, indicating that the parameter correlations properly accounted for prevent overestimation of uncertainties. The corresponding \( 2\sigma \) intervals of the optimal incident energies (OIE) are predicted to be \( 238.1 \)–\( 240.2 \) MeV, \( 222.8 \)–\( 225.1 \) MeV, and \( 227.1 \)–\( 229.3 \) MeV, respectively. Notably, for the \( ^{50}\mathrm{Ti}+^{249}\mathrm{Bk} \) reaction, the experimentally used energy at GSI (indicated by the black arrow) lies well above the predicted OIE, and the corresponding ERCS at that energy is only \( 158^{+102}_{-140} \) fb (at \( 2\sigma \)), with a lower limit of 18 fb—far below the experimental sensitivity of 65 fb. This strongly suggests that the unsuccessful synthesis attempt at GSI may be attributed to an inappropriately chosen incident energy, and that using the predicted optimal energy could significantly enhance the production probability. These results demonstrate the predictive power and practical value of the Bayesian uncertainty quantification framework for guiding future superheavy element experiments.

Intermediate energy HICs are mainly used for understanding the collision mechanism, equation of state (EoS), and in-medium NN cross sections at intermediate energies by comparing the data to the calculations.  
The observables usually used are directed flow, elliptic flow, particle yield, and nuclear stopping power, and the transport models are Quantum Molecular Dynamics (QMD) and Boltzmann-Uehling-Uhlenbeck (BUU) type models.  
In Ref.~\cite{BALi25}, the authors combines the IBUU transport model, Gaussian Process (GP) emulators, and Bayesian inference to analyze the FOPI proton directed- and elliptic-flow excitation functions in Au+Au collisions. The analysis indicates that the in-medium baryon–baryon scattering cross sections gradually approach their free-space values with increasing beam energy, while the nuclear equation of state evolves from relatively soft to stiffer behavior, suggesting a hardening of dense nuclear matter at higher densities and temperatures. In the work of Ref. \cite{Wei:2024obb}, a ML-based method was proposed to study the in-medium correction factor $F$ which is defined as the ratio of the nucleon-nucleon cross section in the nuclear medium to that in the free space. 
It is found that LightGBM can recognize information about the $F$ factor with high accuracy even for event-by-event data. Furthermore, by using an interpretability analysis method in ML, the features that have the greatest effect on the prediction of $F$ are identified. The Bayesian method was also used to infer the signal of the liquid-gas phase transition by measuring the strength of non-Gaussian distribution of the interested observables~\cite{XYWang2025}, and it was verified in ImQMD model. However, these models are often computationally intensive and time-consuming. 

\begin{figure*}
    \centering
    \includegraphics[width=0.7\textwidth]{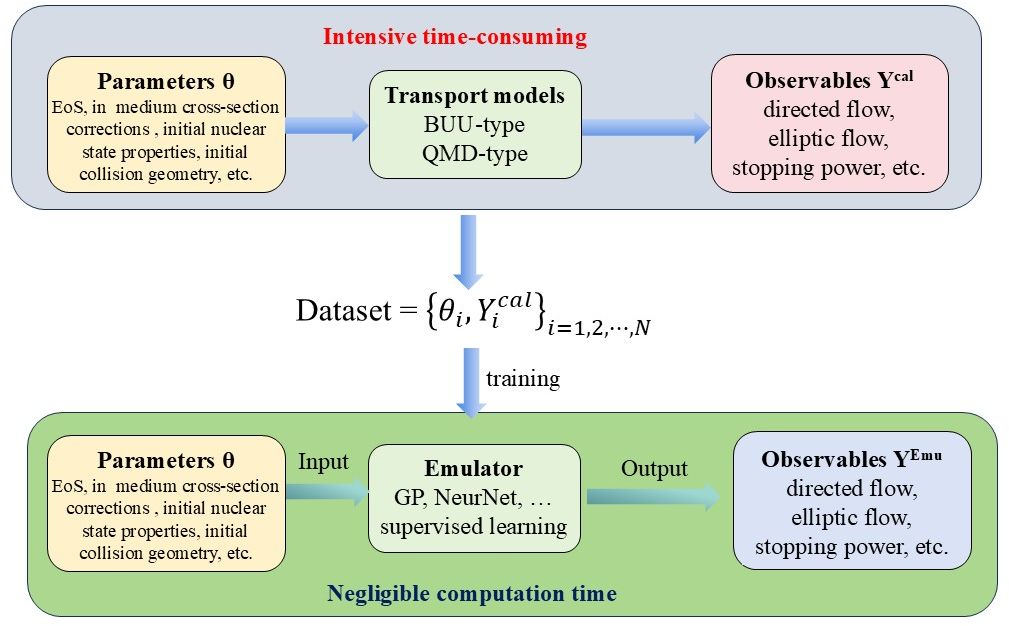}
    \caption{Flowchart of the construction of emulator for transport model.}
    \label{fig:Emulator-flowchart}
\end{figure*}

To address this challenge, emulators that can be used to fast surrogate models trained on a limited set of real transport model simulations have been introduced. These emulators leverage ML techniques to approximate the input–output mapping of transport models, enabling rapid predictions and uncertainty quantification. The flowchart on how to construct emulators of transport models is shown in Fig.\ref{fig:Emulator-flowchart}. In a recent work of Ref. \cite{wei}, three ML algorithms, Gaussian processes (GP), multi-task neural networks (MTNN), and random forests, are used to train emulators based on the simulations of UrQMD transport model. 
In the work of Ref.\cite{Cox:2024mkz}, DNN and GP were used as emulators of an isospin-dependent
Boltzmann-Uehling-Uhlenbeck (IBUU) transport model. 
They found that DNN can emulate the IBUU simulator’s prediction very efficiently even with small training datasets and with
an accuracy about ten times higher than the GP. 

Studying nuclear structure through HICs is a frontier topic in nuclear physics because some final-state observables retain sensitivity to the internal structure and orientation of the colliding nuclei~\cite{Jia:2022ozr,STAR:2024wgy}. Ref.~\cite{Yang:2023djv} trained a convolutional-orientation-filter network on IBUU events for ultra-central U+U collisions at 1~GeV/nucleon. The model combines two-dimensional momentum--rapidity distributions with scalar features such as charged-particle multiplicities and classifies the simulated initial orientation. Because both the labels and events originate from the IBUU model, its robustness to the choice of event generator and to experimental detector effects remains to be established.

The EoS constraints obtained from heavy-ion collisions can be complemented by astrophysical observations of neutron stars. Physics-constrained inverse-TOV methods, including deterministic and Bayesian neural networks, have been developed to infer the speed of sound and pressure--density relation from neutron-star observables while enforcing constraints such as causality and low-density chiral-effective-field-theory predictions~\cite{DeepLearningNSEoS2024,PhysicsInformedBNN2024}. Bayesian analyses can jointly use mass--radius measurements, tidal deformabilities from gravitational-wave events, and maximum-mass constraints from radio pulsars, with Gaussian-process emulators accelerating the required model evaluations~\cite{BayesianNSEoS2024,MicroscopicNSConstraints2025}. Recent ML and Bayesian frameworks further combine these multi-messenger data with heavy-ion constraints on the symmetry energy and high-density EoS, and explore possible hadron--quark phase transitions~\cite{HadronQuarkCrossover2025,SymbolicRegressionNS2024}. This provides a unified strategy for constraining dense nuclear matter from laboratory experiments to neutron-star interiors.

At relativistic energies, ML has been used to classify QCD phases and phase-transition patterns, infer initial conditions, and accelerate Bayesian extraction of transport properties such as the shear and bulk viscosities $\eta/s$ and $\zeta/s$ and the jet-transport coefficient $\hat q$~\cite{Zhou2024QCDML,He:2023zin}. Point-cloud and symmetry-aware networks are natural representations for variable-size final-state particle sets, where correlations rather than an arbitrary ordering carry the relevant physics. The QCD-focused review of Zhou \textit{et al.} provides comprehensive context~\cite{Zhou2024QCDML}, while the living bibliography of Feickert and Nachman provides broader particle-physics context~\cite{Feickert2021LivingReview}. In all cases, a network trained on one event generator can inherit its assumptions; therefore, robustness across generators and comparison with real data are more demanding tests than accuracy on held-out simulated events.

\subsection{AI for experimental physics and techniques}

As experimental facilities in high-energy and nuclear physics evolve toward higher luminosity and greater complexity, traditional data-processing and system-control methods face unprecedented challenges. Representative facilities include the Facility for Rare Isotope Beams (FRIB) at MSU, the China Accelerator Facility for Superheavy Elements (CAFE), and the High-Intensity Heavy-Ion Accelerator Facility (HIAF) at IMP. In this context, AI is no longer merely a supplementary tool but has become a critical enabling technology for accelerator control, high-fidelity simulation, detector optimization, real-time triggering, and massive data compression. These applications span multiple stages of the experimental workflow, from facility operation and detector design to event simulation and front-end data acquisition.

At the level of facility operation, AI-assisted optimization and control methods have already been deployed at several accelerator facilities. At FRIB, customized Bayesian optimization has been applied to operational beam-tuning tasks for diverse ion species~\cite{Hwang2026}. At the SECAR recoil separator, Bayesian optimization first improved beam alignment and ion-optical settings~\cite{Miskovich2022SECAR}. At CAFE, ML has been integrated into online accelerator control, while supervised classifiers have been tested for fault identification in the superconducting radio-frequency cavities of CAFE2~\cite{ChenXiaolong2024SCPMA,Yang2025NST}.

\begin{figure*}
    \centering
    \includegraphics[width=0.8\textwidth]{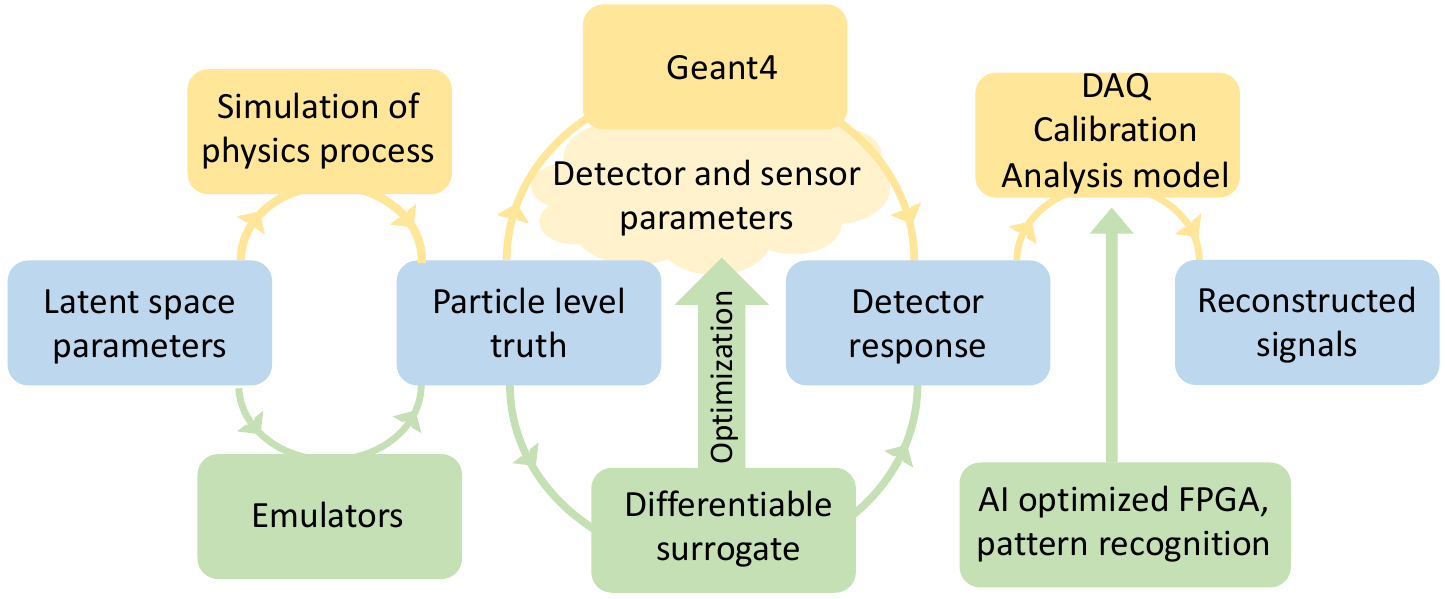}
    \caption{Overview of the detector simulation and reconstruction pipeline.}
    \label{fig:experiment}
\end{figure*}

Beyond facility operation, AI is also reshaping detector simulation and design. Fig.~\ref{fig:experiment} illustrates the integration of deep-learning models into the standard experimental physics workflow. While physics emulators rapidly generate particle-level truth from latent parameters, a differentiable surrogate can replace the non-differentiable Geant4 simulation. This critical substitution establishes a continuous gradient path, allowing the direct optimization of detector and sensor parameters based on downstream task performance. Accordingly, the methodology of detector design has evolved from traditional parameter optimization toward physics-driven, end-to-end differentiable design~\cite{Cisbani_2020,DORIGO2023100085,Nguyen_2026,WangJINST2026,Chadeeva:2025ppo,Wang:2025azr,Fan:2024dcu,Tian:2024yfz,Li:2022tvg}.

Generative models, including variational autoencoders, generative adversarial networks, diffusion models, and normalizing flows, are increasingly used to accelerate computationally expensive event simulations in nuclear and particle physics. In high-energy nuclear physics, normalizing flows combined with diffusion models have enabled fast event-by-event heavy-ion collision simulations, providing promising surrogates for transport models and facilitating large-scale Bayesian analyses~\cite{FastHIC2025}. Related developments in particle physics have applied continuous normalizing flows and flow-matching methods to phase-space sampling and unweighted event generation~\cite{FlowMatching2025,RegFlow2025}. Although these methods may also benefit nuclear-reaction simulations, their performance in conventional low- and intermediate-energy nuclear physics remains to be systematically validated.

In the domain of data acquisition and triggering, AI is rapidly transitioning from an offline analysis tool to a core component of real-time, front-end decision-making systems. Deep-learning models are increasingly optimized for deployment on front-end electronics such as field-programmable gate arrays (FPGAs)~\cite{Duarte_2018,Huang2026,Govorkova2026,Aarrestad_2021}. Consequently, trigger systems are now capable of executing AI inference on microsecond timescales, enabling real-time event filtering and adaptive data compression~\cite{Burazin2024}. For instance, in streaming-readout architectures, AI can dynamically adjust compression strategies according to the local signal density at the moment of data generation, significantly reducing bandwidth requirements~\cite{Huang2026,Rossi2025,Kvapil2024,Ding2024NST,HeLei2024NST}.

Although detector optimization, generative event simulation, and FPGA-based triggering have so far been developed more extensively in high-energy nuclear physics and particle physics, direct nuclear-physics applications are already emerging in accelerator control and heavy-ion simulation. Together, these methods provide a general framework for improving the computational efficiency and autonomy of future nuclear-physics facilities, while their applicability to conventional low- and intermediate-energy nuclear physics experiments remains to be systematically validated.

\section{Emerging trends of AI and ML for nuclear physics}
\label{ML-methodology}

Over the past few years, ML and AI have evolved remarkably, transitioning from basic inferences from data to highly versatile and reliable applications.
To further advance ML and AI applications in nuclear physics,  the integration of physics and AI, the usage of LLM, and quantum ML are promising directions, which have already spurred some impressive developments.

\subsection{Physics informed machine learning}
Despite  booming ML applications, nuclear physics often confronts small datasets, so that
it was not deemed as a good playground for machine learning.
Moreover, nuclear reaction data are generally incomplete, noisy and discrepant,
while nuclear structure data have better accuracies but heavily rely on
non-trivial quantum effects.
To overcome these issues in purely data-driven machine learning, physics-informed, or physics-guided
or physics-constrained machine learning has been proposed~\cite{PIML2021}.
Indeed, without embedding physical information, it would be impossible to exploit
 maximum values of scarce data, for instance, the gravitational wave observations.

\begin{figure*}[t]
    \centering
    \includegraphics[width=0.8\textwidth]{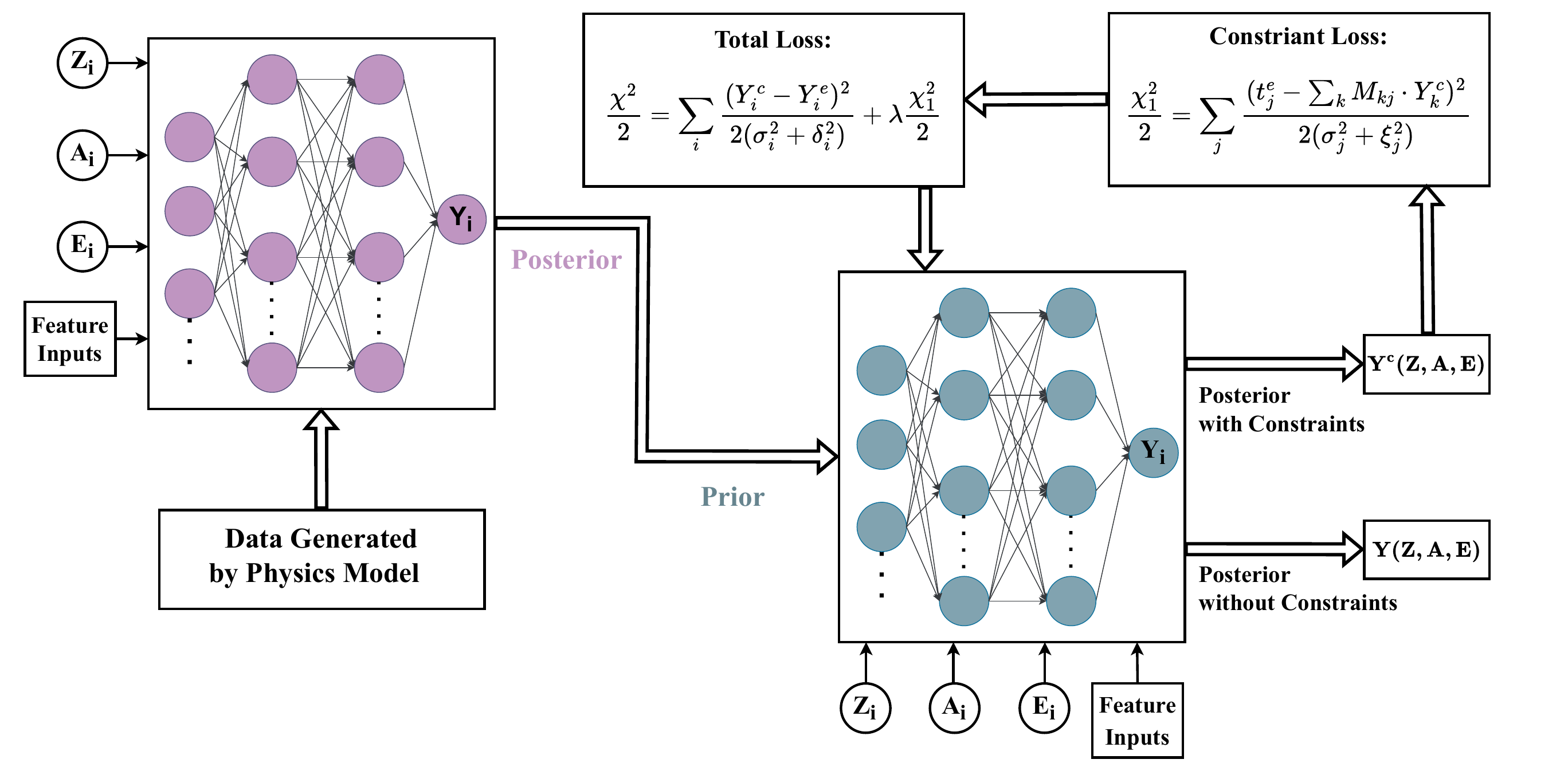}
    \caption{Physics-model-informed Bayesian machine learning for independent fission-yield evaluation, with and without physics constraints. Physics-model-generated data train informative priors for the evaluation of measured data, while heterogeneous cumulative yields constrain the energy dependence through a conversion matrix. Figure taken from Ref.~\cite{liu2026PIBM}.}
    \label{fig:physics_informed_bayesian}
\end{figure*}

There are several typical strategies to inform machine learning by using feature inputs including empirical shell and odd-even effects,
 learning the residual error of physics models, or penalizing learning with constraints.
It is known that the physics informed neural network (PINN) has been very successful in solving differential equations~\cite{PIML2021}.
However, to inform neural networks with comprehensive physics knowledge is another matter and is particularly needed in nuclear physics.
The physics informed machine learning is a more board concept compared to PINN. 
In principle, Bayesian machine learning has the advantage of incorporating physics priors.
It is a promising choice to inform the priors of Bayesian machine learning by physics models, which contain
comprehensive conservation laws and quantum effects consistently, to exploit the maximum values of sparse and imperfect nuclear data~\cite{liu2026PIBM}.
A representative physics-model-informed Bayesian workflow is shown in Fig.~\ref{fig:physics_informed_bayesian}: model-generated data define informative priors, while measured cumulative yields provide an additional physics constraint on sparse independent fission-yield data.
In addition, model mixing can also incorporate physics by learning the residuals of different models. The Gaussian process approach is versatile for performing model mixing and pre-training priors. Transfer learning can be used to pretrain physics informed priors for general neutral networks. Compard to Gaussian process regression that learns physics models by limited residuals, the physics informed priors can effectively integrate physics knowledge and non-local sparse data correlations. With physics informed priors, the learning convergence can be achieved rapidly. In addition, the overfitting issues can also be avoided with sufficient model generated data. 

\subsection{AI PDE Solver in nuclear physics}

Partial differential equations (PDEs) constitute the theoretical backbone of modern nuclear physics, underpinning the description of both nuclear structure and reaction dynamics. 
However, their intrinsic nonlinearity, high dimensional phase space, and complex initial and boundary conditions make them extremely challenging to solve with conventional numerical approaches, for example, extreme computational costs, the high trial-and-error cost of inverse problems, and practical constraints such as missing boundary or initial conditions.

\begin{figure*}[t]
		\centering
		\includegraphics[width=0.6\textwidth]{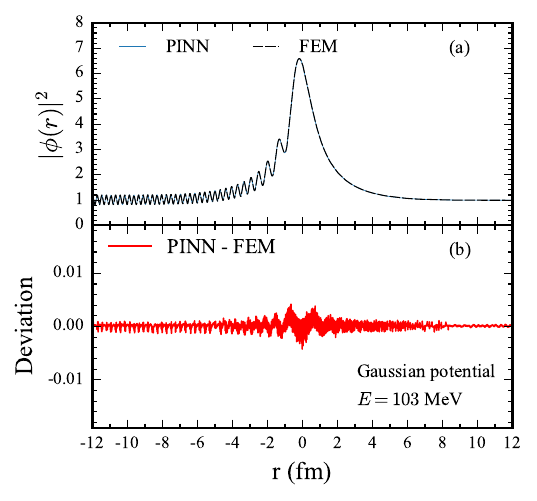}
		\caption{ 
			 Squared modulus of wavefunctions \(( |\phi(r)|^2 )\) 	(a) and  the deviations (b) calculated using PINN and FEM under the one-dimensional Gaussian potential at incident energy (E = 103  \text{MeV}) ~\cite{Wen2025}. The solid lines represent the PINN results, while dashed lines denote the FEM solutions.} 
		\label{fig:pinn}
\end{figure*}

PINN have been applied for the first time to solve the multi-dimensional Schr\"odinger equation for the quantum tunneling problem in Ref. \cite{Wen2025}. The physical constraints, including the coupled-channel equations,
boundary derivative conditions, and flux conservation, are integrated directly into the loss
function to satisfy the underlying physics. 
The framework demonstrates good agreement with high-accuracy finite element method (FEM)
calculations, as shown in Fig.\ref{fig:pinn}. A notable practical advantage of this approach is its
capacity for transfer learning, where pre-trained neural networks can be saved and fine-tuned
for calculations at adjacent energy points, significantly reducing the required training epochs for batch calculations. Future developments of PINNs in nuclear theory are expected to explore adaptive sampling strategies that further
alleviate the curse of dimensionality. 

Recent advances in AI have opened new avenues for solving partial differential equations (PDEs). In particular, operator-learning methods, such as DeepONets\cite{Lu2021DeepONet} and Fourier Neural Operators (FNOs)\cite{Li2021FNO}, learn mappings between functional inputs and outputs, enabling fast predictions of system evolution across a wide range of physical conditions. More recently, the field has begun to move beyond task-specific models toward multi-operator learning and foundation-model architectures. These approaches aim to construct general-purpose PDE solvers capable of handling different classes of nuclear dynamics, from mean-field and density-functional descriptions to transport phenomena, while simultaneously supporting forward simulations, parameter inference, and inverse problems within a unified framework. Furthermore, the development of large-scale scientific foundation models tailored to nuclear systems may enable real-time simulations of heavy-ion collisions, efficient uncertainty quantification, and data-driven discovery of emergent phenomena in nuclear many-body systems. These advances have the potential to bridge microscopic theory and experimental observables more effectively, opening new avenues for AI-driven nuclear science.
	
\subsection{Symbolic regression and automated model discovery}
Symbolic regression (SR) searches over both functional form and numerical coefficients and therefore differs from fitting a prescribed empirical formula. It should not, however, be equated automatically with the discovery of a physical law. In nuclear applications, an equation learned from measured systematics is first an interpretable phenomenological model. A stronger claim requires stable recovery of invariant structure, validation on genuinely unobserved nuclei or reactions, and consistency with dimensions, symmetries, thresholds, conservation laws, and uncertainties in the nuclear data. AI-Feynman reduces a general symbolic search by testing symmetries and separability~\cite{AIFeynman}, but its success on generic benchmarks is not itself a nuclear-physics result.

\begin{figure*}[t]
    \centering
    \includegraphics[width=0.55\textwidth]{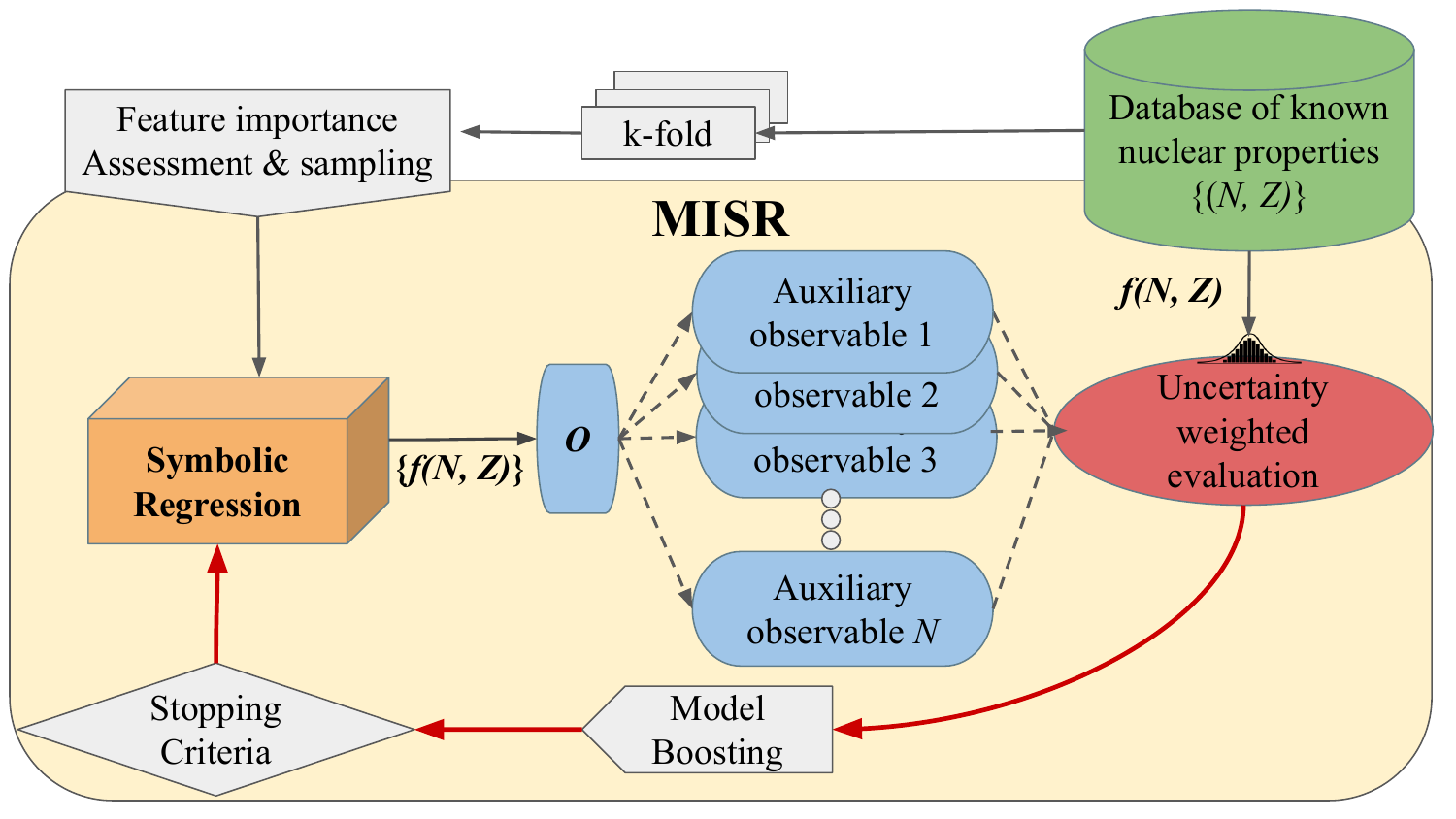}
    \caption{Inner pipeline of multi-objective iterated symbolic regression (MISR). Symbolic models are iteratively refined using auxiliary nuclear observables, uncertainty-weighted evaluation, feature assessment, and model boosting. Figure taken from Ref.~\cite{Munoz2025NuclearSymbolic}.}
    \label{fig:symbolic_regression_misr}
\end{figure*}

A direct nuclear-structure demonstration is the multi-objective iterated symbolic-regression study of Munoz, Udrescu, and Garcia Ruiz~\cite{Munoz2025NuclearSymbolic}. 
The iterative relation among symbolic regression, auxiliary observables, uncertainty weighting, and model boosting in the nuclear MISR framework is summarized in Fig.~\ref{fig:symbolic_regression_misr}.
Applied to AME2020 binding energies and measured charge radii for nuclei with $12\leq Z\leq 50$, it generated compact functions of proton and neutron numbers and selected nuclear-structure descriptors. For the ten-term binding-energy model, the quoted MAE and RMSE were 0.78 and 0.99~MeV, respectively; for charge radii, both the training and test MAE were 0.009~fm. The leading terms recovered familiar liquid-drop- and $A^{1/3}$-like behavior, while additional terms described shell-region corrections. Combining the discovered binding-energy expression with Duflo--Zuker predictions through automatic relevance determination reduced the training and test MAE to 0.389 and 0.411~MeV and was then used to estimate nuclear-stability limits from one- and two-nucleon separation energies. This is substantive automated construction of a nuclear phenomenology, but not autonomous discovery of the nuclear interaction: the input descriptors included isospin asymmetry, valence-nucleon counts, the Casten factor, and prescribed magic numbers, and the reported validation used a uniformly sampled 20\% holdout rather than withholding complete isotopic chains or nuclear regions.

SR has also produced genuinely nuclear but more limited decay systematics. Cheng \textit{et al.} derived compact data-driven expressions for $\alpha$ decay and proton emission~\cite{Cheng2024DecaySR}. Shree and Balasubramaniam fitted 22 measured cluster-decay modes and obtained a readable half-life relation before extrapolating it to candidate emitters~\cite{Shree2025ClusterSR}; a later study combined ML and SR for $\alpha$ decay of superheavy nuclei~\cite{Shree2026NPA}. These studies show that SR can compress known tunnelling and Coulomb-barrier trends into analytical formulas. They do not yet establish new decay laws because the samples are small, physically motivated composite variables are supplied in advance, and independent out-of-domain validation is limited. Likewise, KAN-based binding-energy work can expose liquid-drop-like contributions~\cite{LiuH2025PRC}, but a learnable analytic representation is not by itself autonomous law discovery.

\subsection{Transferable representations and foundation models for nuclear physics}

A model should not be called a nuclear-physics foundation model merely because it is large or uses a transformer. The operational test is whether a shared pretrained representation can be adapted to several nuclear tasks, observables, or instruments with less labelled data or computation than comparable models trained separately from scratch; task-specific representation-learning studies are treated below as precursors rather than as foundation models. By this criterion, nuclear physics has credible proto-foundation models and transferable-representation demonstrations, but no model has yet shown broad, experimentally validated transfer across nuclear structure, reactions, and detector systems.

\begin{figure*}[t]
    \centering
    \includegraphics[width=0.5\textwidth]{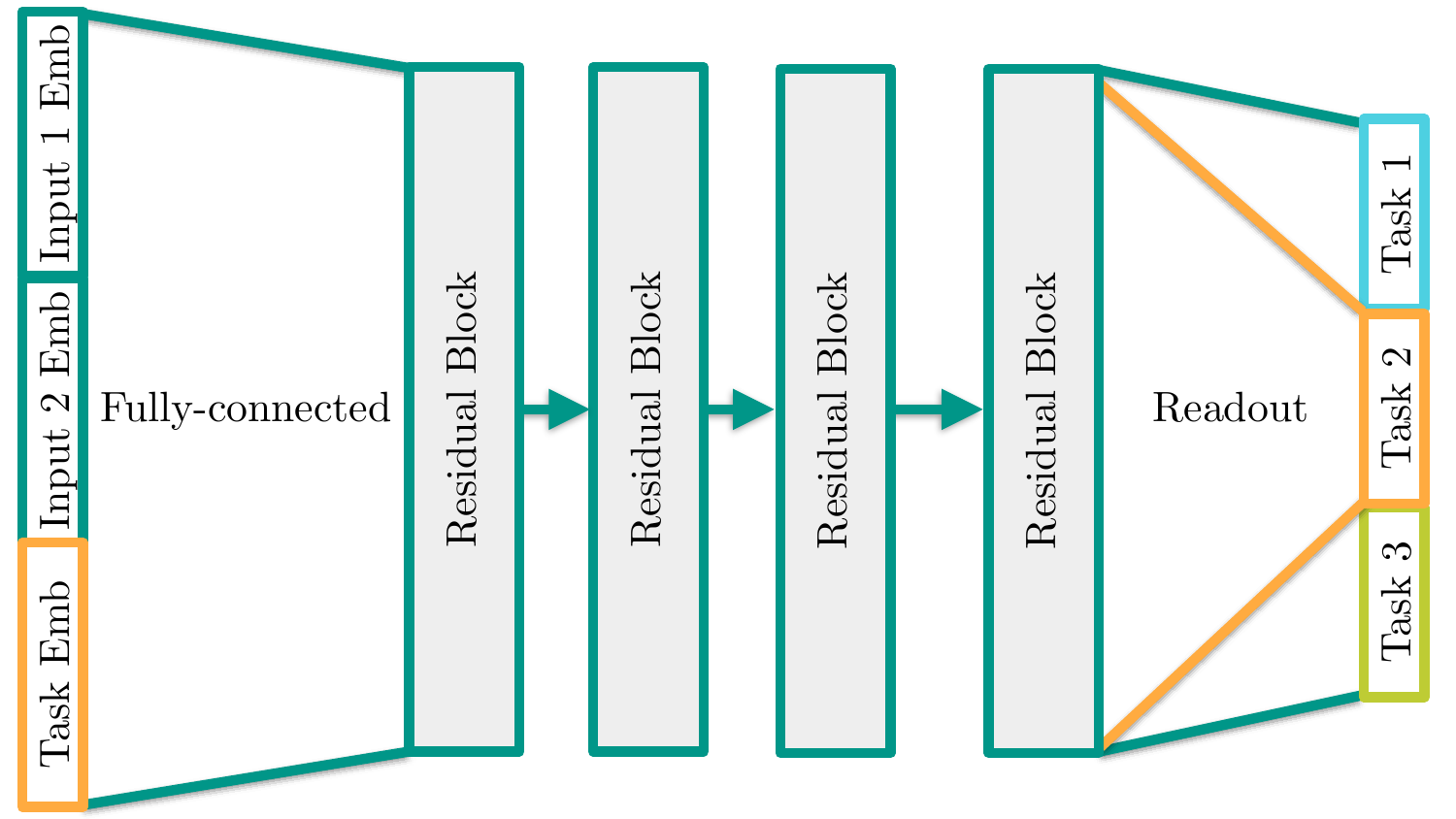}
    \caption{NuCLR multi-task-with-embeddings architecture. Proton-number, neutron-number, and task embeddings are concatenated and processed by shared residual blocks; task-specific readout heads then predict different nuclear observables. Sharing the nuclide embeddings across tasks encourages a task-independent nuclear representation. Figure taken from Ref.~\cite{Kitouni2023NuCLR}.}
    \label{fig:nuclr_mte}
\end{figure*}

Direct nuclear examples are emerging at several scales. The NuCLR workshop study co-learns binding and separation energies, decay quantities, and charge radii through a shared nuclide representation, whose latent space reproduces shell-related regularities~\cite{Kitouni2023NuCLR}. 
Its multi-task-with-embeddings architecture, shown in Fig.~\ref{fig:nuclr_mte}, uses shared proton and neutron embeddings together with a task embedding to learn task-independent nuclear information.
For reaction data, learned latent representations combined with graph neural networks have imputed withheld regions of neutron-induced cross sections on the nuclear chart and recovered some magic-number structure~\cite{Choi2025NuclearCrossSections}.
At the detector level, a recent preprint reports that a sparse encoder pretrained on GADGET II events transferred to a distinct AT-TPC dataset~\cite{Wheeler2025TPCEmbeddings}. This is direct cross-instrument representation transfer on low-energy nuclear data, while remaining a limited two-detector proof of concept.

The strongest current proto-foundation demonstrations involve nuclear facilities with collider-like or sequence-valued readout. A next-token mixture-of-experts model for the future EIC hpDIRC used one backbone for Cherenkov-hit generation, pion--kaon identification, and noise filtering~\cite{Giroux2026ReadoutFM}; a subsequent GlueX DIRC preprint reused the architecture, rather than a cross-detector pretrained checkpoint, and reported particle-identification and hit-noise AUC values of 0.952 and 0.971~\cite{Fanelli2026GlueXFM}. Both studies use Geant4 simulation rather than collision data. FM4NPP scales the idea to more than 11 million fully simulated sPHENIX TPC events and frozen-backbone adapters for track finding, particle identification, and noise tagging~\cite{park2026fmnpp}. Its sample consists of $pp$ collisions at $\sqrt{s}=200$~GeV, so transfer to the high-occupancy heavy-ion events central to the sPHENIX program and to other nuclear facilities remains unproven. In heavy-ion theory, the HEIDi point-cloud diffusion model generates complete UrQMD Au--Au events with 26 hadron species and an approximately two-order-of-magnitude speedup~\cite{Kuttan2025PointCloudDiffusion}. This is direct high-energy nuclear evidence, but it emulates one transport generator under restricted conditions, retains discrepancies in low-$p_T$ and conservation-sensitive fluctuations, and is explicitly a step toward rather than a completed foundation model.

Most named physics foundation models remain imported evidence. Collider models such as OmniJet-$\alpha$, OmniLearn, HEP-JEPA, EveNet, and Bumblebee demonstrate masked or generative pretraining for jets and collider-event particle sets~\cite{Birk_2024,knmd-f5jm,bardhan2025,hsu2026,wildridge2024bumblebee}, not validation on nuclei. An informative bridge is transfer from a collider-pretrained representation to simulated MINERvA neutrino--nucleus events, where pretraining improved available-energy regression and pion-final-state classification relative to comparably sized or identically architected models trained from scratch~\cite{krzmanc2026}; this supports transfer of geometric and kinematic priors but remains simulation-based neutrino evidence. Poseidon, GPhyT, PhysiX, and Walrus demonstrate transfer among classical continuum or PDE simulations~\cite{herde2024poseidonefficientfoundationmodels,wiesner2026physicsfoundationmodel,nguyen2025physixfoundationmodelphysics,mccabe2025}; none has yet been benchmarked on nuclear TDDFT, transport, reaction dynamics, or nuclear Hamiltonians. AION-1 is an astronomy model~\cite{parker2025}, and ColliderML is a collider benchmark dataset rather than a nuclear foundation model~\cite{elitez2025collider}; they are therefore not counted as nuclear demonstrations.

The nuclear impact of this trend should ultimately be assessed through benchmarkable transfer. Candidate models must be tested on unseen nuclei, observables, experimental runs, or facilities, with particular attention to calibrated uncertainties, domain shifts, physical symmetries, and conservation laws. These tests, rather than parameter count or model naming, will determine whether foundation-model technology becomes a genuine nuclear-physics capability.

\subsection{Quantum computing and quantum machine learning for nuclear physics}

Although quantum simulations and quantum computing in nuclear physics have their own limitations, they are still considered emerging fields that could address many nuclear physics problems. Several long-term plans and white papers have already considered and promoted these new research avenues, such as Refs.~\cite{Carlson2018,Cloet2019NPQI} and the white paper~\cite{Beck2023QISNP}. A pioneering application of quantum algorithms to nuclear many-body problem is for deuteron in Ref.~\cite{Dumitrescu2018}, and many efforts have followed by using the different quantum algorithm, such as VQE and FQE ~\cite{Cervia2022PPNP,Jiang:2026gci,BNPeng2026,DBZhang2021}.


The field of quantum machine learning (QML) has grown significantly in the past five years~\cite{Wang:2026vti,Zhang:2025rzl,Fang:2024ple}, and several theoretical proposals, as well as experimental realizations, have been produced on platforms such as superconducting circuits, quantum photonics, and trapped ions. QML models inspired by biology, i.e., quantum biomimetics, have been proposed. However, the use of QML protocols for the analysis of nuclear physics is a relatively unexplored field so far, although certain examples of use already exist in condensed matter or quantum chemistry. There are several proposals in nuclear physics that QML can provide computational advantage in the near future\cite{Ramos2023}: 
determining the phase/shape in schematic nuclear models, calculating the ground state energy of a nuclear shell model-type Hamiltonian, and identifying particles or determining trajectories in nuclear physics experiments.

\section{Summary and outlook}
\label{summary}

Artificial intelligence (AI) and machine learning (ML) have rapidly emerged as transformative tools in nuclear physics, profoundly reshaping the paradigms of theoretical modeling, experimental analysis, detector optimization, and scientific discovery. Taking advantage of advances in deep learning, probabilistic inference, generative modeling, and large-scale computing infrastructures, AI has evolved from a supplementary numerical technique into a powerful interdisciplinary framework capable of addressing many long-standing challenges in nuclear science.

This review summarizes recent progress of AI and ML applications across a broad range of topics in nuclear physics. ML methods have significantly improved the predictive power for nuclear masses, charge radii, decay half-lives, level densities, and spectroscopic observables. These methods can improve prediction and interpretation. However, strict extrapolation tests remain difficult. In nuclear many-body theory, neural-network quantum states, variational neural wavefunctions, and AI-assisted energy density functional approaches have demonstrated remarkable potential for solving ab initio quantum many-body problems with reduced computational cost while maintaining high precision. In nuclear reactions and heavy-ion collisions, AI-based surrogate models and emulators have become increasingly important for accelerating transport simulations, uncertainty quantification, parameter inference, and inverse problems. ML has also shown strong capabilities in fission-yield evaluation, resonance reconstruction, fusion cross-section prediction, and equation-of-state extraction from experimental observables. Particularly notable is the growing trend toward physics-informed Bayesian learning, which incorporates conservation laws, quantum effects, and theoretical priors directly into the learning framework, thereby overcoming the limitations of purely data-driven approaches under sparse and imperfect nuclear datasets. AI technologies are also profoundly transforming experimental nuclear physics. Deep learning-based trigger systems, differentiable detector simulations, FPGA-deployed neural networks, and transformer-based event reconstruction methods are becoming essential components of next-generation facilities.

Beyond current applications, several emerging directions may fundamentally reshape future nuclear physics research. Physics-informed machine learning provides a promising framework for embedding symmetries, conservation laws, and microscopic theoretical constraints into AI architectures. AI-driven PDE solvers, operator learning approaches, and neural differential equation methods may revolutionize simulations of quantum many-body dynamics, transport equations, and heavy-ion evolution by dramatically reducing computational complexity. Symbolic regression can build interpretable empirical models. Autonomous discovery of new nuclear laws has not yet been demonstrated. Meanwhile, the development of physics foundation models and tokenized scientific datasets represents a major paradigm shift toward general-purpose AI systems for physics. Future foundation models trained on large-scale multimodal datasets may provide unified representations that span detector signals, theoretical simulations, and experimental observables, enabling transferable reasoning capabilities across multiple nuclear physics tasks. Integration of large language models with symbolic reasoning, simulation tools, and first-principles calculations may further accelerate theoretical derivations, model construction, and automated scientific workflows.

Looking ahead, the future of AI for nuclear physics will likely move beyond isolated applications toward deeply integrated scientific ecosystems that combine data, theory, simulation, experiments, and autonomous reasoning within unified frameworks. The convergence of AI, high-performance computing, differentiable programming, quantum computing, and first-principles nuclear theory may eventually establish a new paradigm for scientific discovery in nuclear physics. Rather than replacing physicists, AI is expected to become an increasingly powerful collaborator that enhances human creativity, accelerates discovery cycles, and enables the exploration of physical regimes previously beyond computational and experimental reach. Such developments will not only deepen our understanding of nuclear matter and fundamental interactions but also may drive technological innovations in energy, medicine, astrophysics, national security, and quantum science.

Despite the significant opportunities offered by AI in nuclear physics, several challenges must be carefully addressed. ML models often suffer from poor out-of-distribution generalization, limiting their reliability beyond training data. Models trained on simulations may inherit theoretical assumptions and systematic biases, leading to simulator bias and unreliable inference unless rigorously validated against experimental data. In areas involving rare nuclear processes, limited experimental information restricts the reliability of purely data-driven approaches. Furthermore, LLM may generate hallucinated or uninterpretable results, requiring continuous human expert oversight and robust validation. AI applications in nuclear science also raise dual-use concerns. Addressing these challenges requires advances in uncertainty quantification, physics-informed learning, interpretability, reproducibility, and responsible AI practices to ensure that AI serves as a reliable and trustworthy tool for nuclear physics research.

\section*{Acknowledgements}
We are thankful for the support of the National Key R\&D Program of China under Grant Nos. 2023YFA1606601, 2023YFA1606402, and the National Natural Science Foundation of China with Nos. 12375109, 12275359, 12475187, 12335008, 12475118, and 12547102, and the STCSM under Grant No.~23590780100 and ~23JC1400200. 

\bibliographystyle{unsrt}
\bibliography{myref}

@article{WangYF_2026,
  title = {Artificial intelligence empowers particle physics and nuclear physics: From fundamental research to major applications},
  author = { Wang, Y. F. and He, Y. and  Huang, Y. and others  },
  journal = { Bulletin of Chinese Academy of Sciences},
  volume = {41},
  pages = {1089--1102},
  year = {2026},
  doi={10.3724/j.issn.1000-3045.20260415007}
}

@article{ZhangS_2017,
  title = {Nuclear cluster structure effect on elliptic and triangular flows in heavy-ion collisions},
  author = {Zhang, S. and Ma,  Y. G. and Chen, J. H. and He, W. B. and Zhong, C.},
  journal = {Phys. Rev. C},
  volume = {95},
  pages = {064904},
  year = {2017},
doi={10.1103/PhysRevC.95.064904}
}

@article{Ma_2023,
  title = {Influence of Nuclear Structure in Relativistic Heavy-Ion Collisions},
  author = { Ma,  Y. G. and Zhang, S.},
  journal = {A Chapter in Handbook of Nuclear Physics, Springer, Singapore.},
  volume = {},
  pages = {},
  year = {2023},
doi={10.1007/978-981-19-6345-2_5}
}

@article{STAR_2024,
  title = {Imaging shapes of atomic nuclei in high-energy nuclear collisions},
  author = {Abdulhamid,  M. I.  and Aboona,  B. E. and Adam, J. and others },
  journal = {Nature},
  volume = {635},
  pages = {67},
  year = {2024},
doi={10.1038/s41586-024-08097-2}
}

@article{HeJJ_2021,
  title = {Machine-learning-based identification for initial clustering structure in relativistic heavy-ion collisions},
  author = {He, J. J. and He, W. B. and Ma, Y. G.  and  Zhang, S. },
  journal = {Phys. Rev. C},
  volume = {104},
  pages = {044902},
  year = {2021},
doi={10.1103/PhysRevC.104.044902}
}

@article{JiaNST,
author = {Jia, J. and Giacalone, G. and Bally, B. and others},
year = {2024},
pages = {220},
title = { Imaging the initial condition of heavy-ion collisions and nuclear 
structure across the nuclide chart},
volume = {35},
journal = {Nuclear Science and Techniques},
doi = {10.1007/s41365-024-01589-w}
}

@article{GiacaloneNST,
author = {Giacalone, G.},
year = {2024},
pages = {218},
title = {Beyond axial symmetry: high-energy collisions unveil the ground-state shape of 238U},
volume = {35},
journal = {Nuclear Science and Techniques},
doi = {10.1007/s41365-024-01582-3}
}

@article{SchenkeeNST,
author = {Schenke, B.},
year = {2024},
pages = {115},
title = {Violent collisions can reveal hexadecapole deformation of nuclei},
volume = {35},
journal = {Nuclear Science and Techniques},
doi = {10.1007/s41365-024-01509-y}
}

@article{FangNST,
author = {Ding, M. Q.  and Fang, D. Q.  and Ma, Y. G. },
year = {2024},
pages = {211},
title = {Neutron skin and its effects in heavy-ion collisions},
volume = {35},
journal = {Nuclear Science and Techniques},
doi = {10.1007/s41365-024-01584-1}
}

@article{MaYG_arxiv,
  title = {Frontier Questions and Emerging Directions in Nuclear Science and Technology},
  author = {Ma, Y. G.},
  journal = {Nucl. Sci. Tech.},
  volume = {37},
  pages = {in press},
  year = {2026},
      eprint = "2608.26207",
    archivePrefix = "arXiv",
    primaryClass = "nucl-th",
    year = "2026"
}

@misc{BayesianNSEoS2024,
  author = {Imam, Sk Md Adil and Patra, N. K.},
  title = {Bayesian Analysis of the Neutron Star Equation of State and Model Comparison: Insights from {PSR J0437+4715}, {PSR J0614+3329}, and Other Multi-Physics Data},
  year = {2025},
  eprint = {2509.07109},
  archivePrefix = {arXiv},
  primaryClass = {nucl-th},
  url = {https://arxiv.org/abs/2509.07109}
}

@article{Catacora,
  author = {Catacora-Rios, M. and Beyer, K. and Giuliani, P. and Godbey, K. and Furnstahl, R. J. and Nunes, F. M.},
  title = {Wavefunction-based emulation of coupled-channels scattering with nonaffinely parametrized interactions},
  journal = {Physical Review C},
  volume = {113},
  pages = {044623},
  year = {2026},
  doi = {10.1103/tgf9-f2st}
}

@article{Cheng2024DecaySR,
  author = {Cheng, Junhao and Wang, Binglin and Zhang, Wenyu and Duan, Xiaojun and Yu, Tongpu},
  title = {Study {$\alpha$} decay and proton emission based on data-driven symbolic regression},
  journal = {Computer Physics Communications},
  volume = {304},
  pages = {109317},
  year = {2024},
  doi = {10.1016/j.cpc.2024.109317}
}

@article{Choi2025NuclearCrossSections,
  author = {Choi, Hongjun and Mitra, Sinjini and Brodksy, Jason and Glatt, Ruben and Holmbeck, Erika and Liu, Shusen and Schunck, Nicolas and Sieverding, Andre and Wendt, Kyle},
  title = {Learning nuclear cross sections across the chart of nuclides with graph neural networks},
  journal = {Physical Review C},
  volume = {112},
  pages = {044601},
  year = {2025},
  doi = {10.1103/4jd4-bnyh}
}

@article{DeepLearningNSEoS2024,
  author = {Fujimoto, Yuki and Fukushima, Kenji and Murase, Koichi},
  title = {Mapping neutron star data to the equation of state using the deep neural network},
  journal = {Physical Review D},
  volume = {101},
  pages = {054016},
  year = {2020},
  doi = {10.1103/PhysRevD.101.054016}
}

@article{Drischler2021,
  author = {Drischler, C. and Quinonez, M. and Giuliani, P. G. and Lovell, A. E. and Nunes, F. M.},
  title = {Toward emulating nuclear reactions using eigenvector continuation},
  journal = {Physics Letters B},
  volume = {823},
  pages = {136777},
  year = {2021},
  doi = {10.1016/j.physletb.2021.136777}
}

@article{Drischler2023,
  author = {Drischler, C. and Melendez, J. A. and Furnstahl, R. J. and Garcia, A. J. and Zhang, Xilin},
  title = {{BUQEYE} guide to projection-based emulators in nuclear physics},
  journal = {Frontiers in Physics},
  volume = {10},
  pages = {1092931},
  year = {2023},
  doi = {10.3389/fphy.2022.1092931}
}

@misc{Fanelli2026GlueXFM,
  author = {Fanelli, Cristiano and Giroux, James and Granger, Cole and Stevens, Justin},
  title = {Application of a Mixture of Experts-based Foundation Model to the {GlueX DIRC} Detector},
  year = {2026},
  eprint = {2604.24775},
  archivePrefix = {arXiv},
  primaryClass = {physics.data-an}
}

@misc{FastHIC2025,
  author = {Omana Kuttan, Manjunath and Zhou, Kai and Steinheimer, Jan and Stoecker, Horst},
  title = {Ultra fast, event-by-event heavy-ion simulations for next generation experiments},
  year = {2025},
  eprint = {2502.16330},
  archivePrefix = {arXiv},
  primaryClass = {nucl-th},
  url = {https://arxiv.org/abs/2502.16330}
}

@misc{Feickert2021LivingReview,
  author = {Feickert, Matthew and Nachman, Benjamin},
  title = {A Living Review of Machine Learning for Particle Physics},
  year = {2021},
  eprint = {2102.02770},
  archivePrefix = {arXiv},
  primaryClass = {hep-ph}
}

@misc{FlowMatching2025,
  author = {Bothmann, Enrico and Janssen, Timo and Knobbe, Max and Schmitzer, Bernhard and Sinz, Fabian},
  title = {Efficient many-jet event generation with Flow Matching},
  year = {2025},
  eprint = {2506.18987},
  archivePrefix = {arXiv},
  primaryClass = {hep-ph},
  url = {https://arxiv.org/abs/2506.18987}
}

@article{Furnstahl,
  author = {Furnstahl, R. J. and Garcia, A. J. and Millican, P. J. and Zhang, Xilin},
  title = {Efficient emulators for scattering using eigenvector continuation},
  journal = {Physics Letters B},
  volume = {809},
  pages = {135719},
  year = {2020},
  doi = {10.1016/j.physletb.2020.135719}
}

@article{Giroux2026ReadoutFM,
  author = {Giroux, J. and Fanelli, C.},
  title = {Towards foundation models for experimental readout systems combining discrete and continuous data},
  journal = {Machine Learning: Science and Technology},
  volume = {7},
  number = {1},
  pages = {015031},
  year = {2026},
  doi = {10.1088/2632-2153/ae3d81},
  eprint = {2505.08736},
  archivePrefix = {arXiv}
}

@misc{HadronQuarkCrossover2025,
  author = {Grundler, Xavier and Li, Bao-An},
  title = {Bayesian Constraints on the Neutron Star Equation of State with a Smooth Hadron-Quark Crossover},
  year = {2026},
  eprint = {2602.06696},
  archivePrefix = {arXiv},
  primaryClass = {nucl-th},
  url = {https://arxiv.org/abs/2602.06696}
}

@misc{Kitouni2023NuCLR,
  author = {Kitouni, Ouail and Nolte, Niklas and Trifinopoulos, Sokratis and Kantamneni, Subhash and Williams, Mike},
  title = {{NuCLR}: Nuclear Co-Learned Representations},
  year = {2023},
  eprint = {2306.06099},
  archivePrefix = {arXiv},
  primaryClass = {nucl-th}
}

@article{Kuttan2025PointCloudDiffusion,
  author = {Omana Kuttan, Manjunath and Zhou, Kai and Steinheimer, Jan and Stoecker, Horst},
  title = {Toward a foundation model for heavy-ion collision experiments based on point-cloud diffusion},
  journal = {Physical Review C},
  volume = {112},
  pages = {L051902},
  year = {2025},
  doi = {10.1103/6ndd-d1nl}
}

@misc{MicroscopicNSConstraints2025,
  author = {Semposki, A. C. and Drischler, C. and Furnstahl, R. J. and Phillips, D. R.},
  title = {Microscopic constraints for the equation of state and structure of neutron stars: a Bayesian model mixing framework},
  year = {2025},
  eprint = {2505.18921},
  archivePrefix = {arXiv},
  primaryClass = {nucl-th},
  url = {https://arxiv.org/abs/2505.18921}
}

@article{Munoz2025NuclearSymbolic,
  author = {Munoz, Jose M. and Udrescu, Silviu M. and Garcia Ruiz, Ronald F.},
  title = {Discovering nuclear models from symbolic machine learning},
  journal = {Communications Physics},
  volume = {8},
  number = {1},
  pages = {101},
  year = {2025},
  doi = {10.1038/s42005-025-02023-2},
  eprint = {2404.11477},
  archivePrefix = {arXiv}
}

@article{Odell,
  author = {Odell, D. and Giuliani, P. and Beyer, K. and Catacora-Rios, M. and Chan, M. Y.-H. and Bonilla, E. and Furnstahl, R. J. and Godbey, K. and Nunes, F. M.},
  title = {{ROSE}: A reduced-order scattering emulator for optical models},
  journal = {Physical Review C},
  volume = {109},
  pages = {044612},
  year = {2024},
  doi = {10.1103/PhysRevC.109.044612}
}

@misc{PhysicsInformedBNN2024,
  author = {Baker, J. D. and Bertulani, C. A. and Lobato, R. V.},
  title = {A Physics Informed Bayesian Neural Network for the Neutron Star Equation of State},
  year = {2026},
  eprint = {2604.24949},
  archivePrefix = {arXiv},
  primaryClass = {nucl-th},
  url = {https://arxiv.org/abs/2604.24949}
}

@misc{RegFlow2025,
  author = {Bothmann, Enrico and Janssen, Timo and Knobbe, Max and Schmitzer, Bernhard and Sinz, Fabian},
  title = {Monte Carlo Event Generation with Continuous Normalizing Flows},
  year = {2026},
  eprint = {2604.03511},
  archivePrefix = {arXiv},
  primaryClass = {hep-ph},
  url = {https://arxiv.org/abs/2604.03511}
}

@article{Shree2025ClusterSR,
  author = {{Madhumitha Shree}, S. and Balasubramaniam, M.},
  title = {Empirical relations using symbolic regression models for cluster decay half-lives},
  journal = {Physical Review C},
  volume = {111},
  number = {6},
  pages = {064605},
  year = {2025},
  doi = {10.1103/PhysRevC.111.064605}
}

@misc{SymbolicRegressionNS2024,
  author = {Imam, Sk Md Adil and Saxena, Prafulla and Malik, Tuhin and Patra, N. K. and Agrawal, B. K.},
  title = {Calibrating global behaviour of equation of state by combining nuclear and astrophysics inputs in a machine learning approach},
  year = {2024},
  eprint = {2407.08553},
  archivePrefix = {arXiv},
  primaryClass = {nucl-th},
  url = {https://arxiv.org/abs/2407.08553}
}

@misc{Wheeler2025TPCEmbeddings,
  author = {Wheeler, Tyler and Kuchera, Michelle P. and Ramanujan, Raghuram and Krupp, Ryan and Wrede, Chris and Ravishankar, Saiprasad and Cross, Connor L. and Heung, Hoi Yan Ian and Jones, Andrew J. and Votaw, Benjamin},
  title = {Sparse Methods for Vector Embeddings of {TPC} Data},
  year = {2025},
  eprint = {2511.11221},
  archivePrefix = {arXiv},
  primaryClass = {cs.LG}
}

@article{Zhou2024QCDML,
  author = {Zhou, Kai and Wang, Lingxiao and Pang, Long-Gang and Shi, Shuzhe},
  title = {Exploring {QCD} matter in extreme conditions with machine learning},
  journal = {Progress in Particle and Nuclear Physics},
  volume = {135},
  pages = {104084},
  year = {2024},
  doi = {10.1016/j.ppnp.2023.104084}
}

@article{Miskovich2022SECAR,
  author = {Miskovich, S. A. and Montes, F. and Berg, G. P. A. and Blackmon, J. and Chipps, K. A. and Couder, M. and Deibel, C. M. and Hermansen, K. and Hood, A. A. and Jain, R. and Ruland, T. and Schatz, H. and Smith, M. S. and Tsintari, P. and Wagner, L.},
  title = {Online Bayesian optimization for a recoil mass separator},
  journal = {Physical Review Accelerators and Beams},
  volume = {25},
  number = {4},
  pages = {044601},
  year = {2022},
  doi = {10.1103/PhysRevAccelBeams.25.044601}
}

@article{Melendez2022MOR,
  author = {Melendez, J. A. and Drischler, C. and Furnstahl, R. J. and Garcia, A. J. and Zhang, Xilin},
  title = {Model reduction methods for nuclear emulators},
  journal = {Journal of Physics G: Nuclear and Particle Physics},
  volume = {49},
  number = {10},
  pages = {102001},
  year = {2022},
  doi = {10.1088/1361-6471/ac83dd}
}

@article{Zhang2023GCMML,
  author = {Zhang, X. and Lin, W. and Yao, J. M. and Jiao, C. F. and Romero, A. M. and Rodr{\'i}guez, T. R. and Hergert, H.},
  title = {Optimization of the generator coordinate method with machine-learning techniques for nuclear spectra and neutrinoless double-beta decay},
  journal = {Physical Review C},
  volume = {107},
  pages = {024304},
  year = {2023},
  doi = {10.1103/PhysRevC.107.024304}
}

@article{Belley2024NMEUQ,
  author = {Belley, A. and others},
  title = {{Ab initio} uncertainty quantification of neutrinoless double-beta decay in $^{76}$Ge},
  journal = {Physical Review Letters},
  volume = {132},
  pages = {182502},
  year = {2024},
  doi = {10.1103/PhysRevLett.132.182502}
}

@article{Zhang2025SPCDFT,
  author = {Zhang, X. and Wang, C. C. and Ding, C. R. and Yao, J. M.},
  title = {Subspace-projected multireference covariant density functional theory},
  journal = {Physical Review C},
  volume = {112},
  pages = {L021302},
  year = {2025},
  doi = {10.1103/4cnl-5dnm}
}

@book{book,
    author = {Artificial Intelligence Advisory Research Group of the Academic Divisions of the Chinese Academy of Sciences},
    title = {Artificial Intelligence for Science -- The Disciplinary System of Artifical Intelligence},
    publisher = {Science Press},
    year = {2026}
}

@article{ChenXiaolong2024SCPMA,
author = {Chen, Xiaolong and Wang, Zhijun and He, Yuan and Zhao, Hong and Su, Chunguang and Liu, Shuhui and Chen, Weilong and Zhao, Xiaoying and Qi, Xin and Sun, Kunxiang and Jin, Chao and Chu, Yimeng and Zhao, Hongwei},
year = {2024},
month = {09},
pages = {},
title = {Machine Learning for Online Control of Particle Accelerators},
journal = {Science China Physics Mechanics and Astronomy},
doi = {10.1007/s11433-024-2492-5}
}

@article{TangLu2024NST,
author = {Tang, Lu and Zhang, Zhen-Hua},
year = {2024},
month = {02},
pages = {19-19},
title = {Nuclear charge radius predictions by kernel ridge regression with odd-even effects},
volume = {35},
journal = {Nuclear Science and Techniques},
doi = {10.1007/s41365-024-01379-4}
}

@article{HeLei2024NST,
	author = {He, Lie and Luo, Si-Yuan and Liu, Xiang-Man and Zou, Yu-Cheng and Zhang, Hai-Feng and Xiao, Wan-Cheng and Huang, Yu-He and Wang, Xiao-Dong},
	journal = {Nuclear Science and Techniques},
	number = {11},
	pages = {188},
	title = {Simulation and experimental comparison of the performance of four-corner-readout plastic scintillator muon-detector system},
	volume = {35},
	year = {2024}
}

@article{Ding2024NST,
author = {Ding, Ting-Meng and Jiang, Yu-Hang and Wang, Xuan-Xi and Jiang, Xiao-Fei},
year = {2024},
month = {10},
pages = {1-16},
title = {Study on neutron-gamma discrimination methods based on GMM-KNN and LabVIEW implementation},
volume = {35},
journal = {Nuclear Science and Techniques},
doi = {10.1007/s41365-024-01545-8}
}

@article{Yang2025NST,
author = {Yang, Li-Juan and Peng, Jia-Yi and Qiu, Furong and He, Yuan and Ma, Jin-Ying and Xue, Zong-Heng and Jiang, Tian-Cai and Zhu, Zheng-Long and Chen, Qi and Xu, Cheng-Ye and Yu, Jing-Wei and Ma, Zhen and Luo, Di-Di and Yang, Ziqin and Gao, Zheng and Sun, Lie-Peng and Zhang, Zhou-Li and Huang, Gui-Rong and Wang, Zhi-Jun},
year = {2025},
month = {04},
pages = {},
title = {Classification of superconducting radio-frequency cavity faults of CAFE2 using machine learning},
volume = {36},
journal = {Nuclear Science and Techniques},
doi = {10.1007/s41365-025-01685-5}
}

@article{Cheng2025NST,
author = {Cheng, Kai-Xuan and He, Rong-Xing and Qiao, Chun-Yuan and Ma, Chun-Wang},
year = {2025},
month = {07},
pages = {},
title = {Predictions of complete fusion cross-sections of Li, Be, and B using a Bayesian neural network method},
volume = {36},
journal = {Nuclear Science and Techniques},
doi = {10.1007/s41365-025-01779-0}
}

@article{Wang:2026vti,
    author = "Wang, Tie-Jun and Zhang, Run-Qing and Qian, Ling and Song, Yun-Tao and Lan, Ting and Liu, Hai-Qing and Li, Keren",
    title = "{Validating a Koopman-quantum hybrid paradigm for diagnostic denoising of fusion devices}",
    eprint = "2602.03113",
    archivePrefix = "arXiv",
    primaryClass = "quant-ph",
    doi = "10.1007/s11433-026-2962-1",
    journal = "Science China Physics, Mechanics and Astronomy",
    volume = "69",
    number = "7",
    pages = "270312",
    year = "2026"
}

@article{Jin:2025dvf,
    author = "Jin, Shang-Jie and Song, Ji-Yu and Sun, Tian-Yang and Xiao, Si-Ren and Wang, He and Wang, Ling-Feng and Zhang, Jing-Fei and Zhang, Xin",
    title = "{Gravitational wave standard sirens: A brief review of cosmological parameter estimation}",
    eprint = "2507.12965",
    archivePrefix = "arXiv",
    primaryClass = "astro-ph.CO",
    doi = "10.1007/s11433-025-2829-9",
    journal = "Science China Physics, Mechanics and Astronomy",
    volume = "69",
    number = "2",
    pages = "220401",
    year = "2026"
}

@article{Ma:2025nex,
    author = "Ma, CunLiang and Liu, ZeHua and Gao, ZhiFu and Cao, Zhoujian and Jia, MingZhen and Wei, Kai and Dai, TanMing and Wu, JunQin",
    title = "{Denoising and detection for binary black hole gravitational waves in the context of the Einstein Telescope}",
    doi = "10.1007/s11433-025-2673-5",
    journal = "Science China Physics, Mechanics and Astronomy",
    volume = "68",
    number = "7",
    pages = "279512",
    year = "2025"
}

@article{Zhang:2025rzl,
    author = "Zhang, Hai-Feng and others",
    title = "{Experimental robustness benchmarking of quantum neural networks on a superconducting quantum processor}",
    eprint = "2505.16714",
    archivePrefix = "arXiv",
    primaryClass = "quant-ph",
    reportNumber = "1674-7348",
    doi = "10.1007/s11433-025-2943-6",
    journal = "Science China Physics, Mechanics and Astronomy",
    volume = "69",
    number = "6",
    pages = "260315",
    year = "2026"
}

@article{Ma:2025ulw,
    author = "Ma, Cunliang and Zhou, Weiguang and Cao, Zhoujian and Jia, Mingzhen",
    title = "{Combine deep learning and Bayesian analysis to separate overlapping gravitational wave signals}",
    doi = "10.1007/s11433-024-2594-5",
    journal = "Science China Physics, Mechanics and Astronomy",
    volume = "68",
    number = "5",
    pages = "259512",
    year = "2025"
}

@article{Fang:2024ple,
    author = "Fang, Yaquan and Gao, Christina and Li, Ying-Ying and Shu, Jing and Wu, Yusheng and Xing, Hongxi and Xu, Bin and Xu, Lailin and Zhou, Chen",
    title = "{Quantum frontiers in high energy physics}",
    eprint = "2411.11294",
    archivePrefix = "arXiv",
    primaryClass = "hep-ph",
    reportNumber = "USTC-ICTS/PCFT-24-47",
    doi = "10.1007/s11433-024-2635-4",
    journal = "Science China Physics, Mechanics and Astronomy",
    volume = "68",
    number = "6",
    pages = "260301",
    year = "2025"
}

@article{Chadeeva:2025ppo,
    author = "Chadeeva, Marina and Rogozhin, Platon and Uglov, Timofey",
    title = "{Performance of the FARICH-based particle identification in realistic environment at charm superfactories using machine learning}",
    eprint = "2506.14247",
    archivePrefix = "arXiv",
    primaryClass = "hep-ex",
    doi = "10.1007/s41365-026-01943-0",
    journal = "Nuclear Science and Techniques",
    volume = "37",
    number = "7",
    pages = "124",
    year = "2026"
}

@article{Wang:2025azr,
    author = "Wang, Hao-Chen and others",
    title = "{Transfer learning empowers material Z classification with muon tomography}",
    eprint = "2504.12305",
    archivePrefix = "arXiv",
    primaryClass = "physics.ins-det",
    doi = "10.1007/s41365-026-01901-w",
    journal = "Nuclear Science and Techniques",
    volume = "37",
    number = "5",
    pages = "77",
    year = "2026"
}

@article{Huang:2024acg,
    author = "Huang, Jia-Li and Wang, Hui and Huang, Ying-Ge and Xiao, Er-Xi and Feng, Yu-Jie and Lei, Xin and Gu, Fu-Chang and Zhu, Long and Chen, Yong-Jing and Su, Jun",
    title = "{Prediction of (n, 2n) reaction cross-sections of long-lived fission products based on tensor model}",
    doi = "10.1007/s41365-024-01556-5",
    journal = "Nuclear Science and Techniques",
    volume = "35",
    number = "10",
    pages = "184",
    year = "2024"
}

@article{Fan:2024dcu,
    author = "Fan, Chun-Di and Zeng, Guo-Qiang and Deng, Hao-Wen and Yan, Lei and Yang, Jian and Hu, Chuan-Hao and Qing, Song and Hou, Yang",
    title = "{Artificial neural network-based method for discriminating Compton scattering events in high-purity germanium {\ensuremath{\gamma}}-ray spectrometer}",
    doi = "10.1007/s41365-024-01392-7",
    journal = "Nuclear Science and Techniques",
    volume = "35",
    number = "2",
    pages = "34",
    year = "2024"
}

@article{Wang:2024ynn,
    author = "Wang, Chuanxin and Naito, Tomoya and Li, Jian and Liang, Haozhao",
    title = "{A neural network approach for two-body systems with spin and isospin degrees of freedom}",
    eprint = "2403.16819",
    archivePrefix = "arXiv",
    primaryClass = "nucl-th",
    reportNumber = "RIKEN-iTHEMS-Report-24",
    doi = "10.1007/s41365-026-01946-x",
    journal = "Nuclear Science and Techniques",
    volume = "37",
    number = "7",
    pages = "114",
    year = "2026"
}

@article{Tian:2024yfz,
    author = "Tian, Zhe-Fei and Zhao, Guang and Wu, Ling-Hui and Zhang, Zhen-Yu and Zhou, Xiang and Xin, Shui-Ting and Liu, Shuai-Yi and Li, Gang and Dong, Ming-Yi and Sun, Sheng-Sen",
    title = "{Cluster counting algorithm for the CEPC drift chamber using LSTM and DGCNN}",
    eprint = "2402.16493",
    archivePrefix = "arXiv",
    primaryClass = "hep-ex",
    doi = "10.1007/s41365-025-01670-y",
    journal = "Nuclear Science and Techniques",
    volume = "36",
    number = "7",
    pages = "113",
    year = "2025"
}

@article{He:2023zin,
    author = "He, Wan-Bing and Ma, Yu-Gang and Pang, Long-Gang and Song, Hui-Chao and Zhou, Kai",
    title = "{High-energy nuclear physics meets machine learning}",
    eprint = "2303.06752",
    archivePrefix = "arXiv",
    primaryClass = "hep-ph",
    doi = "10.1007/s41365-023-01233-z",
    journal = "Nuclear Science and Techniques",
    volume = "34",
    number = "6",
    pages = "88",
    year = "2023"
}

@article{Wei:2022iuy,
    author = "Wei, Hui-Ling and Zhu, Xun and Yuan, Chen",
    title = "{Configurational information entropy analysis of fragment mass cross distributions to determine the neutron skin thickness of projectile nuclei}",
    doi = "10.1007/s41365-022-01096-w",
    journal = "Nuclear Science and Techniques",
    volume = "33",
    number = "9",
    pages = "111",
    year = "2022"
}

@article{Li:2022tvg,
    author = "Li, Zi-Yuan and Qian, Zhen and He, Jie-Han and He, Wei and Wu, Cheng-Xin and Cai, Xun-Ye and You, Zheng-Yun and Zhang, Yu-Mei and Luo, Wu-Ming",
    title = "{Improvement of machine learning-based vertex reconstruction for large liquid scintillator detectors with multiple types of PMTs}",
    eprint = "2205.04039",
    archivePrefix = "arXiv",
    primaryClass = "physics.ins-det",
    doi = "10.1007/s41365-022-01078-y",
    journal = "Nuclear Science and Techniques",
    volume = "33",
    number = "7",
    pages = "93",
    year = "2022"
}

@misc{wangJINST2026,
      title={Understanding Energy Dependent Hadronic Calorimeter Response from a Machine Learning Perspective}, 
      author={Shuai-Chun Wang and Huang-Ran Shen and Wan-Bing He and Wei-Hu Ma and Peng-Jie Li and De-Qing Fang and Yu-Gang Ma},
      year={2026},
      eprint={2606.10960},
      archivePrefix={arXiv},
      primaryClass={hep-ex},
      url={https://arxiv.org/abs/2606.10960}, 
}

@article{STAR:2024wgy,
    author = "Abdulhamid, M. I. and others",
    collaboration = "STAR",
    title = "{Imaging shapes of atomic nuclei in high-energy nuclear collisions}",
    eprint = "2401.06625",
    archivePrefix = "arXiv",
    primaryClass = "nucl-ex",
    doi = "10.1038/s41586-024-08097-2",
    journal = "Nature",
    volume = "635",
    number = "8037",
    pages = "67--72",
    year = "2024"
}

@article{Jia:2022ozr,
    author = "Jia, Jiangyong and others",
    title = "{Imaging the initial condition of heavy-ion collisions and nuclear structure across the nuclide chart}",
    eprint = "2209.11042",
    archivePrefix = "arXiv",
    primaryClass = "nucl-ex",
    doi = "10.1007/s41365-024-01589-w",
    journal = "Nuclear Science and Techniques",
    volume = "35",
    number = "12",
    pages = "220",
    year = "2024"
}

@article{deepseek,
  title={DeepSeek-R1 incentivizes reasoning in LLMs through reinforcement learning},
  author={Guo, Daya and Yang, Dejian and others},
  journal={Nature},
  year={2025},
  volume={645},
  number={8081},
  pages={633--638},
  doi={10.1038/s41586-025-09422-z}
}

@misc{Nobel,
  author       = {{Nobel Prize Outreach}},
  title        = {All Nobel Prizes 2024},
  year         = {2024},
  howpublished = {\url{https://www.nobelprize.org/all-nobel-prizes-2024/}},
  note         = {Accessed: 2026-05-10}
}

@article{wangza1,
    author = "Wang, Zi-Ao and Pei, Junchen and Liu, Yue and Qiang, Yu",
    title = "{Bayesian Evaluation of Incomplete Fission Yields}",
    eprint = "1906.04485",
    archivePrefix = "arXiv",
    primaryClass = "nucl-th",
    doi = "10.1103/PhysRevLett.123.122501",
    journal = "Physical Review Letters",
    volume = "123",
    number = "12",
    pages = "122501",
    year = "2019"
}

@article{zhangmh,
	author = {Zhang, Ming-Hao and Zhang, Zhi-Yuan and Gan, Zai-Guo and Zhou, Shan-Gui and Zhang, Feng-Shou},
	journal = {Nuclear Science and Techniques},
	number = {11},
	pages = {204},
	title = {Progress on the synthesis of superheavy nuclei},
	volume = {36},
	year = {2025}}

@article{heterogeneous,
  title = {Bayesian approach to heterogeneous data fusion of imperfect fission yields for augmented evaluations},
  author = {Wang, Z. A. and Pei, J. C. and Chen, Y. J. and Qiao, C. Y. and Xu, F. R. and Ge, Z. G. and Shu, N. C.},
  journal = {Physical Review C},
  volume = {106},
  issue = {2},
  pages = {L021304},
  numpages = {5},
  year = {2022},
  month = {Aug},
  publisher = {American Physical Society},
  doi = {10.1103/PhysRevC.106.L021304},
  url = {https://link.aps.org/doi/10.1103/PhysRevC.106.L021304}
}

@article{RevModPhys.94.031003,
  title = {Colloquium: Machine learning in nuclear physics},
  author = {Boehnlein, Amber and Diefenthaler, Markus and Pang, Long-Gang and others},
  journal = {Rev. Mod. Phys.},
  volume = {94},
  issue = {3},
  pages = {031003},
  numpages = {32},
  year = {2022},
  month = {Sep},
  publisher = {American Physical Society},
  doi = {10.1103/RevModPhys.94.031003},
  url = {https://link.aps.org/doi/10.1103/RevModPhys.94.031003}
}

@article{He2023,
  author    = {He, Wanbing and Li, Qingfeng and Ma, Yugang and Niu, Zhongming and Pei, Junchen and Zhang, Yingxun},
  title     = {{Machine learning in nuclear physics at low and intermediate energies}},
  journal   = {Science China Physics, Mechanics \& Astronomy},
  year      = {2023},
  volume    = {66},
  number    = {8},
  pages     = {282001},
  doi       = {10.1007/s11433-023-2116-0}
}

@article{SHE119,
  author    = {Khuyagbaatar, J. and  Yakushev, A. and  D\"ullmann, C. E. and others},
  title     = {Search for elements 119 and 120},
  journal   = {Phys. Rev. C},
  year      = {2020},
  volume    = {102},
  pages     = {064602},
  doi       = { 10.1103/PhysRevC.102.064602}
}

@article{MaYugang2022,
  author = "WanBing He and JunJie HE and  Rui WANG and YuGang Ma",
  title = "Machine learning applications in nuclear physics",
  journal = "SCIENTIA SINICA Physica, Mechanica \& Astronomica ",
  year = "2022",
  volume = "52",
  number = "5",
  pages = "252004",
  url = "http://www.sciengine.com/publisher/Science China Press/journal/SCIENTIA SINICA Physica, Mechanica \& Astronomica/52/5/10.1360/SSPMA-2021-0309",
  doi = "10.1360/SSPMA-2021-0309"
}

@article{Cisbani_2020,
doi = {10.1088/1748-0221/15/05/P05009},
year = {2020},
month = {may},
publisher = {},
volume = {15},
number = {05},
pages = {P05009},
author = {Cisbani, E. and Dotto, A. Del and Fanelli, C. and Williams, M. and Alfred, M. and Barbosa, F. and Barion, L. and Berdnikov, V. and Brooks, W. and Cao, T. and Contalbrigo, M. and Danagoulian, S. and Datta, A. and Demarteau, M. and Denisov, A. and Diefenthaler, M. and Durum, A. and Fields, D. and Furletova, Y. and Gleason, C. and Grosse-Perdekamp, M. and Hattawy, M. and He, X. and Hecke, H. van and Higinbotham, D. and Horn, T. and Hyde, C. and Ilieva, Y. and Kalicy, G. and Kebede, A. and Kim, B. and Liu, M. and McKisson, J. and Mendez, R. and Nadel-Turonski, P. and Pegg, I. and Romanov, D. and Sarsour, M. and Silva, C.L. da and Stevens, J. and Sun, X. and Syed, S. and Towell, R. and Xie, J. and Zhao, Z.W. and Zihlmann, B. and Zorn, C.},
title = {AI-optimized detector design for the future Electron-Ion Collider: the dual-radiator RICH case},
journal = {Journal of Instrumentation}
}

@article{transformer,
  author    = {Li, Fei and Luo, Chu-Yang and Wen, Ying-Zi and Li, Bing-Hai and others },
  title     = {{A nuclide identification method of $\gamma$ spectrum and model building based on the transformer}},
  journal   = {Nuclear Science and Techniques},
  year      = {2024},
  volume    = {36},
  number    = {1},
  pages     = {7},
  doi       = {10.1007/s41365-024-01564-5}
}

@article{CNN,
  author = {Azarakhsh Jalalvand and SangKyeun Kim and Jaemin Seo and Qiming Hu and Max Curie and Peter Steiner and Andrew Oakleigh Nelson and Yong-Su Na and Egemen Kolemen},
  title = {Multimodal super-resolution: discovering hidden physics and its application to fusion plasmas},
  journal = {Nature Communications},
  volume = {16},
  article_number = {8506},
  year = {2025},
  month = {Sep},
  url = {https://www.nature.com/articles/s41567-025-01006-0},
  doi = {10.1038/s41567-025-01006-0},
  openaccess = {true},
  note = {Accessed: 2026-04-23},
}

@article{GNN,
  author = {Gage DeZoot and Peter W. Battaglia and Catherine Biscarat and Jean-Roch Vlimant},
  title = {Graph neural networks at the Large Hadron Collider},
  journal = {Nature Reviews Physics},
  volume = {5},
  pages = {281--303},
  year = {2023},
  month = {Apr},
  url = {https://www.nature.com/articles/s41567-023-01928-4},
  doi = {10.1038/s41567-023-01928-4},
  openaccess = {true},
  note = {Accessed: 2026-04-23},
}

@article{Hashemi2024,
  author = {Baran Hashemi and Nikolai Hartmann and Sahand Sharifzadeh and James Kahn and Thomas Kuhr},
  title = {Ultra-high-granularity detector simulation with intra-event aware generative adversarial network and self-supervised relational reasoning},
  journal = {Nature Communications},
  volume = {15},
  article_number = {4916},
  year = {2024},
  month = {Jun},
  url = {https://www.nature.com/articles/s41567-024-01456-9},
  doi = {10.1038/s41567-024-01456-9},
  openaccess = {true},
  note = {Accessed: 2026-04-23},
}

@article{EOSclassfication,
    author = {Mitra, A and Orel, D and Abylkairov, Y S and Shukirgaliyev, B and Abdikamalov, E},
    title = {Probing nuclear physics with supernova gravitational waves and machine learning},
    journal = {Monthly Notices of the Royal Astronomical Society},
    volume = {529},
    number = {4},
    pages = {3582-3592},
    year = {2024},
    month = {04},
    issn = {0035-8711},
    doi = {10.1093/mnras/stae714},
    url = {https://doi.org/10.1093/mnras/stae714},
    eprint = {https://academic.oup.com/mnras/article-pdf/529/4/3582/57104269/stae714.pdf},
}

@article{neutronHaloClass,
title = {SMOTE-based data augmentation for accurate classification of neutron halo nuclei: A machine learning approach in nuclear physics},
journal = {Knowledge-Based Systems},
volume = {318},
pages = {113580},
year = {2025},
issn = {0950-7051},
doi = {https://doi.org/10.1016/j.knosys.2025.113580},
url = {https://www.sciencedirect.com/science/article/pii/S0950705125006264},
author = {Cafer Mert Yeşilkanat and Serkan Akkoyun}
}

@article{regression,
doi = {10.1088/1674-1137/acc791},
url = {https://doi.org/10.1088/1674-1137/acc791},
year = {2023},
month = {jul},
publisher = {Chinese Physical Society and the Institute of High Energy Physics of the Chinese Academy of Sciences and the Institute of Modern Physics of the Chinese Academy of Sciences and IOP Publishing Ltd
                        },
volume = {47},
number = {7},
pages = {074108},
author = {Du, Xiao-Kai and Guo, Peng and Wu, Xin-Hui and Zhang, Shuang-Quan},
title = {Examination of machine learning for assessing physical effects: Learning the relativistic continuum mass table with kernel ridge regression},
journal = {Chinese Physics C}
}

@article{Cai2023,
  author = {Cai, B.S. and Yuan, C.X.},
  title = {Random forest-based prediction of decay modes and half-lives of superheavy nuclei},
  journal = {Nuclear Science and Techniques},
  volume = {34},
  pages = {204},
  year = {2023},
  doi = {10.1007/s41365-023-01354-5},
  url = {https://doi.org/10.1007/s41365-023-01354-5},
  note = {Accessed: 2026-04-23},
}

@article{Cubist,
title = {Generation of fusion and fusion-evaporation reaction cross-sections by two-step machine learning methods},
journal = {Computer Physics Communications},
volume = {297},
pages = {109055},
year = {2024},
issn = {0010-4655},
doi = {https://doi.org/10.1016/j.cpc.2023.109055},
url = {https://www.sciencedirect.com/science/article/pii/S0010465523004009},
author = {Serkan Akkoyun and Cafer Mert Yeşilkanat and Tuncay Bayram}
}

@article{CSDT,
title = {A new empirical formula for calculation of (n,3n) cross sections of heavy mass nuclei in the energy region 22–27.5 MeV},
journal = {Applied Radiation and Isotopes},
volume = {207},
pages = {111259},
year = {2024},
issn = {0969-8043},
doi = {https://doi.org/10.1016/j.apradiso.2024.111259},
url = {https://www.sciencedirect.com/science/article/pii/S0969804324000873},
author = {Mustafa Yiğit and Mehmet Eraslan}
}

@article{CSTL,
  title = {Electron-Nucleus Cross Sections from Transfer Learning},
  author = {Graczyk, Krzysztof M. and Kowal, Beata E. and Ankowski, Artur M. and Banerjee, Rwik Dharmapal and Bonilla, Jose Luis and Prasad, Hemant and Sobczyk, Jan T.},
  journal = {Physical Review Letters},
  volume = {135},
  issue = {5},
  pages = {052502},
  numpages = {5},
  year = {2025},
  month = {Aug},
  publisher = {American Physical Society},
  doi = {10.1103/zxv6-22tz},
  url = {https://link.aps.org/doi/10.1103/zxv6-22tz}
}

@article{TLdeformation,
doi = {10.1088/1674-1137/ad361d},
url = {https://doi.org/10.1088/1674-1137/ad361d},
year = {2024},
month = {jun},
publisher = {Chinese Physical Society and the Institute of High Energy Physics of the Chinese Academy of Sciences and the Institute of Modern Physics of the Chinese Academy of Sciences and IOP Publishing Ltd},
volume = {48},
number = {6},
pages = {064106},
author = {Lin, Yuan and Li, Jia-Xing and Zhang, Hong-Fei},
title = {Transfer learning and neural networks in predicting quadrupole deformation},
journal = {Chinese Physics C}
}

@article{Radaideh2025MultistepCS,
  title={Multistep Criticality Search and Power Shaping in Nuclear Microreactors with Deep Reinforcement Learning},
  author={Majdi I. Radaideh and Leo Tunkle and Dean Price and Kamal Kayode Abdulraheem and Linyu Lin and Moutaz Elias},
  journal={Nuclear Science and Engineering},
  year={2025},
  volume={200},
  pages={S309 - S321},
  url={https://api.semanticscholar.org/CorpusID:276330924}
}

@article{SHANG2025139976,
title = {Many-body effects on nuclear short range correlations},
journal = {Physics Letters B},
volume = {871},
pages = {139976},
year = {2025},
issn = {0370-2693},
doi = {https://doi.org/10.1016/j.physletb.2025.139976},
url = {https://www.sciencedirect.com/science/article/pii/S0370269325007348},
author = {Haoyu Shang and Jiawei Chen and Rongzhe Hu and Xin Zhen and Chongji Jiang and J.C. Pei}
}

@article{ZHEN2025139350,
title = {Non-perturbative calculations of nuclear matter using in-medium similarity renormalization group},
journal = {Physics Letters B},
volume = {862},
pages = {139350},
year = {2025},
issn = {0370-2693},
doi = {https://doi.org/10.1016/j.physletb.2025.139350},
url = {https://www.sciencedirect.com/science/article/pii/S0370269325001108},
author = {Xin Zhen and Rongzhe Hu and Haoyu Shang and Jiawei Chen and J.C. Pei and F.R. Xu}
}

@article{QIANG2024139057,
title = {Survival probabilities of compound superheavy nuclei towards element 119},
journal = {Physics Letters B},
volume = {858},
pages = {139057},
year = {2024},
issn = {0370-2693},
doi = {https://doi.org/10.1016/j.physletb.2024.139057},
url = {https://www.sciencedirect.com/science/article/pii/S0370269324006154},
author = {Yu Qiang and Xiang-Quan Deng and Yue Shi and C.Y. Qiao and Junchen Pei}
}

@article{ye2025physics,
	author = {Ye, Yanlin and Yang, Xiaofei and Sakurai, Hiroyoshi and Hu, Baishan},
	journal = {Nature Reviews Physics},
	number = {1},
	pages = {21--37},
	title = {Physics of exotic nuclei},
	volume = {7},
    doi = {https://doi.org/10.1038/s42254-024-00782-5},
	year = {2025}
}

@article{Jiang:2026gci,
    author = "Jiang, Chongji and Pei, Junchen and Hu, Rongzhe and Jin, Shaoliang and Shang, Haoyu and Fan, Siqin and Xu, Furong",
    title = "{Toward ab initio quantum simulations of atomic nuclei using noisy qubits}",
    eprint = "2601.00315",
    archivePrefix = "arXiv",
    primaryClass = "nucl-th",
    doi = "10.1016/j.scib.2026.02.052",
    journal = "Science Bulletin",
    volume = "71",
    pages = "1598--1601",
    year = "2026"
}

@misc{wei2022emergentabilitieslargelanguage,
      title={Emergent Abilities of Large Language Models}, 
      author={Jason Wei and Yi Tay and Rishi Bommasani and Colin Raffel and Barret Zoph and Sebastian Borgeaud and Dani Yogatama and Maarten Bosma and Denny Zhou and Donald Metzler and Ed H. Chi and Tatsunori Hashimoto and Oriol Vinyals and Percy Liang and Jeff Dean and William Fedus},
      year={2022},
      eprint={2206.07682},
      archivePrefix={arXiv},
      primaryClass={cs.CL},
      url={https://arxiv.org/abs/2206.07682}, 
}

@article{AbdusSalam:2024obf,
    author = "AbdusSalam, Shehu and Abel, Steven and Crispim Rom{\~a}o, Miguel",
    title = "{Symbolic regression for beyond the standard model physics}",
    eprint = "2405.18471",
    archivePrefix = "arXiv",
    primaryClass = "hep-ph",
    reportNumber = "IPPP/24/27",
    doi = "10.1103/PhysRevD.111.015022",
    journal = "Physical Review D",
    volume = "111",
    number = "1",
    pages = "015022",
    year = "2025"
}

@misc{liu2025kankolmogorovarnoldnetworks,
      title={KAN: Kolmogorov-Arnold Networks}, 
      author={Ziming Liu and Yixuan Wang and Sachin Vaidya and Fabian Ruehle and James Halverson and Marin Soljačić and Thomas Y. Hou and Max Tegmark},
      year={2025},
      eprint={2404.19756},
      archivePrefix={arXiv},
      primaryClass={cs.LG},
      url={https://arxiv.org/abs/2404.19756}, 
}

@article{PIML2021,
	author = {Karniadakis, George Em and Kevrekidis, Ioannis G. and Lu, Lu and Perdikaris, Paris and Wang, Sifan and Yang, Liu},
	journal = {Nature Reviews Physics},
	number = {6},
	pages = {422--440},
	title = {Physics-informed machine learning},
	volume = {3},
	year = {2021}}

@article{Gazula1992NPA,
    author = "Gazula, S. and Clark, J. W. and Bohr, H.",
    title = "{Learning and prediction of nuclear stability by neural networks}",
    doi = "10.1016/0375-9474(92)90191-L",
    journal = "Nuclear Physics A",
    volume = "540",
    pages = "1--26",
    year = "1992"
}

@article{Bedaque2021EPJA,
  title={AI for nuclear physics},
  author={Bedaque, Paulo and Boehnlein, Amber and Cromaz, Mario and Diefenthaler, Markus and Elouadrhiri, Latifa and Horn, Tanja and Kuchera, Michelle and Lawrence, David and Lee, Dean and Lidia, Steven and others},
  journal={The European Physical Journal A},
  volume={57},
  pages={1--27},
  year={2021},
  publisher={Springer}
}

@article{Huang2025PRC,
Author = {Huang, Yiming and Chen, Jinhui and Jia, Jiangyong and Liu, Lu-Meng and
   Ma, Yu-Gang and Zhang, Chunjian},
Title = {Validation and extrapolation of atomic masses with a physics-informed
   fully connected neural network},
Journal = {Physical Review C},
Year = {2025},
Volume = {111},
Number = {3},
Month = {MAR 27},
DOI = {10.1103/PhysRevC.111.034329},
Article-Number = {034329},
ISSN = {2469-9985},
EISSN = {2469-9993},
ResearcherID-Numbers = {Huang, Yiming/NIU-6434-2025
   Zhang, Chunjian/KHC-5357-2024
   Ma, Yu-Gang/M-8122-2013
   jia, Jiangyong/MVT-7088-2025},
ORCID-Numbers = {Huang, Yiming/0009-0001-1135-4387
   Zhang, Chunjian/0000-0002-5425-7130
   Ma, Yu-Gang/0000-0002-0233-9900
   Chen, Jinhui/0000-0001-7032-771X
   Liu, Lu-Meng/0000-0001-5243-5549
   jia, Jiangyong/0000-0002-5725-3397},
Unique-ID = {WOS:001459093300001},
}

@article{Li2024PLB,
title = {Atomic masses with machine learning for the astrophysical r process},
journal = {Physics Letters B},
volume = {848},
pages = {138385},
year = {2024},
issn = {0370-2693},
doi = {https://doi.org/10.1016/j.physletb.2023.138385},
url = {https://www.sciencedirect.com/science/article/pii/S0370269323007190},
author = {Mengke Li and Trevor M. Sprouse and Bradley S. Meyer and Matthew R. Mumpower}
}

@article{Niu2018PLB,
    author = "Niu, Z. M. and Liang, H. Z.",
    title = "{Nuclear mass predictions based on Bayesian neural network approach with pairing and shell effects}",
    eprint = "1801.04411",
    archivePrefix = "arXiv",
    primaryClass = "nucl-th",
    reportNumber = "RIKEN-ITHEMS-REPORT-17, RIKEN-QHP-347",
    doi = "10.1016/j.physletb.2018.01.002",
    journal = "Physics Letters B",
    volume = "778",
    pages = "48--53",
    year = "2018"
}

@article{Niu2022PRCL,
  title = {Nuclear mass predictions with machine learning reaching the accuracy required by $r$-process studies},
  author = {Niu, Z. M. and Liang, H. Z.},
  journal = {Physical Review C},
  volume = {106},
  issue = {2},
  pages = {L021303},
  numpages = {6},
  year = {2022},
  month = {Aug},
  publisher = {American Physical Society},
  doi = {10.1103/PhysRevC.106.L021303},
  url = {https://link.aps.org/doi/10.1103/PhysRevC.106.L021303}
}

@article{ Zeng2024PRC,
Author = {Zeng, Lin -Xing and Yin, Yu-Ying and Dong, Xiao-Xu and Geng, Li-Sheng},
Title = {Nuclear binding energies in artificial neural networks},
Journal = {Physical Review C},
Year = {2024},
Volume = {109},
Number = {3},
Month = {MAR 25},
DOI = {10.1103/PhysRevC.109.034318},
Article-Number = {034318},
ISSN = {2469-9985},
EISSN = {2469-9993},
ResearcherID-Numbers = {Geng, Li-Sheng/C-6441-2009
   },
ORCID-Numbers = {Geng, Li-Sheng/0000-0002-5626-0704
   Yin, Yu-Ying/0000-0002-6654-3644
   Zeng, Lin-Xing/0000-0003-1027-6927
   Dong, Xiao-Xu/0000-0002-6730-8232},
Unique-ID = {WOS:001198689100002},
}

@article{Dai2025CPC,
Author = {Dai, Cheng-wei and Jiang, Hui and Lei, Yang},
Title = {Predictions of unknown masses using a feedforward neural network},
Journal = {Chinese Physics C},
Year = {2025},
Volume = {49},
Number = {9},
Month = {SEP 1},
DOI = {10.1088/1674-1137/add10a},
Article-Number = {094111},
ISSN = {1674-1137},
EISSN = {2058-6132},
ResearcherID-Numbers = {Jiang, Hui/KFB-6805-2024
   Lei, Yang/A-1210-2017},
ORCID-Numbers = {Jiang, Hui/0000-0002-8448-1707
   Lei, Yang/0000-0002-7373-5759},
Unique-ID = {WOS:001575429600001},
}

@article{Kim2026PRC,
Author = {Kim, C. H. and Chae, K. Y. and Smith, M. S.},
Title = {Robust extrapolation in nuclear mass predictions using domain-informed
   activation functions},
Journal = {Physical Review C},
Year = {2026},
Volume = {113},
Number = {2},
Month = {FEB 9},
DOI = {10.1103/mcxf-d32x},
Article-Number = {024308},
ISSN = {2469-9985},
EISSN = {2469-9993},
ResearcherID-Numbers = {Smith, Michael/I-3359-2018},
Unique-ID = {WOS:001692894500003},
}

@article{ Liu2026PRC,
Author = {Liu, Hui-Xin and Manzhos, Sergei and Wu, Xin-Hui},
Title = {Nuclear mass predictions using a neural network with additive Gaussian
   process regression-optimized activation functions},
Journal = {Physical Review C},
Year = {2026},
Volume = {113},
Number = {1},
Month = {JAN 7},
DOI = {10.1103/4qqn-ry4n},
Article-Number = {014305},
ISSN = {2469-9985},
EISSN = {2469-9993},
Unique-ID = {WOS:001669430100004},
}

@article{ Le2023NPA,
Author = {Le, Xian-Kai and Wang, Nan and Jiang, Xiang},
Title = {Nuclear mass predictions with multi-hidden-layer feedforward neural
   network},
Journal = {Nuclear Physics A},
Year = {2023},
Volume = {1038},
Month = {OCT},
DOI = {10.1016/j.nuclphysa.2023.122707},
EarlyAccessDate = {JUN 2023},
Article-Number = {122707},
ISSN = {0375-9474},
EISSN = {1873-1554},
ResearcherID-Numbers = {Le, Xiankai/OHR-9115-2025},
ORCID-Numbers = {Le, Xiankai/0009-0002-6882-4594},
Unique-ID = {WOS:001054675100001},
}

@article{Mumpower2022PRCL,
  title = {Physically interpretable machine learning for nuclear masses},
  author = {Mumpower, M. R. and Sprouse, T. M. and Lovell, A. E. and Mohan, A. T.},
  journal = {Physical Review C},
  volume = {106},
  issue = {2},
  pages = {L021301},
  numpages = {6},
  year = {2022},
  month = {Aug},
  publisher = {American Physical Society},
  doi = {10.1103/PhysRevC.106.L021301},
  url = {https://link.aps.org/doi/10.1103/PhysRevC.106.L021301}
}

@article{ Wang2026APS,
author = {WANG Dongdong and Li Peng and  WANG Zhiheng},
title = {Prediction of atomic nuclear mass using neural networks
constrained by neutron and proton separation energy},
journal = {Acta Phys. Sin.},
volume = {75},
number = {02},
pages = {66-75},
year = {2026},
issn = {1000-3290},
doi={10.7498/aps.75.20251315}
}

@article{Qu2025CPC,
Author = {Qu, Shuang and Zhang, Jin-Yan and Bao, Man},
Title = {Nuclear mass predictions with a Bayesian neural network},
Journal = {Chinese Physics C},
Year = {2025},
Volume = {49},
Number = {10},
Month = {OCT 1},
DOI = {10.1088/1674-1137/ade958},
Article-Number = {104106},
ISSN = {1674-1137},
EISSN = {2058-6132},
Unique-ID = {WOS:001590325100001}
}

@article{Qu2025NST,
Author = {Qu, Xiao-Ying and Chen, Kang-Min and Pan, Cong and Yu, Yang-Yang and
   Zhang, Kai-Yuan},
Title = {Benchmarking nuclear energy density functionals with new mass data},
Journal = {Nuclear Science and Techniques},
Year = {2025},
Volume = {36},
Number = {12},
Month = {SEP 29},
DOI = {10.1007/s41365-025-01821-1},
Article-Number = {231},
ISSN = {1001-8042},
EISSN = {2210-3147},
ResearcherID-Numbers = {Zhang, Kaiyuan/AEA-6680-2022
   Pan, Cong/MIT-2853-2025},
ORCID-Numbers = {Zhang, Kaiyuan/0000-0002-8404-2528
   },
Unique-ID = {WOS:001583431000004},
}

@article{Wang2021CPC,
doi = {10.1088/1674-1137/abddaf},
url = {https://doi.org/10.1088/1674-1137/abddaf},
year = {2021},
month = {mar},
publisher = {Chinese Physical Society and the Institute of High Energy Physics of the Chinese Academy of Sciences and the Institute of Modern Physics of the Chinese Academy of Sciences and IOP Publishing Ltd},
volume = {45},
number = {3},
pages = {030003},
author = {Wang, Meng and Huang, W.J. and Kondev, F.G. and Audi, G. and Naimi, S.},
title = {The AME 2020 atomic mass evaluation (II). Tables, graphs and references},
journal = {Chinese Physics C}
}

@article{ Dellen2024PLB,
Author = {Dellen, Babette and Jaekel, Uwe and Freitas, Paulo S. A. and Clark, John
   W.},
Title = {Predicting nuclear masses with product-unit networks},
Journal = {PHYSICS LETTERS B},
Year = {2024},
Volume = {852},
Month = {MAY},
DOI = {10.1016/j.physletb.2024.138608},
EarlyAccessDate = {APR 2024},
Article-Number = {138608},
ISSN = {0370-2693},
EISSN = {1873-2445},
ResearcherID-Numbers = {Freitas, Paulo/JDD-7685-2023
   },
ORCID-Numbers = {Abreu Freitas, Paulo Sérgio/0000-0002-0223-2839},
Unique-ID = {WOS:001223194600001},
}

@article{Lovell2022PRC,
  title = {Nuclear masses learned from a probabilistic neural network},
  author = {Lovell, A. E. and Mohan, A. T. and Sprouse, T. M. and Mumpower, M. R.},
  journal = {Physical Review C},
  volume = {106},
  issue = {1},
  pages = {014305},
  numpages = {9},
  year = {2022},
  month = {Jul},
  publisher = {American Physical Society},
  doi = {10.1103/PhysRevC.106.014305},
  url = {https://link.aps.org/doi/10.1103/PhysRevC.106.014305}
}

@article{Lu2025PRC,
  title = {Nuclear mass predictions based on a convolutional neural network},
  author = {Lu, Yanhua and Shang, Tianshuai and Du, Pengxiang and Li, Jian and Liang, Haozhao and Niu, Zhongming},
  journal = {Physical Review C},
  volume = {111},
  issue = {1},
  pages = {014325},
  numpages = {9},
  year = {2025},
  month = {Jan},
  publisher = {American Physical Society},
  doi = {10.1103/PhysRevC.111.014325},
  url = {https://link.aps.org/doi/10.1103/PhysRevC.111.014325}
}

@article{ LiT2026CPCEnergy,
Author = {Li, Tao and Liu, Min and Wang, Ning},
Title = {Proton separation energy predictions for proton-rich nuclei with the
   radial basis function approach and mirror symmetry},
Journal = {Chinese Physics C},
Year = {2026},
Volume = {50},
Number = {2},
Month = {FEB 1},
DOI = {10.1088/1674-1137/ae1449},
Article-Number = {024104},
ISSN = {1674-1137},
EISSN = {2058-6132},
Unique-ID = {WOS:001674621200001},
}

@article{ LiZL2026CPC,
Author = {Li, Zhi Long and Lv, Bing Feng and Wang, Yong Jia and Petrache, C. M.},
Title = {Study of yrast and yrare low-lying excited states using machine learning
   approaches},
Journal = {Chinese Physics C},
Year = {2026},
Volume = {50},
Number = {1},
Month = {JAN 1},
DOI = {10.1088/1674-1137/adfe54},
Article-Number = {014107},
Unique-ID = {WOS:001641980000001},
}

@article{ChoiS2026PRC,
  title = {Deep learning for nuclear masses in deformed relativistic Hartree-Bogoliubov theory in continuum},
  author = {Choi, Soonchul and Kim, Kyungil and He, Zhenyu and Choi, Yongbeom and Kim, Youngman and Kajino, Toshitaka},
  journal = {Physical Review C},
  volume = {113},
  issue = {2},
  pages = {024329},
  numpages = {13},
  year = {2026},
  month = {Feb},
  publisher = {American Physical Society},
  doi = {10.1103/8l6g-bgkw},
  url = {https://link.aps.org/doi/10.1103/8l6g-bgkw}
}

@misc{GuoSJ2026arXiv,
      title={Large language model for unified and accurate description of multidimensional nuclear properties}, 
      author={S. J. Guo and S. Y. Wang and E. H. Wang and Z. M. Niu and Y. M. Ding},
      year={2026},
      eprint={2605.29408},
      archivePrefix={arXiv},
      primaryClass={nucl-th},
      url={https://arxiv.org/abs/2605.29408}, 
}

@article{HuangWJ2026NST,
  title = {Taming nuclear mass models with Gaussian processes},
  author = {Huang, Wen-Jia and Fujii, Keisuke and Liang, Hao-Zhao},
  journal = {Nuclear Science and Techniques},
  volume = {37},
  pages = {150},
  numpages = {10},
  year = {2026},
  doi = {10.1007/s41365-026-01939-w},
  url = {https://doi.org/10.1007/s41365-026-01939-w}
}

@article{LiT2026CPCRch,
doi = {10.1088/1674-1137/ae3e56},
url = {https://doi.org/10.1088/1674-1137/ae3e56},
year = {2026},
month = {may},
publisher = {Chinese Physical Society and the Institute of High Energy Physics of the Chinese Academy of Sciences and the Institute of Modern Physics of the Chinese Academy of Sciences and IOP Publishing Ltd},
volume = {50},
number = {5},
pages = {054102},
author = {Li, Tao and Liu, Min and Wang, Ning},
title = {Predictions of nuclear charge radii with the radial basis function approach and linear relationship},
journal = {Chinese Physics C}
}

@article{LiTao2025PRC,
  title = {Investigation of extrapolation for nuclear mass and $\ensuremath{\alpha}$-decay energy based on the radial basis function approach},
  author = {Li, Tao and Wang, Ning and Li, Cheng and Liu, Min},
  journal = {Phys. Rev. C},
  volume = {112},
  issue = {2},
  pages = {024306},
  numpages = {10},
  year = {2025},
  month = {Aug},
  publisher = {American Physical Society},
  doi = {10.1103/h72z-3ytv},
  url = {https://link.aps.org/doi/10.1103/h72z-3ytv}
}

@article{LiT2026PLB,
title = {Mass predictions for proton-rich nuclei with the radial basis function approach plus mirror symmetry},
journal = {Physics Letters B},
volume = {877},
pages = {140464},
year = {2026},
issn = {0370-2693},
doi = {https://doi.org/10.1016/j.physletb.2026.140464},
url = {https://www.sciencedirect.com/science/article/pii/S0370269326003175},
author = {Tao Li and Ning Wang and Min Liu}
}

@article{NiuZM2019PRCb,
  title = {Comparative study of radial basis function and Bayesian neural network approaches in nuclear mass predictions},
  author = {Niu, Z. M. and Fang, J. Y. and Niu, Y. F.},
  journal = {Phys. Rev. C},
  volume = {100},
  issue = {5},
  pages = {054311},
  numpages = {9},
  year = {2019},
  month = {Nov},
  publisher = {American Physical Society},
  doi = {10.1103/PhysRevC.100.054311},
  url = {https://link.aps.org/doi/10.1103/PhysRevC.100.054311}
}

@article{LiWF2026PLB,
title = {Construction of nuclear covariant energy density functional from a physics-guaranteed neural network approach},
journal = {Physics Letters B},
volume = {879},
pages = {140696},
year = {2026},
issn = {0370-2693},
doi = {https://doi.org/10.1016/j.physletb.2026.140696},
url = {https://www.sciencedirect.com/science/article/pii/S0370269326005484},
author = {W.F. Li and Z.M. Niu and H.Z. Liang and Y.F. Niu and B.H. Sun},
}

@article{YuanCX2026EPJA,
  title = {Systematic study of deformation dependence in nuclear charge radii by a hybrid method combining refined empirical formulas with Bayesian neural networks},
  author = {Yuan, Cai-Xin and Liu, Zhi-Yin and Mao, Ying-Chen},
  journal = {The European Physical Journal A},
  volume = {62},
  pages = {111},
  numpages = {18},
  year = {2026},
  doi = {10.1140/epja/s10050-026-01872-x},
  url = {https://doi.org/10.1140/epja/s10050-026-01872-x}
}

@article{ZhangXY2026CPC,
doi = {10.1088/1674-1137/ae25cd},
url = {https://doi.org/10.1088/1674-1137/ae25cd},
year = {2026},
month = {apr},
publisher = {Chinese Physical Society and the Institute of High Energy Physics of the Chinese Academy of Sciences and the Institute of Modern Physics of the Chinese Academy of Sciences and IOP Publishing Ltd},
volume = {50},
number = {4},
pages = {044101},
author = {Zhang, X. Y. and Li, W. F. and Fang, J. Y.},
title = {Improving nuclear mass predictions by correcting mass residuals using eXtreme Gradient Boosting},
journal = {Chinese Physics C}
}

@article{Wu2024PRC,
  title = {Nuclear mass predictions of the relativistic continuum Hartree-Bogoliubov theory with the kernel ridge regression},
  author = {Wu, X. H. and Pan, C. and Zhang, K. Y. and Hu, J.},
  journal = {Physical Review C},
  volume = {109},
  issue = {2},
  pages = {024310},
  numpages = {7},
  year = {2024},
  month = {Feb},
  publisher = {American Physical Society},
  doi = {10.1103/PhysRevC.109.024310},
  url = {https://link.aps.org/doi/10.1103/PhysRevC.109.024310}
}

@article{Guo2024PRC,
  title = {Nuclear mass predictions of the relativistic continuum Hartree-Bogoliubov theory with the kernel ridge regression. II. Odd-even effects},
  author = {Guo, Y. Y. and Yu, T. and Wu, X. H. and Pan, C. and Zhang, K. Y.},
  journal = {Physical Review C},
  volume = {110},
  issue = {6},
  pages = {064310},
  numpages = {9},
  year = {2024},
  month = {Dec},
  publisher = {American Physical Society},
  doi = {10.1103/PhysRevC.110.064310},
  url = {https://link.aps.org/doi/10.1103/PhysRevC.110.064310}
}

@article{Wu2024PRC1,
  title = {Nuclear mass predictions with anisotropic kernel ridge regression},
  author = {Wu, X. H. and Pan, C.},
  journal = {Physical Review C},
  volume = {110},
  issue = {3},
  pages = {034322},
  numpages = {8},
  year = {2024},
  month = {Sep},
  publisher = {American Physical Society},
  doi = {10.1103/PhysRevC.110.034322},
  url = {https://link.aps.org/doi/10.1103/PhysRevC.110.034322}
}

@article{Tian2025PRC,
  title = {Enhanced nuclear mass predictions using anisotropic kernel ridge regression incorporating multiple complementary observables},
  author = {Tian, Junlong and Ma, Pengfei and Wu, Xinhui and Hu, Minghui and Li, Cheng and Wang, Ning},
  journal = {Physical Review C},
  volume = {112},
  issue = {6},
  pages = {064306},
  numpages = {9},
  year = {2025},
  month = {Dec},
  publisher = {American Physical Society},
  doi = {10.1103/9mgr-6mq7},
  url = {https://link.aps.org/doi/10.1103/9mgr-6mq7}
}

@article{Yuksel2024PRC,
  title = {Nuclear mass predictions using machine learning models},
  author = {Y\"uksel, Esra and Soydaner, Derya and Bahtiyar, H\"useyin},
  journal = {Physical Review C},
  volume = {109},
  issue = {6},
  pages = {064322},
  numpages = {11},
  year = {2024},
  month = {Jun},
  publisher = {American Physical Society},
  doi = {10.1103/PhysRevC.109.064322},
  url = {https://link.aps.org/doi/10.1103/PhysRevC.109.064322}
}

@article{ Yuan2024NST,
Author = {Yuan, Zi-Yi and Bai, Dong and Wang, Zhen and Ren, Zhong-Zhou},
Title = {Reliable calculations of nuclear binding energies by the Gaussian
   process of machine learning},
Journal = {Nuclear Science and Techniques},
Year = {2024},
Volume = {35},
Number = {6},
Month = {JUN},
DOI = {10.1007/s41365-024-01463-9},
Article-Number = {105},
ISSN = {1001-8042},
EISSN = {2210-3147},
ResearcherID-Numbers = {Wang, Zhen/GQH-4924-2022
   Bai, Dong/K-2622-2018
   yuan, ziyi/LUY-8406-2024},
ORCID-Numbers = {Bai, Dong/0000-0001-7116-721X
   },
Unique-ID = {WOS:001250623500014},
}

@article{Ye2025PRC,
  title = {Understanding on prediction differences among theoretical mass models with machine learning techniques},
  author = {Ye, Weihu and Wan, Niu},
  journal = {Physical Review C},
  volume = {111},
  issue = {4},
  pages = {044317},
  numpages = {9},
  year = {2025},
  month = {Apr},
  publisher = {American Physical Society},
  doi = {10.1103/PhysRevC.111.044317},
  url = {https://link.aps.org/doi/10.1103/PhysRevC.111.044317}
}

@article{Jalili2025EPJA,
Author = {Jalili, Amir and Saleki, Ziba and Luo, Y. A. and Pan, Feng and Chen, Ai.
   Xi. and Draayer, Jerry P.},
Title = {Performance of various kernel functions for mass prediction with support
   vector machine},
Journal = {EUROPEAN PHYSICAL JOURNAL A},
Year = {2025},
Volume = {61},
Number = {6},
Month = {JUN 19},
DOI = {10.1140/epja/s10050-025-01610-9},
Article-Number = {143},
ISSN = {1434-6001},
EISSN = {1434-601X},
ORCID-Numbers = {aayer, Jerry/0000-0003-3568-8223
   Jalili, Amir/0000-0002-0280-3427},
Unique-ID = {WOS:001511850300002},
}

@article{Ye2026PRC,
  title = {Simultaneous improvements of nuclear mass and charge radius predictions using multitask Gaussian process approaches},
  author = {Ye, Weihu and Wan, Niu},
  journal = {Physical Review C},
  volume = {113},
  issue = {2},
  pages = {024304},
  numpages = {10},
  year = {2026},
  month = {Feb},
  publisher = {American Physical Society},
  doi = {10.1103/1mgv-jypl},
  url = {https://link.aps.org/doi/10.1103/1mgv-jypl}
}

@article{ LiuGP2025PRC,
Author = {Liu, Guo-Ping and Wang, Hua-Lei and Zhang, Zhen-Zhen and Liu, Min-Liang},
Title = {Model-repair capabilities of tree-based machine-learning algorithms
   applied to theoretical nuclear mass models},
Journal = {Physical Review C},
Year = {2025},
Volume = {111},
Number = {2},
Month = {FEB 6},
DOI = {10.1103/PhysRevC.111.024306},
Article-Number = {024306},
ISSN = {2469-9985},
EISSN = {2469-9993},
ResearcherID-Numbers = {Liu, Guo-Ping/O-3511-2014},
ORCID-Numbers = {Hua-lei, Wang/0000-0001-5173-6980
   Liu, Guo-Ping/0000-0002-0699-2296},
Unique-ID = {WOS:001451343100002},
}

@article{ LiuH2025PRC,
Author = {Liu, Hao and Lei, Jin and Ren, Zhongzhou},
Title = {Kolmogorov-Arnold networks in nuclear binding energy prediction},
Journal = {Physical Review C},
Year = {2025},
Volume = {111},
Number = {2},
Month = {FEB 25},
DOI = {10.1103/PhysRevC.111.024316},
Article-Number = {024316},
ISSN = {2469-9985},
EISSN = {2469-9993},
ResearcherID-Numbers = {Lei, Jin/P-5159-2015},
ORCID-Numbers = {Liu, Hao/0009-0003-6424-4290
   Lei, Jin/0000-0002-2323-2061},
Unique-ID = {WOS:001451308200001},
}

@article{ Guo2025PRC,
Author = {Guo, Jin-Liang and Wang, Hua-Lei and Zhang, Zhen-Zhen and Liu, Min-Liang},
Title = {Probing the refined performance of the categorical-boosting algorithm to
   the Hartree-Fock-Bogoliubov mass model with several Skyrme forces},
Journal = {Physical Review C},
Year = {2025},
Volume = {111},
Number = {5},
Month = {MAY 29},
DOI = {10.1103/PhysRevC.111.054322},
Article-Number = {054322},
ISSN = {2469-9985},
EISSN = {2469-9993},
ORCID-Numbers = {Hua-lei, Wang/0000-0001-5173-6980},
Unique-ID = {WOS:001500430300001},
}

@article{ Wu2024SCPMA,
Author = {Wu, Xin-Hui and Zhao, Pengwei},
Title = {Principal components of nuclear mass models},
Journal = {SCIENCE CHINA-PHYSICS MECHANICS \& ASTRONOMY},
Year = {2024},
Volume = {67},
Number = {7},
Month = {JUL},
DOI = {10.1007/s11433-023-2342-4},
Article-Number = {272011},
ISSN = {1674-7348},
EISSN = {1869-1927},
ResearcherID-Numbers = {Zhao, Pengwei/F-9107-2010
   Wu, Xin-Hui/LLL-7848-2024},
ORCID-Numbers = {Zhao, Pengwei/0000-0001-8243-2381
   Wu, Xin-Hui/0000-0003-0237-5853},
Unique-ID = {WOS:001232254600001},
}

@article{Saito2024PRC,
  title = {Uncertainty quantification of mass models using ensemble Bayesian model averaging},
  author = {Saito, Yukiya and Dillmann, I. and Kr\"ucken, R. and Mumpower, M. R. and Surman, R.},
  journal = {Physical Review C},
  volume = {109},
  issue = {5},
  pages = {054301},
  numpages = {14},
  year = {2024},
  month = {May},
  publisher = {American Physical Society},
  doi = {10.1103/PhysRevC.109.054301},
  url = {https://link.aps.org/doi/10.1103/PhysRevC.109.054301}
}

@article{Zhang2024NPA,
title = {Nuclear mass predictions with the naive Bayesian model averaging method},
journal = {Nucl. Phys.  A},
volume = {1043},
pages = {122820},
year = {2024},
issn = {0375-9474},
doi = {https://doi.org/10.1016/j.nuclphysa.2024.122820},
url = {https://www.sciencedirect.com/science/article/pii/S0375947424000022},
author = {X.Y. Zhang and W.F. Li and J.Y. Fang and Z.M. Niu}
}

@article{ Zhang2024PRC,
Author = {Zhang, X. Y. and Liu, H. R. and Liu, L. L. and Niu, Z. M. and Huang, X.
   L. and Niu, Y. F.},
Title = {Power-moderated mean method in nuclear mass predictions},
Journal = {Physical Review C},
Year = {2024},
Volume = {110},
Number = {4},
Month = {OCT 8},
DOI = {10.1103/PhysRevC.110.044307},
Article-Number = {044307},
ISSN = {2469-9985},
EISSN = {2469-9993},
ResearcherID-Numbers = {Niu, ZhongMing/I-1288-2012
   Niu, Yifei/J-9686-2013},
ORCID-Numbers = {Xiaolong, Huang/0000-0002-9695-0996
   Zhang, Xiaoyan/0009-0001-8376-7076
   },
Unique-ID = {WOS:001331653200002},
}

@article{Wu2026PLB,
title = {Principal components of nuclear mass model residuals},
journal = {Physics Letters B},
volume = {874},
pages = {140262},
year = {2026},
issn = {0370-2693},
doi = {https://doi.org/10.1016/j.physletb.2026.140262},
url = {https://www.sciencedirect.com/science/article/pii/S0370269326001164},
author = {Y. Y. Huang and X. H. Wu}
}

@article{XXDong2023PLB,
    author = "Dong, Xiao-Xu and An, Rong and Lu, Jun-Xu and Geng, Li-Sheng",
    title = "{Nuclear charge radii in Bayesian neural networks revisited}",
    eprint = "2206.13169",
    archivePrefix = "arXiv",
    primaryClass = "nucl-th",
    doi = "10.1016/j.physletb.2023.137726",
    journal = "Physics Letters B",
    volume = "838",
    pages = "137726",
    year = "2023"
}

@article{ZYXian2025PLB,
    author = "Xian, Zhen-Yan and Ya, Yan and An, Rong",
    title = "{Shell quenching in nuclear charge radii based on Monte Carlo dropout Bayesian neural network}",
    eprint = "2410.15784",
    archivePrefix = "arXiv",
    primaryClass = "nucl-th",
    doi = "10.1016/j.physletb.2025.139662",
    journal = "Physics Letters B",
    volume = "868",
    pages = "139662",
    year = "2025"
}

@article{XZhang2024PRC,
    author = "Zhang, X. and He, H. and Qu, G. and Liu, X. and Zheng, H. and Lin, W. and Han, J. and Ren, P. and Wada, R.",
    title = "{Investigation of the difference in charge radii of mirror pairs with deep Bayesian neural networks}",
    doi = "10.1103/PhysRevC.110.014316",
    journal = "Physical Review C",
    volume = "110",
    number = "1",
    pages = "014316",
    year = "2024"
}

@article{ZXYang2023PRC,
    author = "Yang, Zu-Xing and Fan, Xiao-Hua and Naito, Tomoya and Niu, Zhong-Ming and Li, Zhi-Pan and Liang, Haozhao",
    title = "{Calibration of nuclear charge density distribution by back-propagation neural networks}",
    eprint = "2205.15649",
    archivePrefix = "arXiv",
    primaryClass = "nucl-th",
    reportNumber = "RIKEN-iTHEMS-Report-22",
    doi = "10.1103/PhysRevC.108.034315",
    journal = "Physical Review C",
    volume = "108",
    number = "3",
    pages = "034315",
    year = "2023"
}

@article{ZXYang2023PLB,
    author = "Yang, Zu-Xing and Fan, Xiao-Hua and Li, Zhi-Pan and Liang, Haozhao",
    title = "{A Kohn-Sham scheme based neural network for nuclear systems}",
    eprint = "2212.02093",
    archivePrefix = "arXiv",
    primaryClass = "nucl-th",
    reportNumber = "RIKEN-iTHEMS-Report-23",
    doi = "10.1016/j.physletb.2023.137870",
    journal = "Physics Letters B",
    volume = "840",
    pages = "137870",
    year = "2023"
}

@article{ Shang2024PRC,
Author = {Shang, Tian Shuai and Xie, Hui Hui and Li, Jian and Liang, Haozhao},
Title = {Global prediction of nuclear charge density distributions using a deep
   neural network},
Journal = {Physical Review C},
Year = {2024},
Volume = {110},
Number = {1},
Month = {JUL 2},
DOI = {10.1103/PhysRevC.110.014308},
Article-Number = {014308},
ISSN = {2469-9985},
EISSN = {2469-9993},
ResearcherID-Numbers = {Liang, Haozhao/A-6747-2010
   Li, Jian/KEH-0749-2024
   },
ORCID-Numbers = {Shang, Tian-Shuai/0009-0006-1695-6906
   Liang, Haozhao/0000-0002-2950-8559
   Li, Jian/0000-0002-0864-5108
   Xie, Hui Hui/0000-0002-6185-0856},
Unique-ID = {WOS:001263213600005},
}

@article{YYCao2023NST,
	author = {Cao, Ying-Yu and Guo, Jian-You and Zhou, Bo},
	journal = {Nuclear Science and Techniques},
	number = {10},
	pages = {152},
	title = {Predictions of nuclear charge radii based on the convolutional neural network},
	volume = {34},
	year = {2023}}

@article{ Su2023Symmetry,
Author = {Su, Ping and He, Wan-Bing and Fang, De-Qing},
Title = {Progress of Machine Learning Studies on the Nuclear Charge Radii},
Journal = {SYMMETRY-BASEL},
Year = {2023},
Volume = {15},
Number = {5},
Month = {MAY 8},
DOI = {10.3390/sym15051040},
Article-Number = {1040},
Unique-ID = {WOS:000996864600001},
}

@article{WFLi2024PS,
    author = "Li, Weifeng and Zhang, Xiaoyan and Fang, Jiyu",
    title = "{Nuclear charge radius predictions based on eXtreme Gradient Boosting}",
    doi = "10.1088/1402-4896/ad3170",
    journal = "Phys. Scripta",
    volume = "99",
    number = "4",
    pages = "045308",
    year = "2024"
}

@article{ ZLLi2025PRC,
Author = {Li, Zhilong and Wang, Yongjia and Li, Qingfeng and Lv, Bing-feng},
Title = {Machine-learning predictions for the nuclear charge radius: Bayesian
   method versus decision-tree-based algorithm},
Journal = {Physical Review C},
Year = {2025},
Volume = {112},
Number = {1},
Month = {JUL 11},
DOI = {10.1103/vj25-zwd3},
Article-Number = {014312},
Unique-ID = {WOS:001537511200001},
}

@article{ Maheshwari2026PLB,
Author = {Maheshwari, B. and Van Isacker, P.},
Title = {Understanding charge radii with machine learning: Discovering physics
   expressions},
Journal = {PHYSICS LETTERS B},
Year = {2026},
Volume = {872},
Month = {JAN},
DOI = {10.1016/j.physletb.2025.140102},
EarlyAccessDate = {DEC 2025},
Article-Number = {140102},
Unique-ID = {WOS:001651631200001},
}

@article{ LiuJ2025NST,
Author = {Liu, Jian and Tan, Kai-Zhong and Wang, Lei and Gao, Wan-Qing and Shang,
   Tian-Shuai and Li, Jian and Xu, Chang},
Title = {Continuous Bayesian probability estimator in predictions of nuclear
   charge radii},
Journal = {Nuclear Science and Techniques},
Year = {2025},
Volume = {36},
Number = {11},
Month = {AUG 30},
DOI = {10.1007/s41365-025-01792-3},
Article-Number = {215},
Unique-ID = {WOS:001560318400002},
}

@article{ Li2024JPG,
Author = {Li, W. F. and Zhang, X. Y. and Niu, Y. F. and Niu, Z. M.},
Title = {Comparative study of neural network and model averaging methods in
   nuclear β-decay half-life predictions},
Journal = {JOURNAL OF PHYSICS G-NUCLEAR AND PARTICLE PHYSICS},
Year = {2024},
Volume = {51},
Number = {1},
Month = {JAN 1},
DOI = {10.1088/1361-6471/ad0314},
Article-Number = {015103},
Unique-ID = {WOS:001115005500001},
}

@article{ Li2025NST,
Author = {Li, Peng and Niu, Zhong-Ming and Niu, Yi-Fei},
Title = {Selection of abnormal trends in nuclear β-decay half-lives by neural
   network and exploration of the physical mechanisms},
Journal = {Nuclear Science and Techniques},
Year = {2025},
Volume = {36},
Number = {3},
Month = {FEB 13},
DOI = {10.1007/s41365-025-01663-x},
Article-Number = {50},
Unique-ID = {WOS:001421771800001},
}

@article{ Jalili2025PRC,
Author = {Jalili, Amir and Pan, Feng and Luo, Y. A. and Draayer, Jerry P.},
Title = {Nuclear β-decay half-life predictions and r-process
   nucleosynthesis using machine learning models},
Journal = {Physical Review C},
Year = {2025},
Volume = {111},
Number = {3},
Month = {MAR 14},
DOI = {10.1103/PhysRevC.111.034321},
Article-Number = {034321},
Unique-ID = {WOS:001459110300001},
}

@article{ Wen2023APS,
Author = {Wen, Hu-Feng and Shang, Tian-Shuai and Li, Jian and Niu, Zhong-Ming and
   Yang, Dong and Xue, Yong-He and Li, Xiang and Huang, Xiao-Long},
Title = {Prediction of ground-state spin in odd-A nuclei within decision tree},
Journal = {ACTA PHYSICA SINICA},
Year = {2023},
Volume = {72},
Number = {15},
Month = {AUG 5},
DOI = {10.7498/aps.72.20230530},
Article-Number = {152101},
Unique-ID = {WOS:001051263500012},
}

@article{ Liu2024NST,
Author = {Liu, Deng and Noor, Alam A. and Qin, Zhen-Zhen and Lei, Yang},
Title = {Neural network study of the nuclear ground-state spin distribution
   within a random interaction ensemble},
Journal = {Nuclear Science and Techniques},
Year = {2024},
Volume = {35},
Number = {3},
Month = {MAR},
DOI = {10.1007/s41365-024-01424-2},
Article-Number = {64},
ISSN = {1001-8042},
EISSN = {2210-3147},
ResearcherID-Numbers = {Qin, Zhenzhen/M-5789-2017
   Lei, Yang/A-1210-2017},
ORCID-Numbers = {NOOR A, ALAM/0009-0005-4157-3942
   Lei, Yang/0000-0002-7373-5759},
Unique-ID = {WOS:001214204700012},
}

@article{ Gao2024JPG,
Author = {Gao, T. J. and Wang, H. D. and Lu, Jing-Bin and Lu, Yi and Yang, Pei-Yao
   and Qin, M. J.},
Title = {Prediction of the 1st excitation energy of odd-odd nuclei with the
   Bayesian neural network approach},
Journal = {JOURNAL OF PHYSICS G-NUCLEAR AND PARTICLE PHYSICS},
Year = {2024},
Volume = {51},
Number = {8},
Month = {AUG 1},
DOI = {10.1088/1361-6471/ad5196},
Article-Number = {085101},
Unique-ID = {WOS:001251659500001},
}

@article{ Lv2024PLB,
Author = {Lv, B. F. and Li, Z. L. and Wang, Y. J. and Petrache, C. M.},
Title = {Mapping low-lying states and B(E2; 0 1+ →
   2 1+) in even-even nuclei with machine learning},
Journal = {PHYSICS LETTERS B},
Year = {2024},
Volume = {857},
Month = {OCT},
DOI = {10.1016/j.physletb.2024.139013},
EarlyAccessDate = {SEP 2024},
Article-Number = {139013},
Unique-ID = {WOS:001315486600001},
}

@article{ Lv2025PRC,
Author = {Lv, B. F. and Wang, Yongjia and Li, Zhilong and Petrache, C. M.},
Title = {Machine learning for decoding spin-zero and octupole excitations},
Journal = {Physical Review C},
Year = {2025},
Volume = {111},
Number = {6},
Month = {JUN 26},
DOI = {10.1103/vjwy-m9xv},
Article-Number = {064324},
Unique-ID = {WOS:001523563800003},
}

@article{Liu2025NST,
Author = {Liu, Hui and Li, Xin-Xiang and Yuan, Yun and Luo, Wen and Xu, Yi},
Title = {Prediction of the first 2+ states properties for atomic
   nuclei using light gradient boosting machine},
Journal = {Nuclear Science and Techniques},
Year = {2025},
Volume = {36},
Number = {2},
Month = {JAN 9},
DOI = {10.1007/s41365-024-01613-z},
Article-Number = {21},
Unique-ID = {WOS:001394257000003},
}

@article{ Zhang2026EPJP,
Author = {Zhang, Xin and Fortunato, Lorenzo},
Title = {Prediction of 2+energies in even-even nuclei of the Terra
   incognita with Bayesian neural network},
Journal = {EUROPEAN PHYSICAL JOURNAL PLUS},
Year = {2026},
Volume = {141},
Number = {2},
Month = {FEB 8},
DOI = {10.1140/epjp/s13360-026-07362-9},
Article-Number = {121},
Unique-ID = {WOS:001683464000001},
}

@article{ Du2024PRC,
Author = {Du, Peng-Xiang and Shang, Tian-Shuai and Geng, Kun-Peng and Li, Jian and
   Fang, Dong-Liang},
Title = {Inference of parameters for the back-shifted Fermi gas model using a
   feedforward neural network},
Journal = {Physical Review C},
Year = {2024},
Volume = {109},
Number = {4},
Month = {APR 4},
DOI = {10.1103/PhysRevC.109.044325},
Article-Number = {044325},
ISSN = {2469-9985},
EISSN = {2469-9993},
ResearcherID-Numbers = {Du, PengXiang/NKP-4049-2025
   Li, Jian/KEH-0749-2024},
ORCID-Numbers = {Shang, Tian-Shuai/0009-0006-1695-6906
   Du, PengXiang/0009-0003-8404-2418
   Li, Jian/0000-0002-0864-5108},
Unique-ID = {WOS:001224120400004},
}

@article{ Wang2024CPC,
Author = {Wang, Xinyu and Cui, Ying and Tian, Yuan and Zhao, Kai and Zhang,
   Yingxun},
Title = {Uncertainties of nuclear level density estimated using Bayesian neural
   networks},
Journal = {Chinese Physics C},
Year = {2024},
Volume = {48},
Number = {8},
Month = {AUG 1},
DOI = {10.1088/1674-1137/ad47a7},
Article-Number = {084105},
ISSN = {1674-1137},
EISSN = {2058-6132},
ResearcherID-Numbers = {Zhang, Yingxun/AAN-5704-2021
   Zhao, kai/K-2229-2019
   },
ORCID-Numbers = {Cui, Ying/0000-0002-5156-5306},
Unique-ID = {WOS:001253574200001},
}

@article{ Zhao2026ApJ,
Author = {Zhao, Tian Liang and Diao, Yi Xiang and Bao, Xiao Jun},
Title = {Precise Inference of Nuclear Level Density Parameters Far from the
   Stable Line in Nuclear Astrophysics: The Advantages and Verification of
   the Physical-constraint Neural Network},
Journal = {ASTROPHYSICAL JOURNAL},
Year = {2026},
Volume = {1001},
Number = {1},
Month = {APR 10},
DOI = {10.3847/1538-4357/ae518a},
Article-Number = {51},
ISSN = {0004-637X},
EISSN = {1538-4357},
Unique-ID = {WOS:001731873000001},
}

@article{ Jyothish2025PRC,
Author = {Jyothish, K. and Manangode, Govardhan and Kumar, A. K. Rhine},
Title = {Transfer-learning-driven machine-learning model for α-decay half-life
   predictions},
Journal = {Physical Review C},
Year = {2025},
Volume = {112},
Number = {6},
Month = {DEC 8},
DOI = {10.1103/67c1-2dvd},
Article-Number = {064309},
ISSN = {2469-9985},
EISSN = {2469-9993},
ORCID-Numbers = {K, Jyothish/0009-0002-0609-5090
   A K, Rhine Kumar/0000-0002-9340-0796},
Unique-ID = {WOS:001642130300009},
}

@article{ Shree2025EPJAa,
Author = {Shree, S. Madhumitha and Balasubramaniam, M.},
Title = {α-decay half-life predictions for superheavy elements through machine
   learning techniques},
Journal = {EUROPEAN PHYSICAL JOURNAL A},
Year = {2025},
Volume = {61},
Number = {2},
Month = {FEB 16},
DOI = {10.1140/epja/s10050-025-01494-9},
Article-Number = {32},
ISSN = {1434-6001},
EISSN = {1434-601X},
ResearcherID-Numbers = {Balasubramaniam, M./J-5797-2012},
ORCID-Numbers = {, Madhumitha Shree S/0009-0007-1863-2478
   },
Unique-ID = {WOS:001423140800001},
}

@article{ Shree2025EPJAb,
Author = {Madhumitha Shree, S. and Balasubramaniam, M.},
Title = {Kolmogorov-Arnold networks for empirical modeling of α-decay half-lives
   in superheavy nuclei},
Journal = {EUROPEAN PHYSICAL JOURNAL A},
Year = {2025},
Volume = {61},
Number = {12},
Month = {DEC 1},
DOI = {10.1140/epja/s10050-025-01710-6},
Article-Number = {272},
ISSN = {1434-6001},
EISSN = {1434-601X},
ResearcherID-Numbers = {Balasubramaniam, M./J-5797-2012},
Unique-ID = {WOS:001628109300002},
}

@article{ You2025NST,
Author = {You, Hong-Qiang and He, Xiao-Tao and Wu, Ren-Hang and Zhang,
   Shuang-Shuang and Li, Jing-Jing and He, Qing-Hua and Zhang, Hai-Qian},
Title = {Nuclear deformation effects on α-decay half-lives with empirical formula
   and machine learning},
Journal = {Nuclear Science and Techniques},
Year = {2025},
Volume = {36},
Number = {10},
Month = {JUL 24},
DOI = {10.1007/s41365-025-01766-5},
Article-Number = {191},
ISSN = {1001-8042},
EISSN = {2210-3147},
ResearcherID-Numbers = {You, Hongqiang/OIR-8277-2025
   He, Xiao-tao/GWQ-4134-2022
   Qinghua, He/ABF-8560-2020},
ORCID-Numbers = {You, Hongqiang/0009-0004-4403-6934
   },
Unique-ID = {WOS:001539316400002},
}

@article{ Shree2026NPA,
Author = {Shree, S. Madhumitha and Balasubramaniam, M.},
Title = {Machine learning and symbolic regression-based modeling of α-decay
   half-lives for superheavy nuclei},
Journal = {Nuclear Physics A},
Year = {2026},
Volume = {1068},
Month = {APR},
DOI = {10.1016/j.nuclphysa.2026.123327},
EarlyAccessDate = {JAN 2026},
Article-Number = {123327},
ISSN = {0375-9474},
EISSN = {1873-1554},
ResearcherID-Numbers = {Balasubramaniam, M./J-5797-2012},
Unique-ID = {WOS:001674446200002},
}

@article{ Yang2026PRC,
Author = {Yang, Haitao and Li, Xiaopan and Song, Xiefei and Ma, Dianxu and Yu,
   Gongming and Bao, Xiaojun},
Title = {α-decay half-lives of superheavy nuclei with support-vector regression},
Journal = {Physical Review C},
Year = {2026},
Volume = {113},
Number = {1},
Month = {JAN 9},
DOI = {10.1103/q2bb-1cjn},
Article-Number = {014307},
ISSN = {2469-9985},
EISSN = {2469-9993},
ResearcherID-Numbers = {Song, Xiefei/LVR-9775-2024},
Unique-ID = {WOS:001669420100003},
}

@article{ Bairwa2025PS,
Author = {Bairwa, Manish Kumar and Abhishek, R. and Balasubramanian, R. and
   Arumugam, P.},
Title = {Integrating physics insights into machine learning: a case study with
   giant dipole resonance},
Journal = {PHYSICA SCRIPTA},
Year = {2025},
Volume = {100},
Number = {5},
Month = {MAY 1},
DOI = {10.1088/1402-4896/adc5b5},
Article-Number = {056010},
Unique-ID = {WOS:001464688700001},
}

@article{ Mehta2019PRp,
Author = {Mehta, Pankaj and Bukov, Marin and Wang, Ching-Hao and Day, Alexandre G.
   R. and Richardson, Clint and Fisher, Charles K. and Schwab, David J.},
Title = {A high-bias, low-variance introduction to Machine Learning for
   physicists},
Journal = {PHYSICS REPORTS-REVIEW SECTION OF PHYSICS LETTERS},
Year = {2019},
Volume = {810},
Pages = {1-124},
Month = {MAY 30},
DOI = {10.1016/j.physrep.2019.03.001},
ISSN = {0370-1573},
EISSN = {1873-6270},
ResearcherID-Numbers = {Schwab, David/B-7498-2012
   Bukov, Marin/AAD-7407-2022},
ORCID-Numbers = {Bukov, Marin/0000-0002-3688-9599},
Unique-ID = {WOS:000471739400001},
}

@article{Hermann2023NRC,
	author = {Hermann, Jan and Spencer, James and Choo, Kenny and Mezzacapo, Antonio and Foulkes, W. M. C. and Pfau, David and Carleo, Giuseppe and No{\'e}, Frank},
	journal = {Nature Reviews Chemistry},
	number = {10},
	pages = {692--709},
	title = {Ab initio quantum chemistry with neural-network wavefunctions},
	volume = {7},
	year = {2023}}

@article{ Wang2025EPJA,
Author = {Wang, Chuanxin and Naito, Tomoya and Li, Jian and Liang, Haozhao},
Title = {A deep neural network approach to solve the Dirac equation},
Journal = {EUROPEAN PHYSICAL JOURNAL A},
Year = {2025},
Volume = {61},
Number = {7},
Month = {JUL 15},
DOI = {10.1140/epja/s10050-025-01630-5},
Article-Number = {162},
ISSN = {1434-6001},
EISSN = {1434-601X},
ResearcherID-Numbers = {Li, Jian/KEH-0749-2024
   Liang, Haozhao/A-6747-2010
   Naito, Tomoya/H-4962-2018},
ORCID-Numbers = {Naito, Tomoya/0000-0002-0010-3558},
Unique-ID = {WOS:001529238200001},
}

@article{ Du2026CTP,
Author = {Du, Xiao-Kai and Zhang, Shuang-Quan},
Title = {Solving nuclear Dirac Woods-Saxon potential with a physics-informed
   neural network},
Journal = {COMMUNICATIONS IN THEORETICAL PHYSICS},
Year = {2026},
Volume = {78},
Number = {3},
Month = {MAR 1},
DOI = {10.1088/1572-9494/ae1a5c},
Article-Number = {035303},
ISSN = {0253-6102},
EISSN = {1572-9494},
ResearcherID-Numbers = {Zhang, Shuangquan/B-3838-2012
   },
ORCID-Numbers = {Zhang, Shuangquan/0000-0002-9590-1818
   Du, Xiaokai/0000-0001-5977-1326},
Unique-ID = {WOS:001639115700001},
}

@article{Wu2022PRCDFT,
    author = "Wu, X. H. and Ren, Z. X. and Zhao, P. W.",
    title = "{Nuclear energy density functionals from machine learning}",
    eprint = "2105.07696",
    archivePrefix = "arXiv",
    primaryClass = "nucl-th",
    doi = "10.1103/PhysRevC.105.L031303",
    journal = "Physical Review C",
    volume = "105",
    number = "3",
    pages = "L031303",
    year = "2022"
}

@article{ Wu2025CP,
Author = {Wu, X. H. and Ren, Z. X. and Zhao, P. W.},
Title = {Machine learning orbital-free density functional theory resolves shell
   effects in deformed nuclei},
Journal = {COMMUNICATIONS PHYSICS},
Year = {2025},
Volume = {8},
Number = {1},
Month = {AUG 1},
DOI = {10.1038/s42005-025-02234-7},
Article-Number = {316},
ISSN = {2399-3650},
ResearcherID-Numbers = {Wu, Xin-Hui/AAZ-6225-2021
   Zhao, Pengwei/F-9107-2010
   Ren, Zhixiang/IQS-1889-2023},
ORCID-Numbers = {Wu, Xin-Hui/0000-0003-0237-5853
   Zhao, Pengwei/0000-0001-8243-2381
   },
Unique-ID = {WOS:001541736100001},
}

\end{multicols}
\end{document}